\documentclass[aps,showpacs,twocolumn,prd,superscriptaddress,nofootinbib]{revtex4}

\usepackage{amsmath}
\usepackage{epsfig}
\usepackage{graphicx}
\usepackage{hyperref}
\usepackage{url}
\usepackage{color}

\newcommand{\Msun}{\>{\rm M_{\odot}}}
\newcommand{\beq}{\begin{equation}}
\newcommand{\eeq}{\end{equation}}
\newcommand{\bea}{\begin{eqnarray}}
\newcommand{\eea}{\end{eqnarray}}

\newcommand{\imagi}{i}

\def\leq{\raise 0.4ex\hbox{$<$}\kern -0.8em\lower 0.62ex\hbox{$-$}}
\def\geq{\raise 0.4ex\hbox{$>$}\kern -0.7em\lower 0.62ex\hbox{$-$}}
\def\lsim{\raise 0.4ex\hbox{$<$}\kern -0.8em\lower 0.62ex\hbox{$\sim$}}
\def\gsim{\raise 0.4ex\hbox{$>$}\kern -0.7em\lower 0.62ex\hbox{$\sim$}}
\def\appropto{\raise 0.4ex\hbox{$\propto$}\kern -0.7em\lower 0.62ex\hbox{$\sim$}}
\def\pm{\,\raise 0.4ex\hbox{$+$}\kern -0.8em\lower 0.62ex\hbox{$-$}\,}

\begin{document}

\title{Observing complete gravitational wave signals from dynamical capture binaries}
\author{William E. East}
\email{weast@princeton.edu}
\affiliation{Department of Physics, Princeton University, Princeton, New Jersey 08544, USA}
\author{Sean T. McWilliams}
\email{stmcwill@princeton.edu}
\affiliation{Department of Physics, Princeton University, Princeton, New Jersey 08544, USA}
\affiliation{Institute for Strings, Cosmology and Astroparticle Physics, Columbia University, New York, New York 10027, USA}
\author{Janna Levin}
\affiliation{Institute for Strings, Cosmology and Astroparticle Physics, Columbia University, New York, New York 10027, USA}
\affiliation{Department of Physics and Astronomy, Barnard College of Columbia University, 3009 Broadway, New York, New York 10027, USA}
\author{Frans Pretorius}
\affiliation{Department of Physics, Princeton University, Princeton, NJ 08544}

\date{\today}


\begin{abstract}
We assess the detectability of the gravitational wave signals from highly eccentric compact binaries. 
We use a simple model for the inspiral, merger, and ringdown of these systems.
The model is based on mapping the binary to an effective single black hole system
described by a Kerr metric, thereby including certain relativistic effects
such as zoom-whirl-type behavior.
The resultant geodesics
source quadrupolar radiation and, in turn, are evolved under its dissipative effects. 
At the light ring, we attach a merger model that 
was previously developed for quasicircular mergers but also performs well for eccentric mergers with little modification.
We apply this model to determine the detectability of these sources for initial, Enhanced, and Advanced LIGO
across the parameter space of nonspinning close capture compact binaries.  We conclude that, should these systems
exist in nature, the vast majority will be missed by conventional burst searches or by quasicircular waveform templates in the advanced detector
era. Other methods, such as eccentric templates or, more practically, a stacked excess power search, must be developed to avoid
losing these sources.  These systems would also have been missed frequently in the initial 
LIGO data analysis. Thus, previous null coincidence
results with detected gamma-ray bursts cannot exclude the possibility of coincident gravitational wave signals from eccentric binaries.
\end{abstract}


\maketitle

\section{Introduction}
\label{sec:intro}
In dense stellar regions, such as galactic nuclei or globular clusters,
individual black holes (BHs) or neutrons stars (NSs) can
become gravitationally bound
as energy is lost to gravitational radiation during a close passage.
These dynamically captured pairs
may be additional sources for gravitational wave (GW) detectors, as
well as sources of electromagnetic (EM) transients such as short gamma-ray bursts (SGRBs).
Eccentric pairs will be distinguishable from quasicircular coeval binaries,
which are born in a bound system and have had time to circularize before reaching the sensitive bandwidths of
ground-based GW observatories such as LIGO~\cite{Abramovici:1992ah}, 
VIRGO~\cite{Caron:1997hu}, GEO600~\cite{Willke:2002bs}, and KAGRA(LCGT)~\cite{Ohashi:2011zz}.

The primary purpose of this paper is to study
the detectability of sources that retain eccentricity while in the LIGO band (for simplicity, we only employ LIGO sensitivity curves).
Before getting into the details of our model
and results, we briefly review current event rate estimates, high-eccentricity population fractions, possible 
EM counterparts, and the GW detectability of dynamically captured compact objects.

\subsection{Event rates}

Galactic nuclei are a promising setting for the formation of dynamical capture binaries.
Mass segregation around a central massive BH can lead to large densities of stellar mass BHs and stars.  
For example, the Fokker-Planck model used in~\cite{hopman06} suggests that our 
galactic nucleus should have $\sim2000$ BHs and $\sim400$ NSs in the central 0.1~pc.
In ~\cite{O'Leary:2008xt,Kocsis_Levin}, the event rate for the formation of BH-BH binaries from GW capture in this setting was estimated to be roughly
between 0.01 and 1.0 yr$^{-1}$ Gpc$^{-3}$, with corresponding Advanced LIGO detection rates
of $\approx 1-10^2$ yr$^{-1}$.
This rate assumes that the number density $n$ of BHs in galactic nuclei has a scatter
with $\langle n^2 \rangle/\langle n\rangle ^2=30$.  Assuming no scatter would reduce the above rate by a factor of 30.  This also assumes
a number density of contributing galaxies of $0.05$ Mpc$^{-3}$; i.e. it includes all galaxies as contributing
roughly equally.  Lower mass galaxies are not as well understood,
though if a significant number of them have total cluster mass fractions above the $2.5\%$ used in the aforementioned
calculation, this rate would increase.  Other unaccounted for effects, such as steeper profiles from light-dominated 
mass functions~\cite{1538-4357-698-1-L64}, could also potentially increase this rate.
The formation of BH-NS binaries is estimated to be $\sim1\%$ of this rate~\cite{O'Leary:2008xt}.

Dynamical capture binaries may also form in globular clusters (GCs) that undergo core collapse~\cite{fabian75,pooley2003}.
In~\cite{lee2010} binary formation through tidal capture was studied. Using M15 as a prototypical GC, it was calculated that
the NS-NS tidal capture rate would peak at $\sim50$ yr$^{-1}$ 
Gpc$^{-3}$ at $z=0.7$ (falling to $\sim30$ yr$^{-1}$ Gpc$^{-3}$ by $z=0$) for their default model of core collapse.
They also provide a scaling to BH-NS and BH-BH mergers which 
(assuming $M_{\rm BH}=4.5 M_{\odot}$ and a relative fraction of BHs to NSs $f_{\rm BH}/f_{\rm NS}\approx 0.28$) 
gives rates that peak at $\sim70~$yr${}^{-1}$Gpc${}^{-3}$ and
$\sim20~$yr${}^{-1}$Gpc${}^{-3}$ for BH-NS and BH-BH mergers, respectively.  
This scaling does not include complications due to BH ejection~\cite{bh_gc1,bh_gc2,bh_gc3,bh_gc4,Aarseth:2012eg}. 
Also, these calculations do not include the likely reduction in compact object (CO) populations
within the GC due to natal kicks.
In~\cite{Murphy:2011} it was found that including a $5\%$ NS retention fraction when fitting 
simulation results to observations of M15, and assuming no central BH, reduced the estimated
number of NSs in the inner $0.2$ pc by  $\sim 1/4$ compared to a similar study that did not 
include natal kicks~\cite{Dull}. 
The calculated NS-NS merger rate is quite sensitive to the fraction $f$ of NSs in the core, scaling
as $\sim f^2$, which means the aforementioned rates could be too large by an order of magnitude if retention rates
are this low.
However, observations suggest that in some GCs the NS retention fraction could be as high as $20\%$~\citep{Pfahl2002}.
Also, note that the tidal capture cross section used in~\cite{lee2010} is more than an order of magnitude
smaller than the GW capture cross section (discussed in the following section)
for compact objects, and using the latter would increase the rates by the same factor.
In geometric units $G=c=1$ (which, unless otherwise stated, we use throughout), tidal capture is estimated
to occur in~\cite{lee2010} for periapse values
$r_p/M\leq$ 32, 25, and 13 for NS-NS, BH-NS, and BH-BH binaries, respectively.  

In~\cite{grindlay2006} NS-NS binary formation in GCs via exchange interactions was studied, giving a 
merger rate of $\sim2$ yr$^{-1}$ Gpc$^{-3}$. A similar mechanism was explored in~\cite{Clausen:2012zu} for BH-NS systems;
the results depend sensitively on the initial mass fraction of BHs, with more massive BHs leading
to higher event rates. For example, models where the GC contained $M=35_\odot$ BHs lead to 
advanced LIGO detection rates of 0.04--0.7 yr$^{-1}$.
Though in contrast to tidal or GW capture discussed in the previous paragraph,
the mechanisms looked at in both these studies typically produce binaries
with periods of $0.1$ days or longer, and they will effectively circularize before entering the LIGO band.

There is also the possibility that eccentric mergers could result from hierarchical triples 
through the Kozai mechanism.  This has been suggested to occur in BH-BH mergers in GCs~\cite{Wen:2002km,Miller2002,Aarseth:2012eg} and
CO mergers around supermassive BHs in galactic nuclei~\cite{Antonini2012},
as well as in coeval or dynamically formed BH-NS or NS-NS binaries~\cite{Thompson}.
Though the dynamics of these systems will be different from those studied here, they could be similar at late times.
Efforts to understand this mechanism in the general-relativistic regime are ongoing (see e.g.~\cite{Naoz2012}), and 
the event rates of these systems are not well known (though see~\cite{2012arXiv1211.4584K}).

\subsection{Cross sections and high-eccentricity fractions}

We focus on mergers with initial periapse $r_p \lesssim 10 M,$
where $M$ is the total mass, making 
this study complementary to previous studies \cite{O'Leary:2008xt,Kocsis_Levin}. 
As we show later, in this 
regime essentially all mergers occur with non-negligible eccentricity ($e\gtrsim0.2$).
This is also the regime where strong-field effects such as black hole spin and
zoom-whirl behavior can influence the dynamics. To estimate the fraction of 
dynamical capture binaries that retain high eccentricity, we can use Newtonian dynamics
with quadrupolar energy loss following~\cite{Peters:1963ux,Peters:1964zz,1977ApJ...216..610T}. First,
for a hyperbolic orbit with a small velocity at infinity $v\ll1$, the relationship
between impact parameter $b$ and $r_p$ is $r_p \approx b^2v^2/2M$. In other words,
the cross section $\sigma\propto b^2$ scales {\em linearly} with $r_p$. The maximum
pericenter passage that leads to a bound system through gravitational 
radiation loss\footnote{As mentioned earlier, for the COs considered here,
energy lost to tidal effects is much less than GW emission at these separations, so
the latter process determines the cross section. Also, when a bound system is formed, the
fraction that has a semimajor axis large enough to have the binary
tidally unbound by a subsequent interaction with the surrounding cluster potential is insignificant.}
is $r_{p,m} \approx (31 \eta)^{2/7} v^{-4/7} M$, 
where $\eta=m_1 m_2/M^2=q/(1+q)^2$ is the symmetric mass ratio,
with $q$ the mass ratio.
For a galactic nuclear cluster
where $v\approx 1000$ km/s, between $20\%$ and $30\%$ of dynamical capture binaries
(where the range is from $q=1$ to $q=0.1$) 
will have $r_p/M<10$; for a globular cluster with $v\approx 10$ km/s,  this
drops to $1.5\%$--$2.0\%$. 

Although we focus on 
those with small initial periapse, {\it all} dynamical capture binaries 
will have a repeated burst
phase \cite{Kocsis_Levin}. For a large fraction of expected 
binary masses the repeated bursts will be within the Advanced LIGO band. The burst frequency is 
$\nu_b\approx r_p^{-1}(r_p/M)^{-1/2}$; the lowest frequency occurs
at $r_p=r_{p,m}$, which ranges from $(1-100{\rm Hz})/M_{10}$ for
$q=0.1$ encounters in globular clusters to $q=1$
encounters in nuclear clusters, with $M_{10}=M/10 M_\odot$. 
To estimate the percentage of systems that will end with a low-eccentricity 
inspiral phase, if the initial periapse is 
$r_{p,i}$, and we consider the repeated burst phase to end at
a periapse of $r_{p,f}$ with eccentricity
$e_f$, from~\cite{Peters:1964zz} $r_{p,i}\approx 0.57 r_{p,f} (1+e_f) e_f^{-12/19}[1+O(e_f^{2})]$.
For example, if a binary with $e_f<0.1$ by $r_{p,f}=10 M$ can be considered to have
a low eccentricity inspiral phase, then this corresponds to all systems with $r_{p,i}>27M$. For nuclear
clusters, this is between $20\%$ and $40\%$ for $q=1 - 0.1$,
while the corresponding range for globular clusters is $94\%$ and $96\%$.

\subsection{Electromagnetic counterparts}

Binary NS or BH-NS mergers are thought to be progenitors for SGRBs, and may also source
a number of other EM transients~\cite{Metzger2012,Piran2012}. Possibilities
include optical/UV emission on time scales of a day from radioactive
decay of ejected material. (This depends on heavy element opacities. Recent
work using more detailed calculations suggests the time scale may be up 
to a week with emission peaking in the IR~\cite{kasen_kitp_talk}).
Interaction of the outflow with surrounding matter can also produce radio emission on time scales
of weeks to years~\cite{2011Natur.478...82N}. And, for binary NS
mergers, the strong shocks
produced can emit in radio to x rays over a second to day time scales~\cite{Kyutoku:2012fv}.

Simulations of eccentric BH-NS and NS-NS mergers have shown a rich variation in outcome with impact parameter,
with the possibility of large accretion disks as well as ejecta that could undergo the 
$r$ process~\cite{lee2010,bhns_astro_letter,bhns_astro_paper,nsns_astro_letter,Rosswog2012}.   
There is also significant variability in observed SGRBs. It is not implausible
that this may in part be due to a subclass of SGRBs associated with
dynamical capture binaries. Though not conclusive, there is also observational evidence for multiple
SGRB progenitors. Of SGRBs with identified host galaxies, $\sim 25$\%
have offsets of $\gtrsim 15$~kpc from their
hosts~\cite{berger2010},  which would be
consistent with kicked, primordially formed binary COs or with dynamically formed
binaries in globular clusters.  The latter may be preferred for the largest
offsets~\cite{church}, especially if primordial binary COs experience weak
kicks~\cite{dewi}.
X-ray afterglows suggest that different progenitors may be responsible for SGRBs with and without extended
emission~\cite{Norris:2011tt}; simulations of dynamical capture binaries
show that it is more common to get long tidal tails, which could
lead to extended emission as the material falls back to the accretion disk. 
There is also a claim that a high-energy gamma-ray
source observed in Terzan 5 may be the remnant of
a binary CO merger-powered SGRB~\cite{Domainko:2011gv}; if true, this provides
evidence that dense cluster environments can be significant sources of binary CO mergers.

The time scales between close encounters in eccentric mergers may also explain observed delays 
between precursors and SGRBs~\cite{Troja2010}. For example, NS crust cracking on a nonmerging
close encounter could potentially cause flares
that precede the merger by an interval ranging from milliseconds to possibly a few 
seconds~\cite{nsns_astro_letter}.

\subsection{Gravitational wave detectability}

Multimessenger exploration is of course an exciting possibility. 
Even a null GW detection provides astronomical information as it rules out 
compact object mergers as the source of an observed GRB, but only if the detectability
of these types of signals is understood. Given the disparate
nature of the waves from dynamical capture vs coeval mergers, data analysis methods 
designed specifically for each are required for this kind of astronomy.
Methods to search for quasicircular inspiral (of relevance to the majority
of coeval binaries, and a subset of dynamical capture binaries that form 
with a sufficiently large periapse to circularize before merger) have been
the predominant focus of the GW community over the past decades~\cite{Colaboration:2011np}. Comparatively, there is a dearth 
of studies on the detectability of highly eccentric mergers
\footnote{Though see~\cite{Huerta:2013qb} for
a recent study of the efficacy of quasicircular templates
to detect lower eccentricity NS binaries.}.
In~\cite{0004-637X-648-1-411} the single burst from a parabolic close encounter
was studied, while~\cite{O'Leary:2008xt} included the additional signal provided by
subsequent bursts. This repeated burst phase was studied in~\cite{Kocsis_Levin} using 2.5 and
3.5 order post-Newtonian (PN) equations of motion.  It was found that GWs from this phase may
be detectable by Advanced LIGO out to 200--300 Mpc for BH-NS binaries and 300--600 Mpc
for BH-BH binaries. 
Since the PN approximations
begin to break down close to merger, the evolution was only followed
to $r_p=10M$.
To model the last stages of merger requires numerical relativity (NR), and
there have been a 
number of numerical studies of eccentric 
mergers~\cite{Pretorius:2007jn,Hinder:2008kv,Healy:2008js,Gold:2009hr,Gold2011,bhns_astro_letter,bhns_astro_paper,nsns_astro_letter,Gold:2012tk}. 
However, because of the computational
expense of these simulations, it is not possible with current computer resources to follow high-eccentricity 
binaries through multiple close encounters. The challenge is compounded by a large parameter space, including the impact parameter,
mass ratio, BH spin, and NS equation of state. It is thus not reasonable to expect that brute-force numerical
simulations will be able to provide templates before the Advanced LIGO era, even accounting for expected
increases in computer power.

\subsection{Outline of remainder of paper}

To begin to bridge the gap between large periapse PN solutions
and late-time numerical solutions, we introduce a
model for the inspiral, merger, and ringdown of dynamical capture compact binaries.
This model is based on geodesic 
equations of motion in an effective Kerr spacetime, combined with quadrupole radiation (Sec.~\ref{geo_model})
and a version of the Implicit Rotating
Source (IRS) model~\cite{Baker:2008mj,Kelly:2011bp} for the merger and ringdown parts of the GW signal (Sec.~\ref{merge_model}).
(Except for the IRS extension, and the comparable masses, our hybrid is reminiscent of the ``kludge'' 
introduced to study extreme mass ratio inspirals~\cite{Glampedakis:2002cb,Babak:2006uv,Yunes:2007zp,Sopuerta:2011te},
based in part on the ``semirelativistic'' approach of~\cite{1981PThPh..66.1627R}.)
We validate this model through a comparison to full numerical simulations in the strong-field regime (Sec.~\ref{sec_nr_comp})
and to the PN approximation for $r_{\rm p} > 10\,M$ (Sec.~\ref{sec_pn_comp}).

The waveforms we produce here are likely not accurate enough
for optimal template-based detection of multiple-burst events.
Indeed, creating improved accuracy waveforms will probably require
a different approach, for example, using the effective one body (EOB) formalism which 
has recently been extended to generic orbits~\cite{Bini:2012ji} and calibrating it using 
full numerical simulations. 
However, our waveforms capture the relevant 
features with sufficient faithfulness that we can use them to assess the efficacy of existing LIGO search strategies.
We can also use our waveforms to investigate new search strategies
that may be better suited to highly eccentric mergers.
In Sec.~\ref{sec:det} we use this model to evaluate how well
these GW signals could be seen with each generation of the LIGO detectors,
varying impact parameter (equivalently $r_p$), total mass, and mass ratio. We
use various analysis methods: matched filtering with the model templates,
filtering with ringdown templates, and a burst search with sine-Gaussian templates. We
also estimate how well a hypothetical search using incoherent stacking of 
bursts following~\cite{Kalmus} would perform. Though not as optimal as matched filtering, stacking
is likely more robust to timing uncertainties in the burst sequence.
We find that if capture binaries do exist, in many cases their GW signals will be missed by
single-burst or ringdown searches (and, as we argue, quasicircular templates),
whereas these sources would be detectable with a full template or a
stacked burst search. In particular, GRB051103~\cite{Hurley} had a measured distance
of 3.6 Mpc, and no coincident GW signal was found using traditional searches~\cite{Abbott2010,Abadie2010}. However, there is a sizable
region of the parameter space of dynamical capture binaries that existing searches would
have missed. The possibility that the GRB was preceded by an eccentric merger remains a viable possibility.

In Sec.~\ref{sec:conc} we make concluding remarks and comment on the direction of future work.

\section{Waveform Model}
\label{sec:model}
In this section we describe our model for high-eccentricity merger waveforms.
We first look at the inspiral phase in Sec.~\ref{geo_model}, which can be considered
a sequence of GW bursts, each generated at a periapse passage. In Sec.~\ref{sec_nr_comp}
we compare the model expressions we use for the bursts to full numerical
simulations. In Sec.~\ref{merge_model} we discuss the IRS model for the merger and
ringdown phase, and in Sec.~\ref{sec_pn_comp} we present examples of the full signal,
and make further comparisons to PN results for the inspiral phase.

\subsection{Repeated burst phase}
\label{geo_model}
Our objective here is to model the GW signal from an eccentric binary that passes through a series of close
encounters prior to merger. To this end,  we use a prescription based on the equations of motion of a geodesic in 
a Kerr spacetime, coupled with the quadrupole formula for gravitational radiation.  
We identify the mass and total angular momentum
of the binary with the mass and spin parameters
of the effective Kerr spacetime
and the orbital angular momentum and energy with that of the geodesics.
This approach has the advantage of
reproducing the correct orbital dynamics in the Newtonian limit and general-relativistic test particle limit, while 
still incorporating strong-field phenomena such as pericenter precession, frame dragging, and the existence of unstable
orbits and related zoom-whirl dynamics.
For simplicity, in this first study we restrict our attention to equatorial orbits and, for the most part, nonspinning BHs (we compare the IRS model to a merger
involving a spinning BH in Sec.~\ref{merge_model}).

The equations for an equatorial geodesic in a Kerr spacetime with mass $M$ and dimensionless spin $a$ 
can be written in first order form using Boyer-Lindquist coordinates as
\begin{eqnarray}
\dot{\tau} &=& \frac{\Delta}{\tilde{E} R_0^2 - 2 M^2 a \tilde{L}/r} \equiv Q \nonumber \label{Q}, \\
\dot{\phi} &=& \frac{1}{R_0^2}\left[\tilde{L} Q + 2 M^2 a/r\right] \equiv \Omega \nonumber \label{w}, \\
\dot{r} &=& \Delta Q P_r/r^2 \nonumber, \\
\dot{P_r} &=& \frac{1}{r^2 Q}\left[\Omega^2(r^3-M^3 a^2) + M(2 M a \Omega -1)\right] \nonumber \\
& &+\frac{P_r^2 Q}{r^3}\left[M^2 a^2-M r\right]\label{geo_eqn}
\end{eqnarray}
where $R_0=r^2+2M^3 a^2/r+M^2 a^2$, $\Delta=r^2-2Mr+M^2 a^2$, $P_r=r^2\dot{r}/(\Delta Q)$, $\tau$ is proper time, and 
the overdot indicates a derivative with respect to the coordinate time. Here $\tilde{E}$ and $\tilde{L}$
are the energy and angular momentum of the geodesic.

In order to apply these equations to a binary system we go to the 
center-of-mass frame and let $\mathbf{r}$ be the separation between 
the objects.  Then we identify the geodesic parameters $\tilde{E}$ 
and $\tilde{L}$ with the reduced energy and angular momentum of the system
and promote these quantities to time-dependent variables.  To determine
the amount of energy and angular momentum radiated away to gravitational
waves, we use the quadrupole formula
\begin{eqnarray}
\dot{\tilde{E}} &=& -\frac{\mu}{5}\dddot{\mathcal{I}}_{ij}\dddot{\mathcal{I}}_{ij} \nonumber \\
\dot{\tilde{L}} &=& -\frac{2 \mu}{5}\epsilon^{zij}\ddot{\mathcal{I}}_{ik}\dddot{\mathcal{I}}_{jk}
\label{quad_loss}
\end{eqnarray}
where $\mu$ is the reduced mass, $\mathcal{I}_{ij}$ is the reduced quadrupole moment, and 
$\ddot{\mathcal{I}}_{ij}$ and $\dddot{\mathcal{I}}_{ij}$ are 
written in terms of the variables $\{r,\phi,P_r,\tilde{E},\tilde{L}\}$ using~(\ref{geo_eqn}).  
We set $M$ in~(\ref{geo_eqn}) to the total mass (neglecting orbital energy contributions),
and we set $a=\mu\tilde{L}/M^2+a_{BH}$, where $a_{BH}$ is the net spin of
any BHs (though again for this study we focus on nonspinning BHs, where $a_{BH}=0$).
The use of an effective spinning BH spacetime based on total angular momentum
is motivated by~\cite{Pretorius:2007jn}, where it was found that the properties of zoom-whirl-like dynamics exhibited 
in equal mass mergers in full numerical relativity are better approximated by geodesics on the 
effective Kerr spacetime than Schwarzschild spacetime, and it differs from the EOB approach which uses deformations
of the Schwarzschild metric for the merger of nonspinning objects~\cite{Buonanno:1998gg}.  
We note that 
when the orbital angular momentum is large we will have $a>1$. However, this will occur
only when the separation $r$ is also large, so general-relativistic effects are small,
and no unusual behavior arises from exceeding the Kerr limit. 
We numerically integrate the coupled set of equations~(\ref{geo_eqn}) and~(\ref{quad_loss}).

The remaining element is to calculate the observed gravitational radiation, which will depend on the intrinsic source 
parameters (i.e.~the mass, mass ratio, eccentricity, and initial periapse distance), and will also 
vary with sky location and relative orientation of the source to the detector.
At linear order and in the transverse traceless gauge, 
the complex gravitational wave strain $h_{\rm opt}$ a distance $d$ from an optimally oriented source is simply related to changes in the quadrupole moment
through 
\beq
h^{\rm opt} \equiv h_+^{\rm opt} + \imagi h_{\times}^{\rm opt} \equiv \frac{2}{d}\left( \ddot{\mathcal{I}}_{x\,x} + \imagi \ddot{\mathcal{I}}_{y\,x}\right)\,.
\label{eqn:hopt}
\eeq
For general orientations, the emitted strain can be represented through a mode decomposition as
\beq
\bar{h} \equiv h_+ + \imagi h_{\times} = \sum_{\ell=2}^{\infty} \sum_{m=-\ell}^{\ell} h_{\ell m}(t,d) \,_{-2}Y_{\ell m}(\theta,\phi) \, ,
\label{eqn:hlm_def}
\eeq
where $_{-2}Y_{\ell m}$ are the spherical harmonics of spin weight $-2$ \cite{Goldberg:1966uu},
and $\theta$ and $\phi$ are the polar and azimuthal angles of orientation, respectively.
For the comparable mass, nonspinning systems that we are primarily interested in, the quadrupole
(i.e.,~$\ell=2$, $m=\pm 2$) component dominates the strain, so that
\beq
\bar{h} \approx h_{22}(t,d)\,_{-2}Y_{22}(\theta,\phi) + h_{2-2}(t,d)\,_{-2}Y_{2-2}(\theta,\phi).
\label{eqn:hquad_def}
\eeq
This completes the approach for calculating the source waveform that reaches a detector.  In a later section
we will include the sensitivity of the detector in the analysis.

\subsection{Comparison to fully general-relativistic numerical simulations}\label{sec_nr_comp}
To provide some validation for this model we compare several waveforms 
of single high-eccentricity fly-by encounters from full general-relativistic numerical simulations to 
those obtained from the geodesic equation with the quadrupole formula.  
The simulations include a 4:1 mass ratio BH-NS system~\cite{bhns_astro_paper}, an 
equal mass NS-NS system~\cite{nsns_astro_letter}, and an equal mass BH-BH system. 
The NR simulations were all performed using the code described in~\cite{code_paper}.

\begin{figure*}
\includegraphics[width=.4\textwidth]{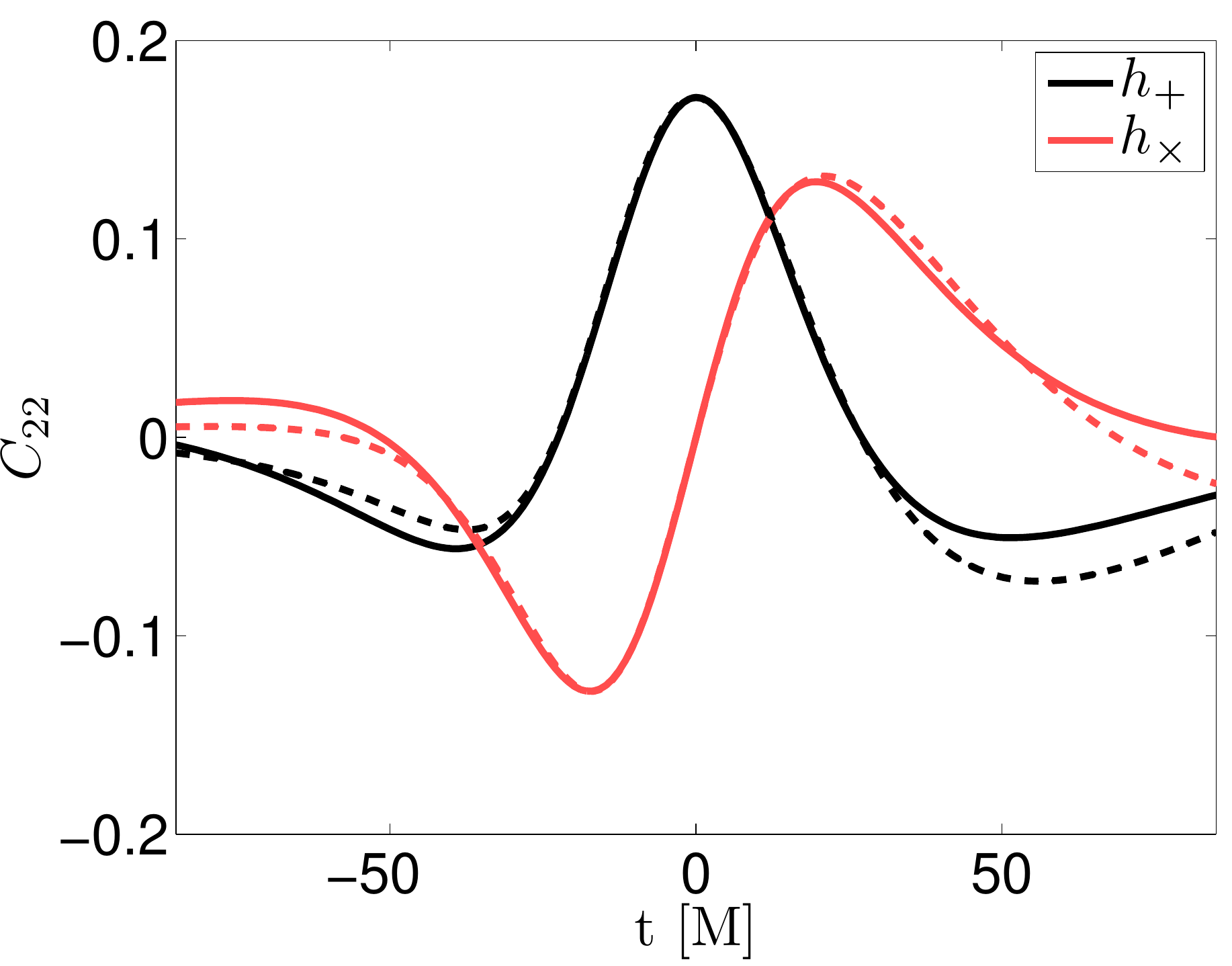}
\includegraphics[width=.4\textwidth]{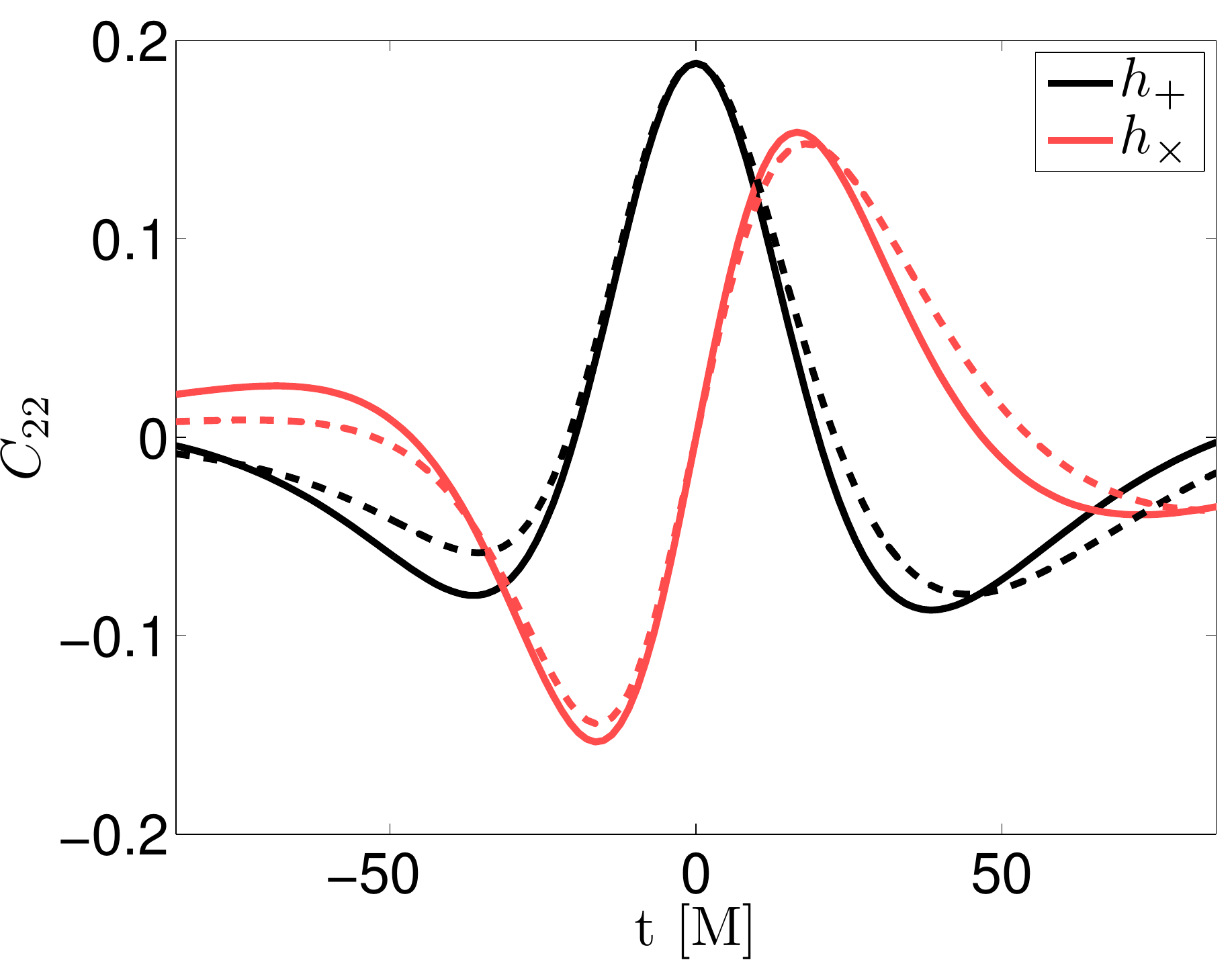}
\includegraphics[width=.4\textwidth]{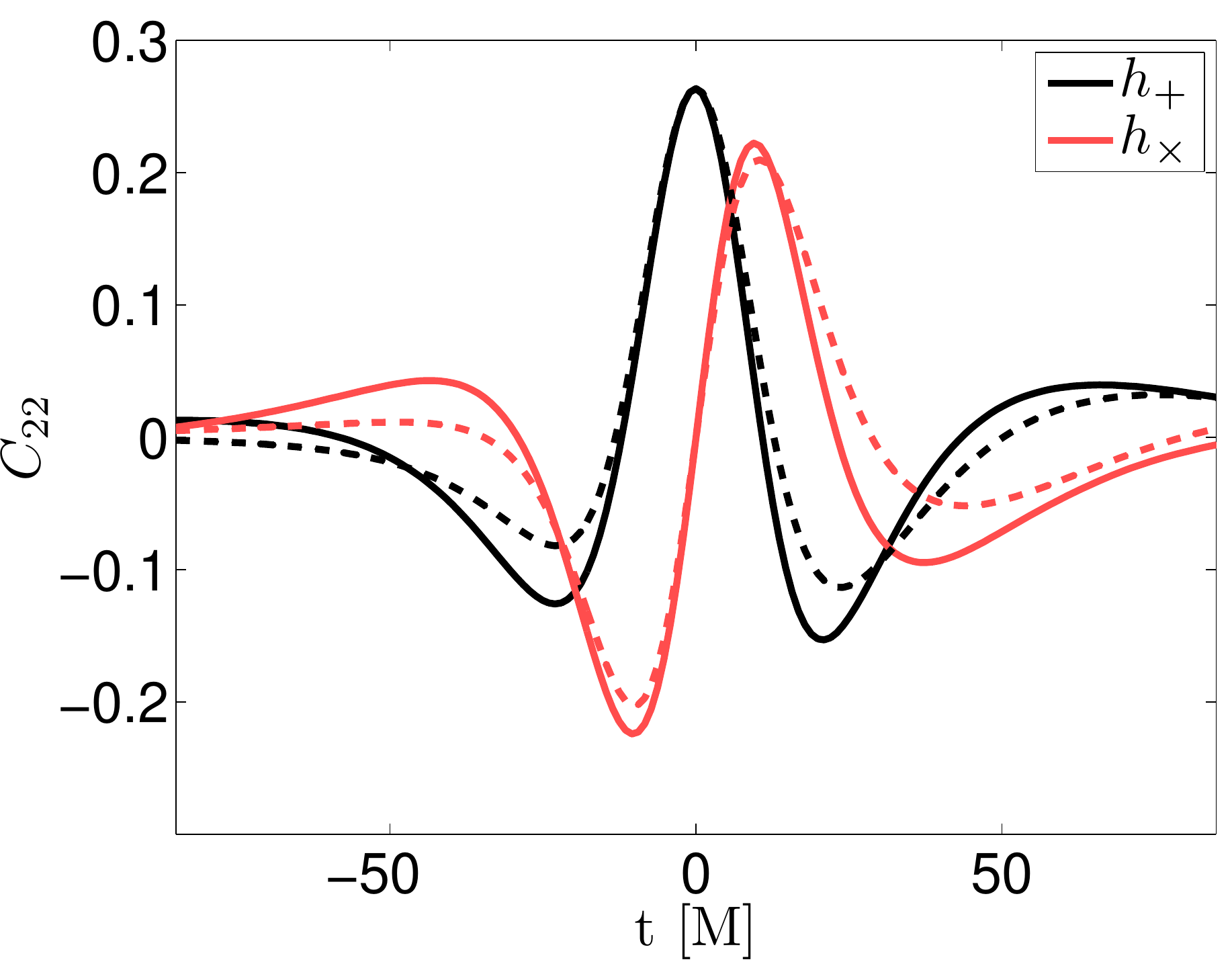}
\includegraphics[width=.4\textwidth]{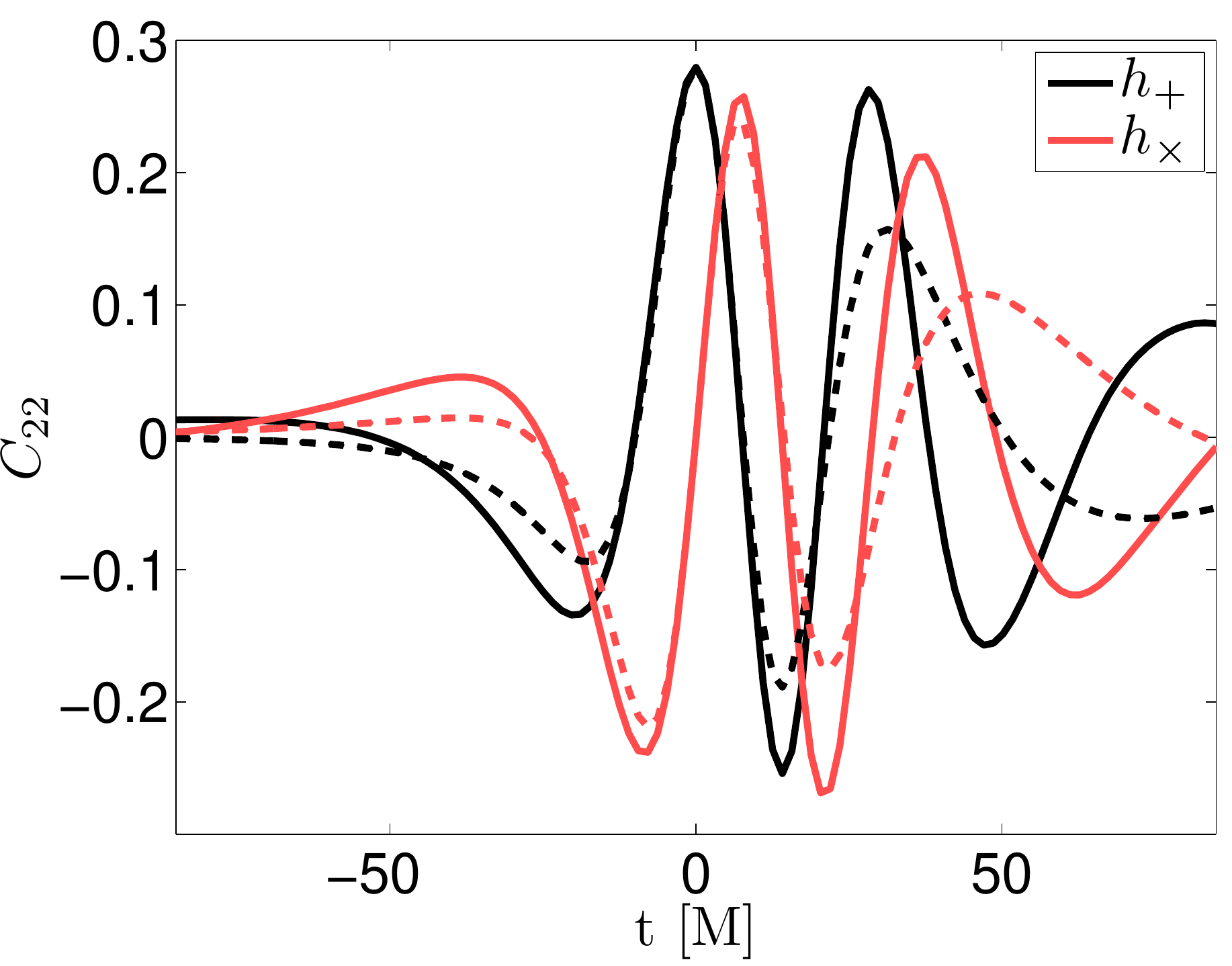}

\caption{
Comparison of the $\ell=2$, $m=2$ component of $\bar{h}$
for fly-by waveforms from 4:1 BH-NS simulations (solid line) and our model with best-fit periapse distance (dashed line).
The approximate effective geodesic orbital parameters of the simulated system (left to right, top to bottom) are 
$(r_p,e)=(8.3,1.0)$, $(8.0,0.8)$, $(5.6,1.0)$, and $(5.0, 1.0)$.
The fit parameters are given in Table~\ref{fit_table}.
\label{nsbh_geo_cmp}
}
\end{figure*}

In Fig.~\ref{nsbh_geo_cmp} we show several such examples from NR simulations of the 4:1 BH-NS system 
alongside corresponding waveforms from our model with best-fit parameters.  
The peak amplitude of the geodesic is scaled to be the same as in the simulations. The fit is performed by finding
the initial orbital parameters that maximize the phase overlap between the waveforms (see e.g.~\cite{Damour:1997ub}). 
In this regime the match between the waveforms is most sensitive to $r_p$ as opposed to $e$.  As can be seen, the 
fly-by waveforms from our model provide a good match to those from simulations. Even close
to the effective innermost stable orbit (ISO) for the BH-NS system (the bottom panels of Fig.~\ref{nsbh_geo_cmp}), where the system
begins to show evidence of whirling behavior, our model is able to approximately capture the shape of
the waveform. 

In Table~\ref{fit_table} we give the fit parameters, amplitude enhancement, and overlap.  
We also show the approximate initial orbital parameters ($r_p$ and $e$) of the simulation obtained by equating a Newtonian estimate of the
reduced orbital energy and angular momentum at the beginning of the simulation with the $\tilde{E}$ and $\tilde{L}$ parameters of the geodesic model
described above. (Note, this is different from the \emph{Newtonian} values for $r_p$ and $e$ used in~\cite{bhns_astro_paper,nsns_astro_letter}.)
For most of the BH-NS systems in Table~\ref{fit_table} we can see that the enhancement required to match the amplitude of our model
to the simulation results is $\sim 4\%$--$11\%$.
This is presumably due to aspects not captured by this simple model, such as finite-size effects,
as well as truncation error from the simulations. 
The one case where the amplitude of the simulation waveform was below the model result
was a simulation with strong whirling behavior (bottom-right panel of Fig.~\ref{nsbh_geo_cmp}) where the NS
had large $f$-mode oscillations excited as described in~\cite{bhns_astro_paper}.  

\begin{figure}
\includegraphics[width=.4\textwidth]{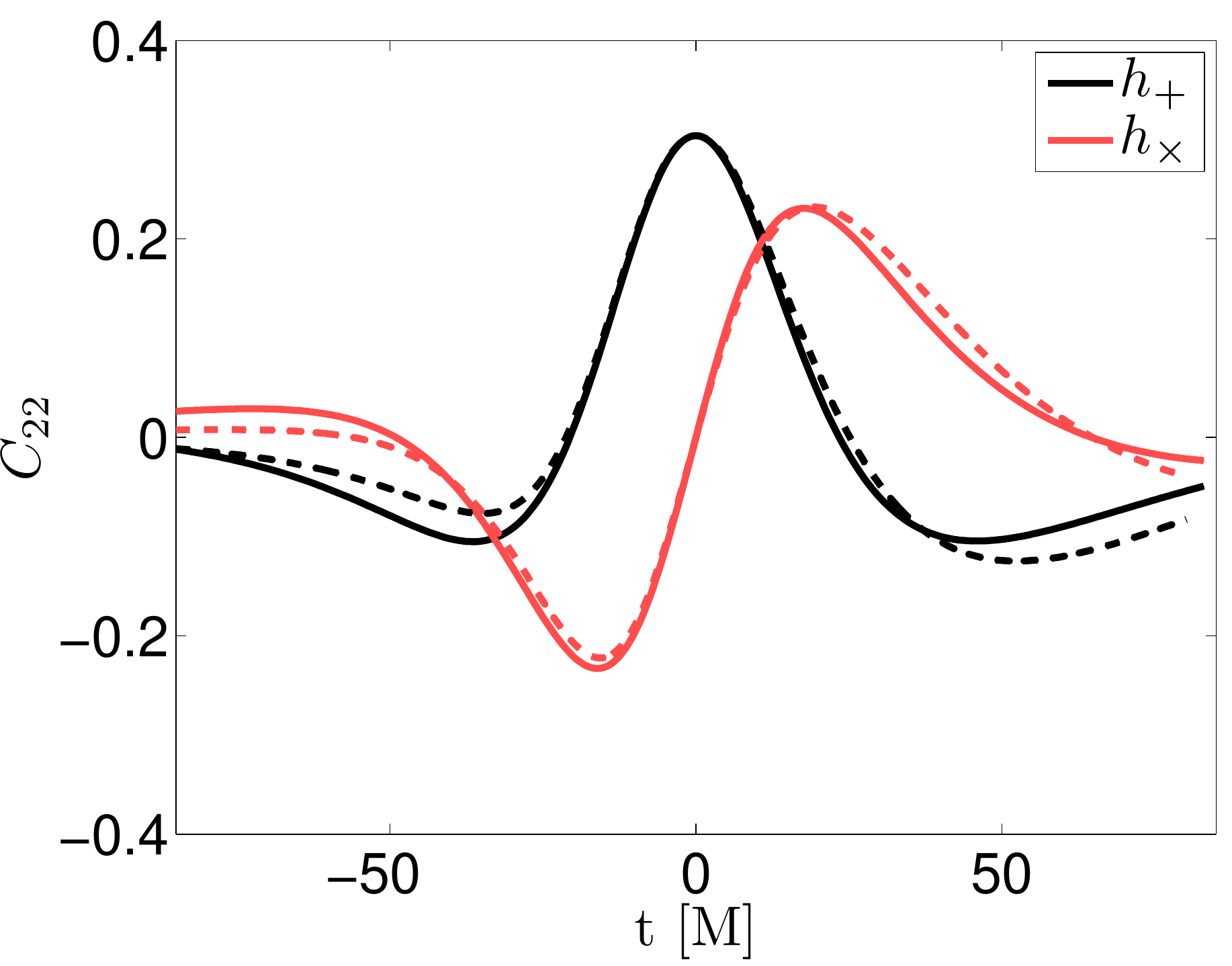}
\includegraphics[width=.4\textwidth]{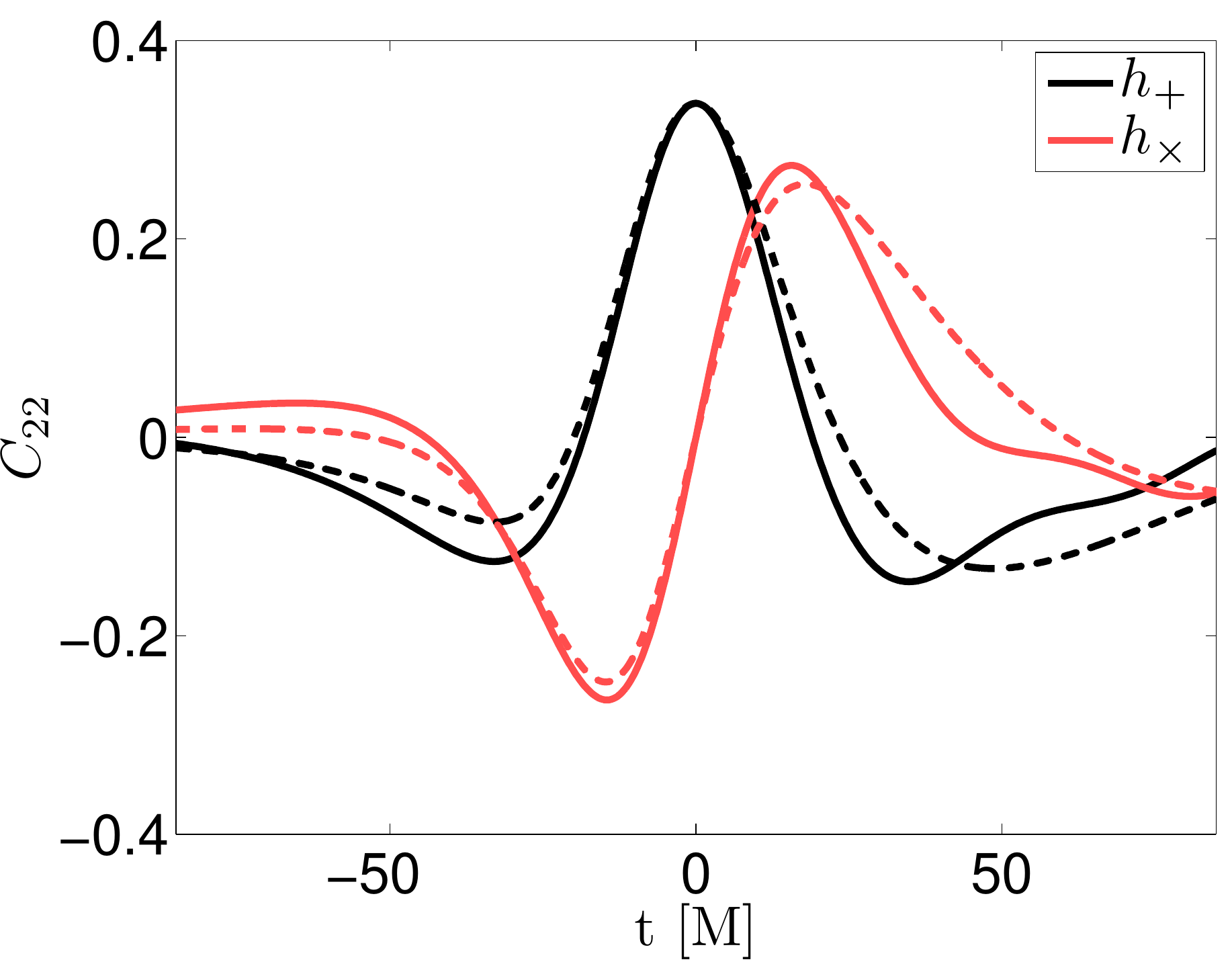}
\caption{
Comparison of the $\ell=2$, $m=2$ component of $\bar{h}$ for 
fly-by waveforms from equal mass BH-BH (top panel) and NS-NS (bottom pannel) simulations (solid lines)
and our model (dotted lines) with best-fit parameters.
The approximate effective geodesic orbital parameters of the simulated systems are $(r_p,e)=(8.7,1.0)$ for both cases.
The fit parameters are given in Table~\ref{fit_table}.
The feature in the waveform after the peak in the NS-NS simulation is from $f$-mode excitation that occurs
during the close encounter.
\label{eq_mass_geo_cmp}
}
\end{figure}

We also compare the geodesic model with an equal mass BH-BH and an equal mass NS-NS system as 
shown in Fig.~\ref{eq_mass_geo_cmp}.  Although one would expect a geodesic approximation
to be most accurate in the limit that one mass is much larger than the other, it still provides good fits for equal masses.  This   
model, however, does not attempt to include finite-size effects (such as the $f$-mode excitation
visible in the later part of the bottom of Fig.~\ref{eq_mass_geo_cmp}), which would be required to address
questions related to measuring the NS equation of state from such GW signals. 

\begin{table}
\begin{center}
\begin{tabular}{ c c c c c c c}
\hline\hline
  & \multicolumn{2}{c}{Simulation$^a$} &  \multicolumn{2}{c}{Fit$^b$} & & \\
Binary  &  $r_p$ & $e$ & $r_p$ &  $e$ & $A^c$ & Overlap$^d$ \\
\hline
NS-BH  & 8.30  & 1.00  & 8.77 & 1.00  &  1.04 & 0.99 \\
NS-BH  & 8.00  & 0.80  & 7.97 & 0.81  &  1.11 & 0.98 \\
NS-BH  & 5.62  & 1.00  & 5.61 & 1.00  &  1.11 & 0.97 \\
NS-BH  & 5.04  & 1.00  & 4.26 & 1.00  &  0.61 & 0.74 \\
BH-BH  & 8.71  & 1.00  & 8.23 & 1.00  &  1.16 & 0.99 \\
NS-NS  & 8.71  & 1.00  & 7.82 & 1.00  &  1.28 & 0.96 \\

\hline
\end{tabular}
\caption{Fit parameters for close-encounter GWs. 
\\
$^a$ Approximate initial parameters of the geodesic model based on the initial orbital energy and angular momentum
of the simulation.
\\
$^b$ Initial parameters of the geodesic model that best fit the simulation data.
\\
$^c$ Amplitude enhancement applied to the waveform from the best-fit geodesic model.
\\
$^d$ Overlap between simulation and the best-fit geodesic model.
}
\label{fit_table} 
\end{center}
\end{table}

\subsection{Merger model}\label{merge_model}
After a binary has evolved through some number of close encounters, it will merge.
In order to include the waveforms resulting from merger, we supplement
the model outlined in Sec.~\ref{geo_model} with a version of the IRS model~\cite{Baker:2008mj,Kelly:2011bp} for the merger and ringdown part 
of the GW signal.  Note that
the IRS assumes the waveform is circularly polarized.
This is not strictly valid for the complete merger-ringdown phase of eccentric binaries,
though as we show below, it does provide a reasonably good approximation to results from
numerical simulations. As with other aspects of our waveform model, this assumption could
be refined in the future, but it is adequate for the purpose of testing the efficacy
of existing search strategies for detecting eccentric binaries.

In particular, we model the phase evolution to asymptotically approach 
the least damped quasinormal mode frequency of the final BH, $\omega_{\rm QNM}$, via
\begin{equation}
\omega(t) =\omega_{\rm QNM}(1-\hat{f})
\end{equation} 
where 
\begin{equation}\label{eqn_irs}
\hat{f}=\frac{c}{2}(1+\frac{1}{\kappa})^{1+\kappa}
\left(1-(1+\frac{1}{\kappa}e^{-2t/b})^{-\kappa}\right). 
\end{equation} 
Here $b=2Q/\omega_{\rm QNM}$ is determined by the quality factor and frequency of the final BH,
and $\kappa$ and $c$ are free parameters of the model.
The amplitude is given, up to an overall factor $A_0$, by
\begin{equation}
A=\frac{A_0}{\omega(t)}\left(\frac{|\dot{\hat{f}}|}{1+\alpha(\hat{f}^2-\hat{f}^4)}\right)^{1/2}
\end{equation}   
where $\dot{\hat{f}}=d\hat{f}/dt$, and $\alpha$ is a free parameter.  We find that $\alpha=72.3/Q^2$
provides a reasonably good fit to our numerical simulations.
 
\begin{figure}
\includegraphics[width=.4\textwidth]{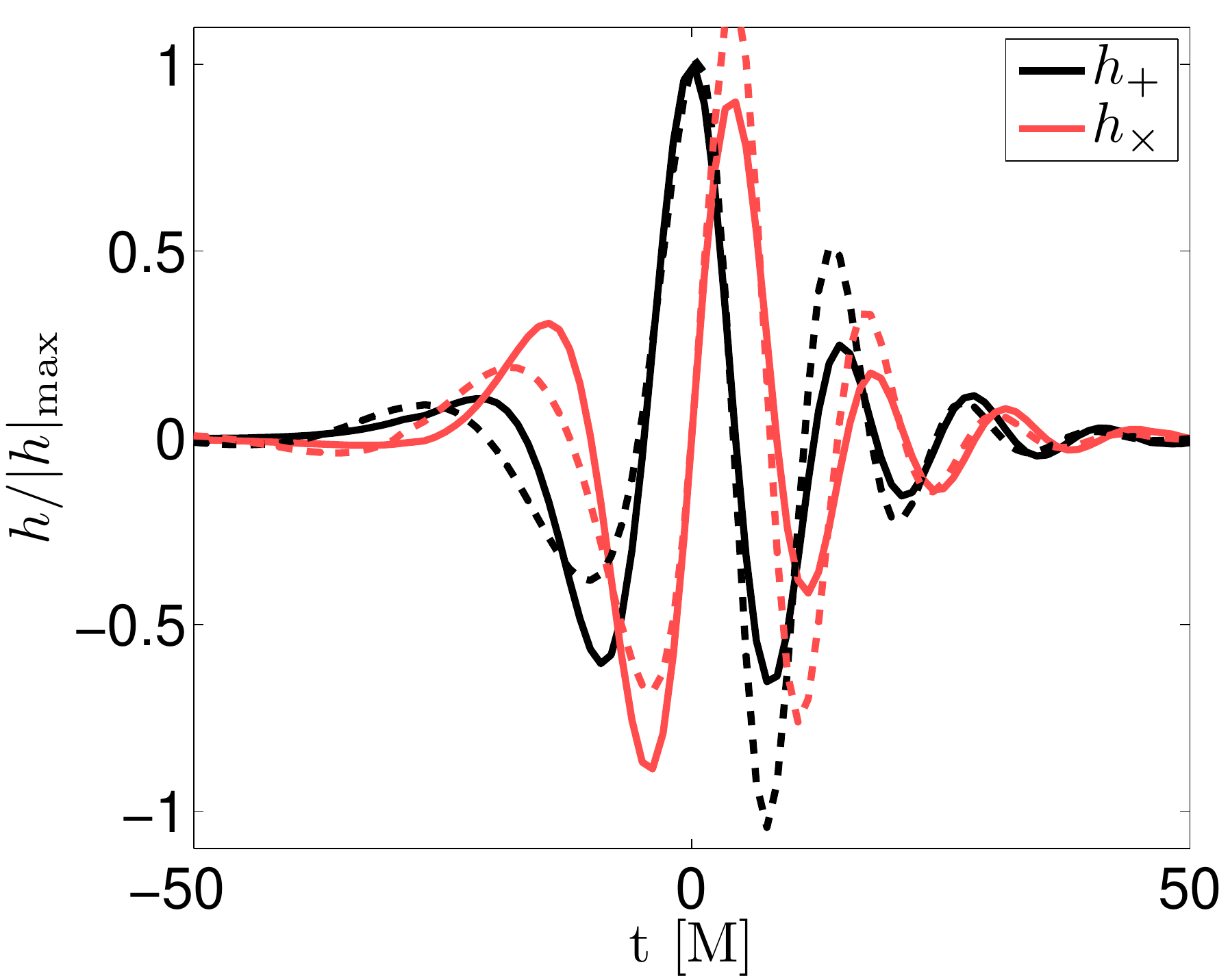}
\includegraphics[width=.4\textwidth]{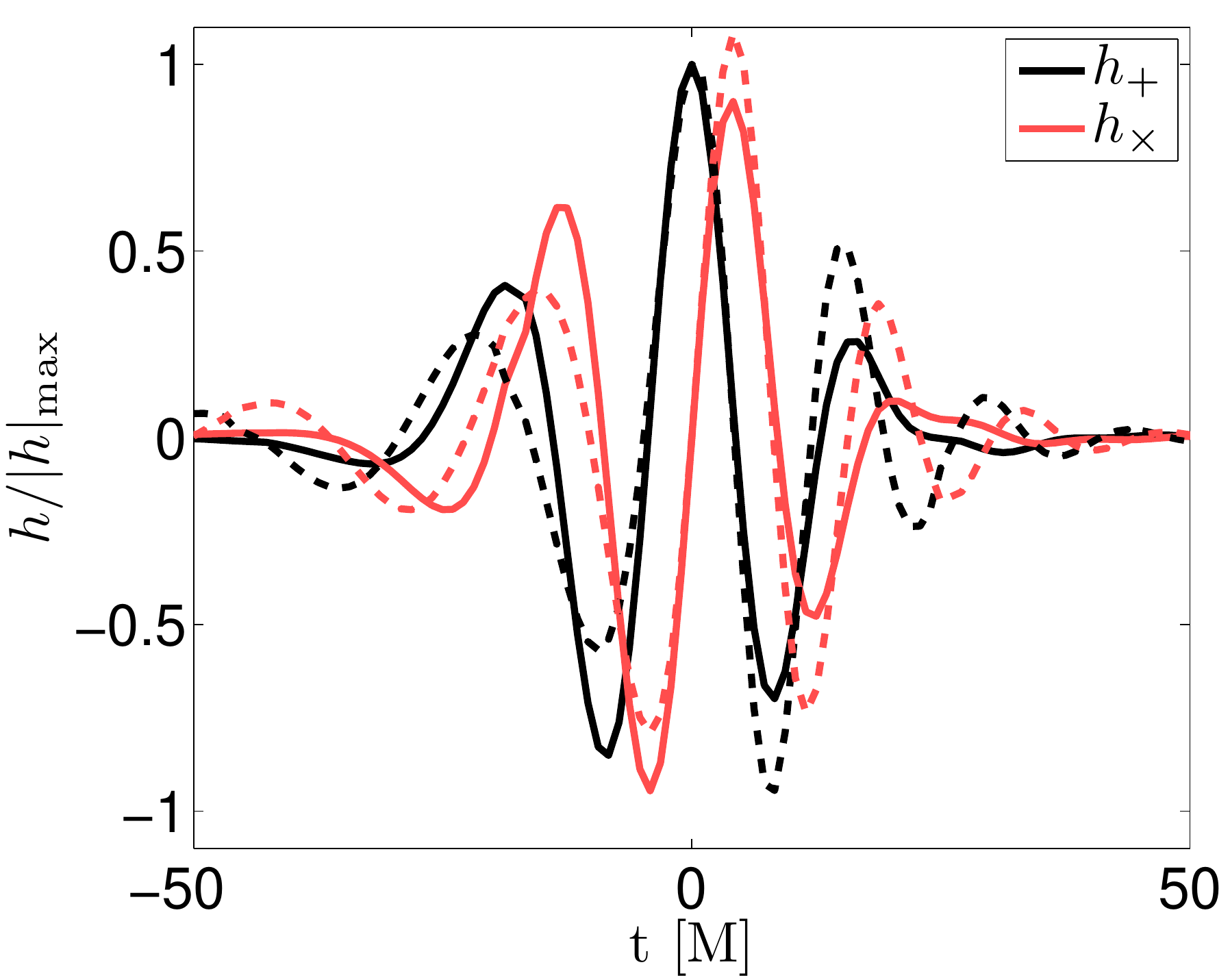}
\caption{
Comparison of the merger GW strain from a 4:1 BH-NS simulation 
(solid lines) and the IRS model (dotted lines) with best-fit parameters.
The top panel shows a case where the initial BH was nonspinning. The bottom panel
shows a case with $a_{\rm BH}=0.5$, which results in more whirling behavior and
tidal disruption of the NS. The best-fit parameters in (\ref{eqn_irs}) are $(\kappa,c)=(0.66,0.28)$ 
(top) and $(0.46,0.18)$ (bottom), and the matches are 0.98 and 0.96, respectively. 
The match is weighted based on the ``whitened" waveforms as described in Sec.~\ref{sec:det} assuming a total mass of 10 $M_{\odot}$. 
\label{nsbh_merger_cmp}
}
\end{figure}

\begin{figure}
\includegraphics[width=.4\textwidth]{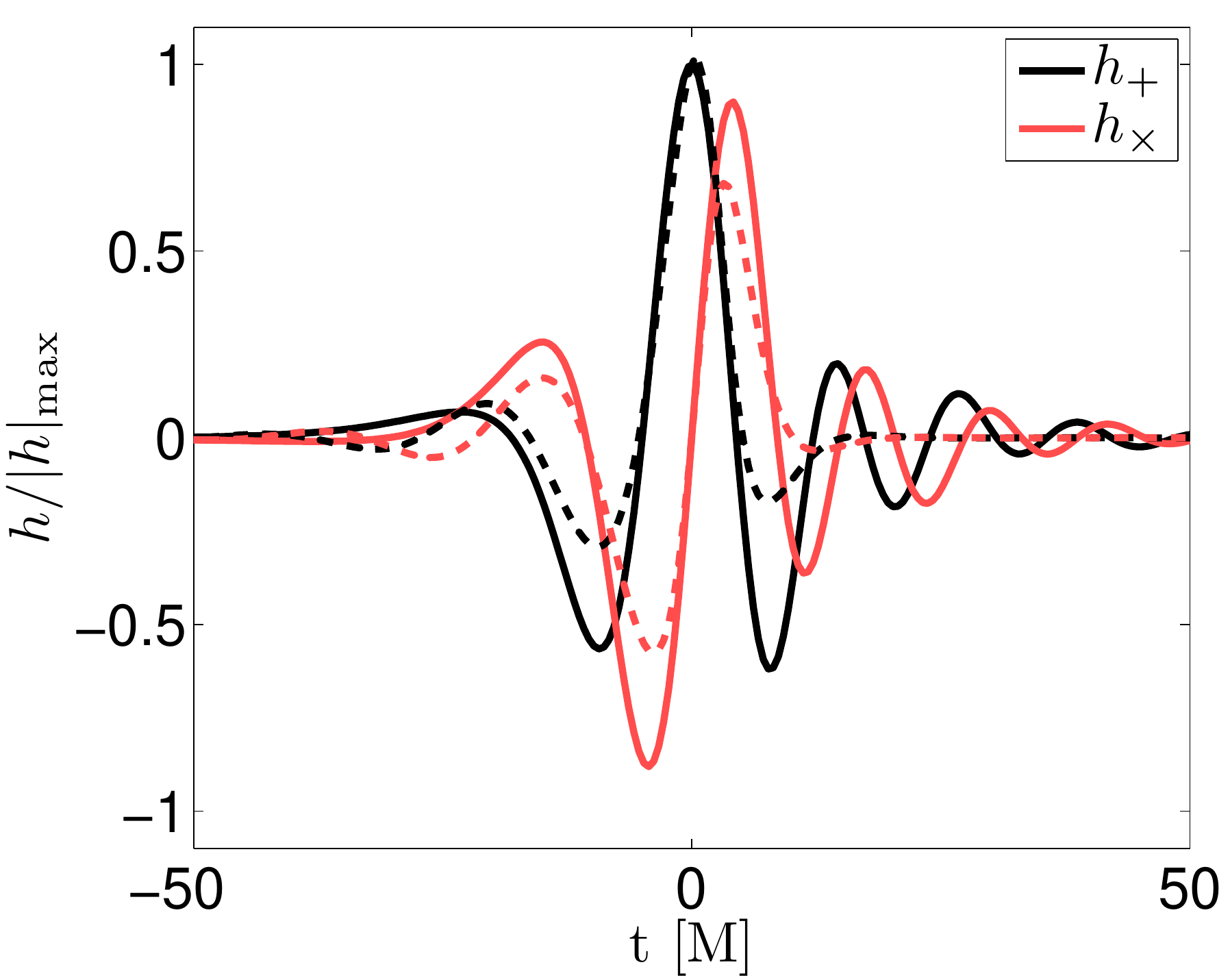}
\includegraphics[width=.4\textwidth]{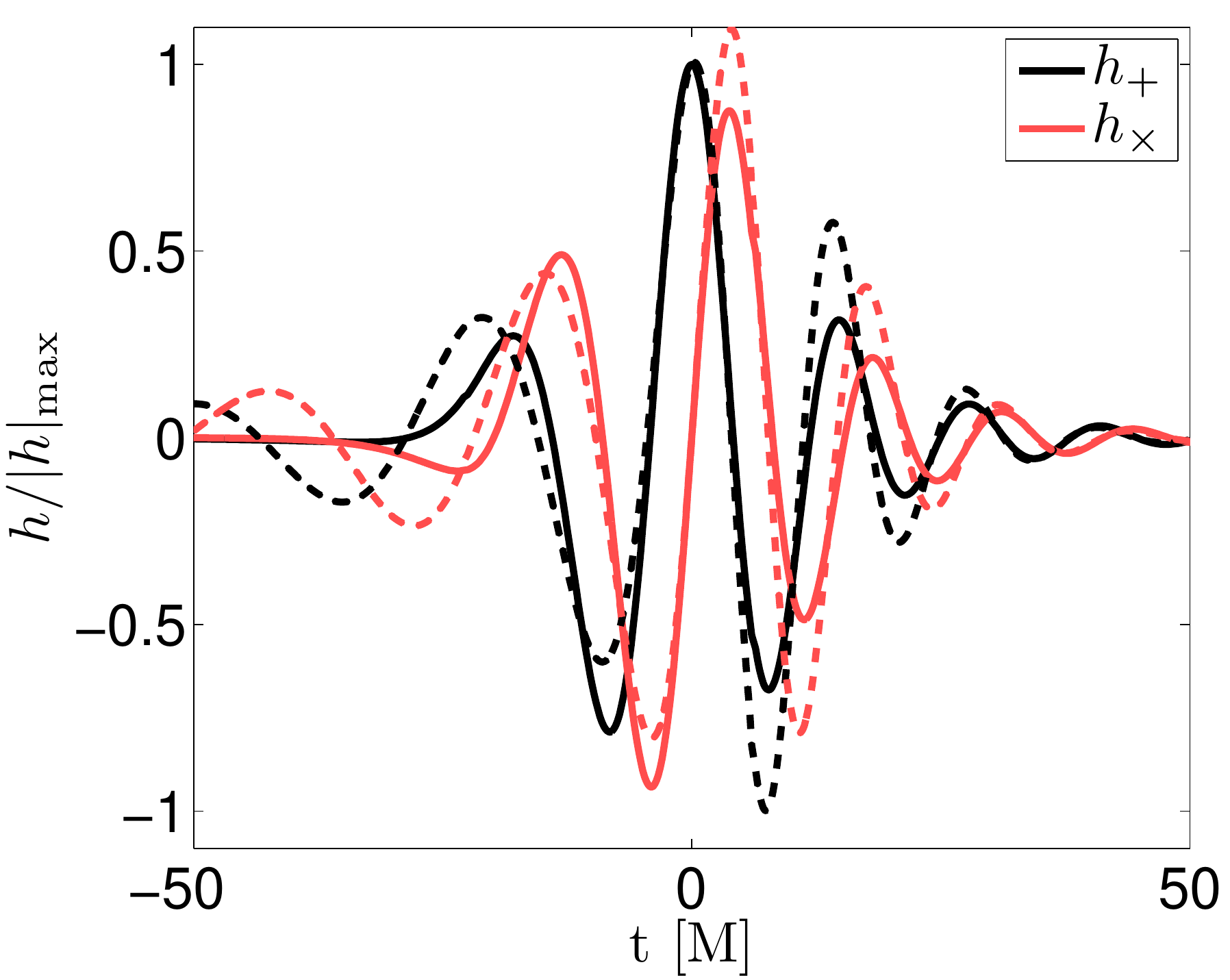}
\caption{
Comparison of merger waveforms from an equal mass BH-BH simulation (top panel) and
NS-NS simulation that forms a BH (bottom panel) with the IRS model (dotted lines) with
best-fit parameters. The best-fit parameters in (\ref{eqn_irs}) are $(\kappa,c)=(0.31,0.03)$
(top) and $(0.36,0.19)$ (bottom), and the matches are 0.98 and 0.97, respectively.
The match is weighted based on the ``whitened" waveforms as described in Sec.~\ref{sec:det} assuming a total mass of 20 $M_{\odot}$ and 2.8 $M_{\odot}$ for the BH-BH and NS-NS binaries,
respectively.
\label{bin_merger_cmp}
}
\end{figure}

In Fig.~\ref{nsbh_merger_cmp} we show a comparison between simulation results of
BH-NS mergers and the best-match IRS model waveforms, where we let $\kappa$ and $c$ be 
fitting parameters. In Fig.~\ref{bin_merger_cmp} we show the same thing for equal
mass NS-NS and BH-BH mergers.  This simple model will not capture disruption or other
matter effects, and best-fit values for $\kappa$ and $c$ will have some dependence on the
parameters of the binary, such as the impact parameter preceding merger.
However, when studying signal detectability we fix $\kappa=0.64$ and $c=0.26$, which empirically provides reasonably good fits 
to a large number of simulated waveforms, and therefore 
provides an adequate representation of a generic eccentric merger.
We attach the IRS part of the waveform to the model from Sec.~\ref{geo_model} when the separation
reaches the light ring of the effective Kerr spacetime. 

\subsection{Model properties and comparison to post-Newtonian}\label{sec_pn_comp}

\begin{figure}
\includegraphics[width=.4\textwidth]{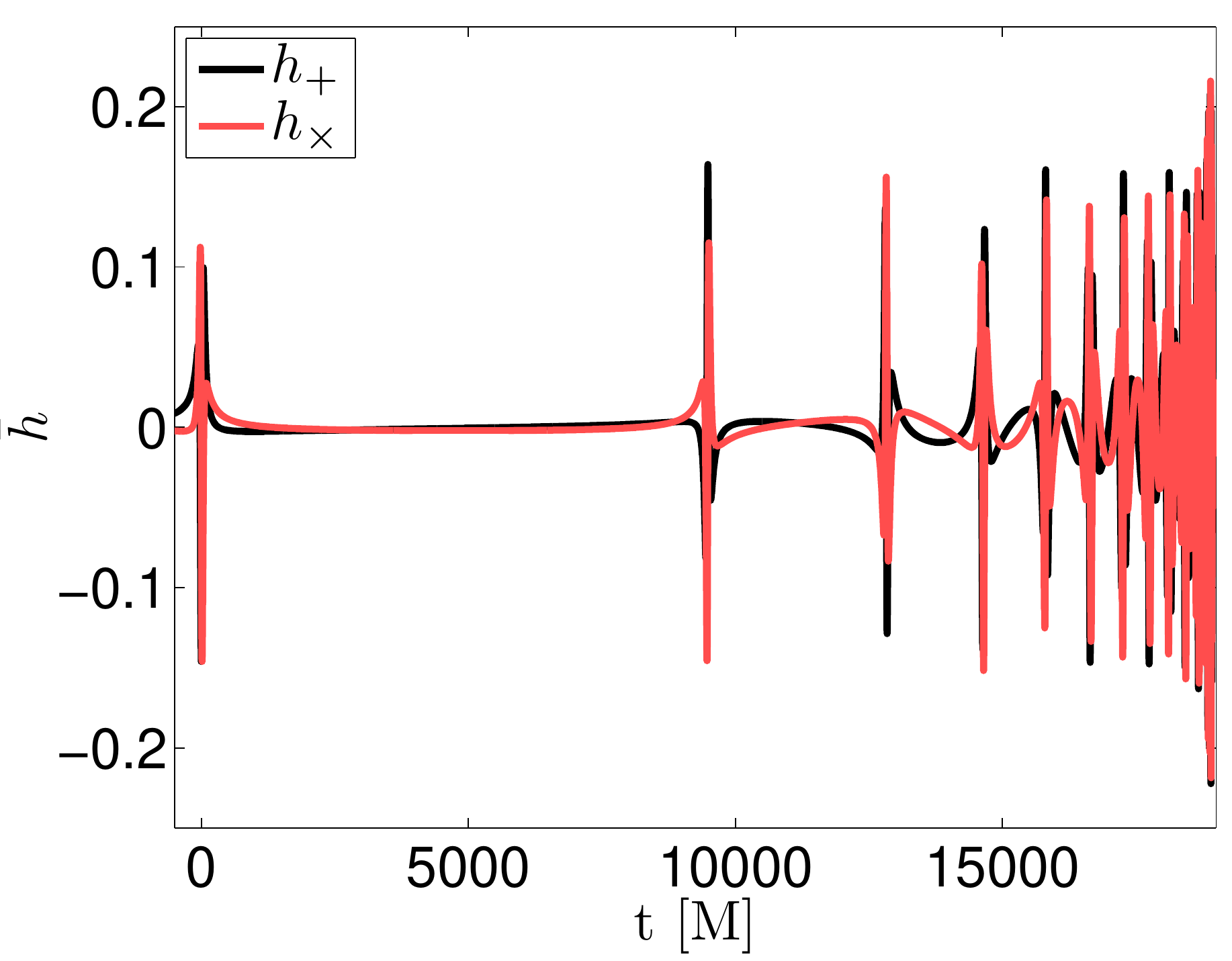}
\includegraphics[width=.4\textwidth]{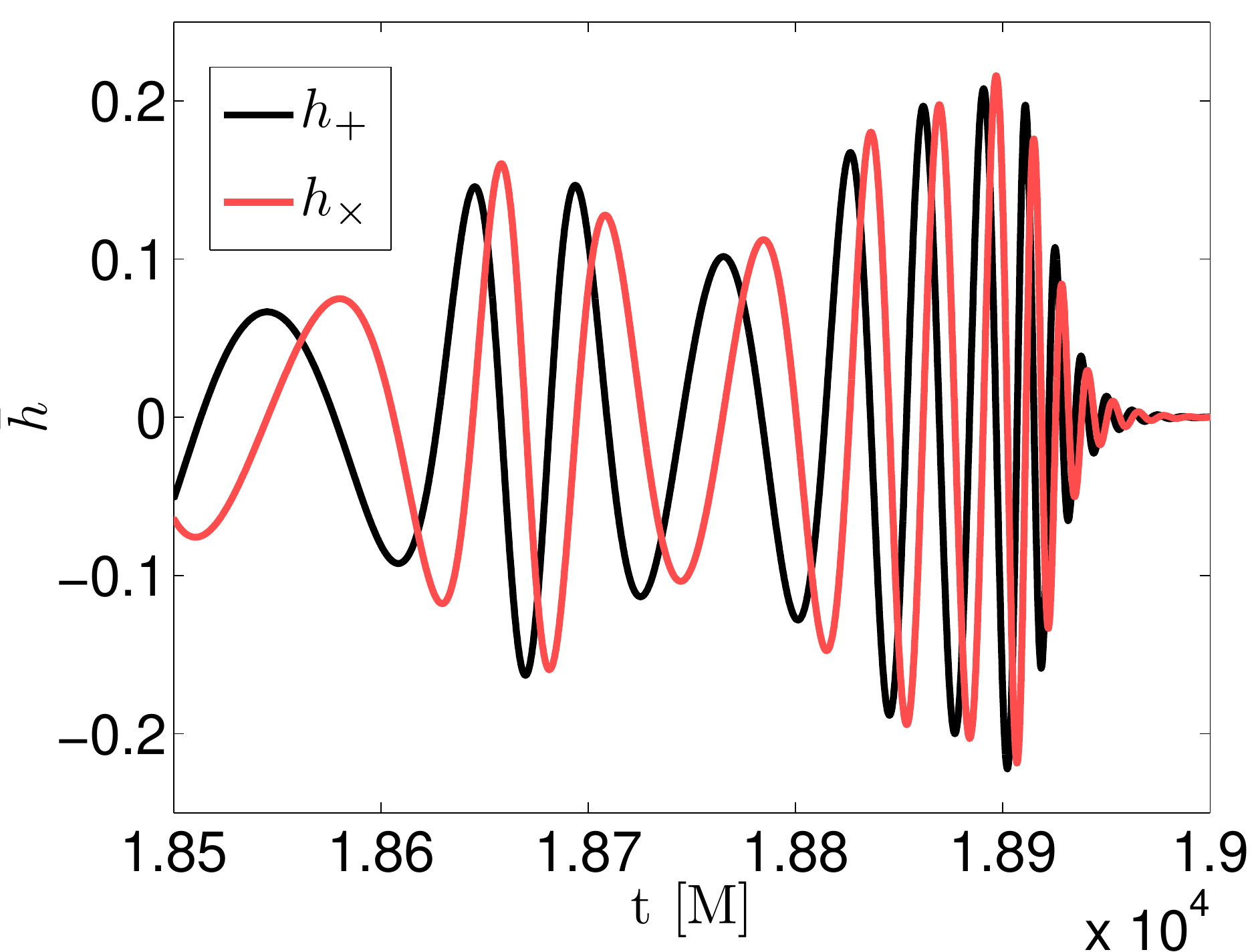}
\caption{
GW strain generated with our model and initial conditions $r_{\rm p}=8\,M$ and $e=1$.
The top panel shows the entire waveform, while the bottom panel shows a zoomed-in view of the end of the waveform.
\label{example_waveform}
}
\end{figure}

\begin{figure}
\includegraphics[width=.4\textwidth, viewport=0 25 521 418]{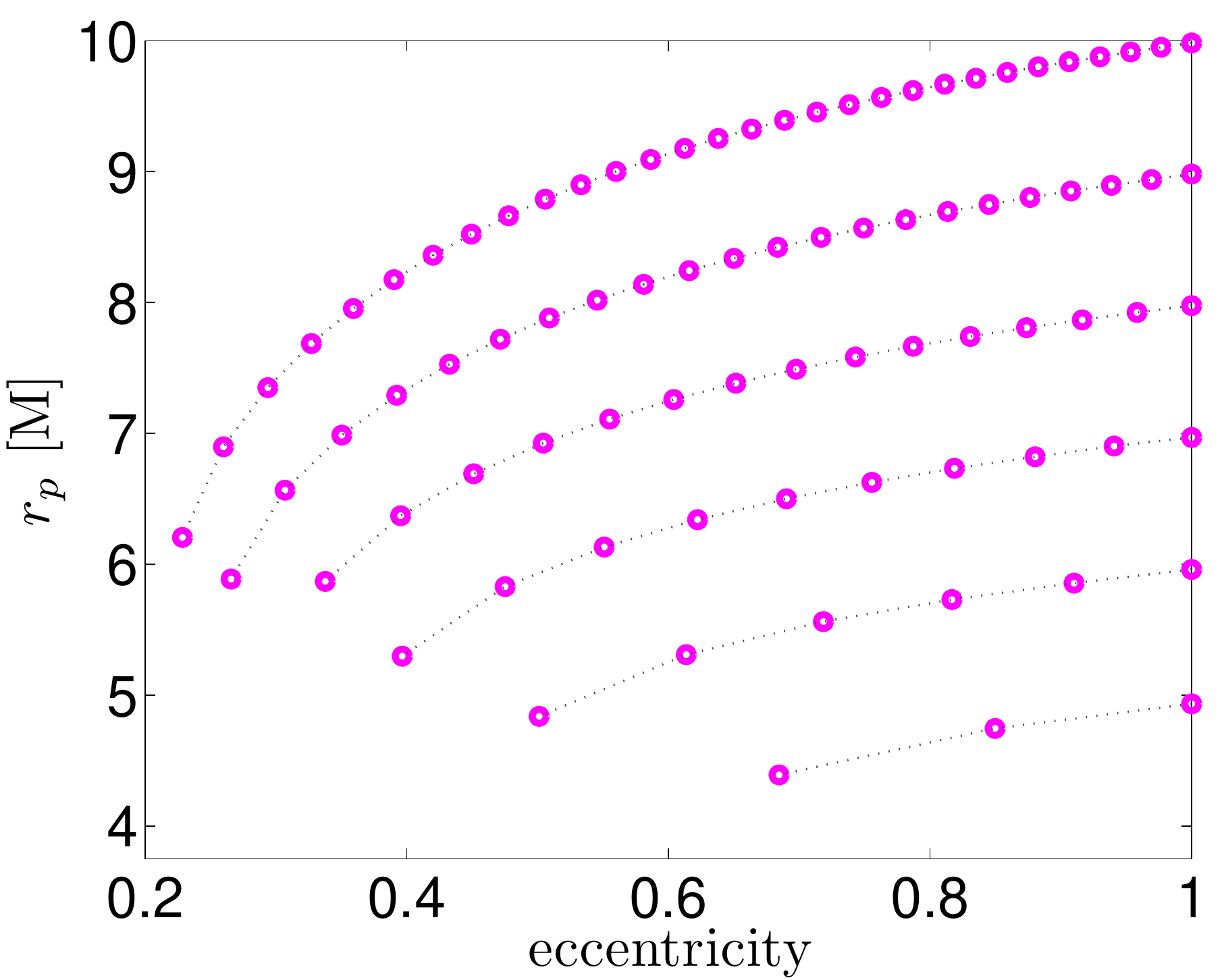}
\includegraphics[width=.4\textwidth]{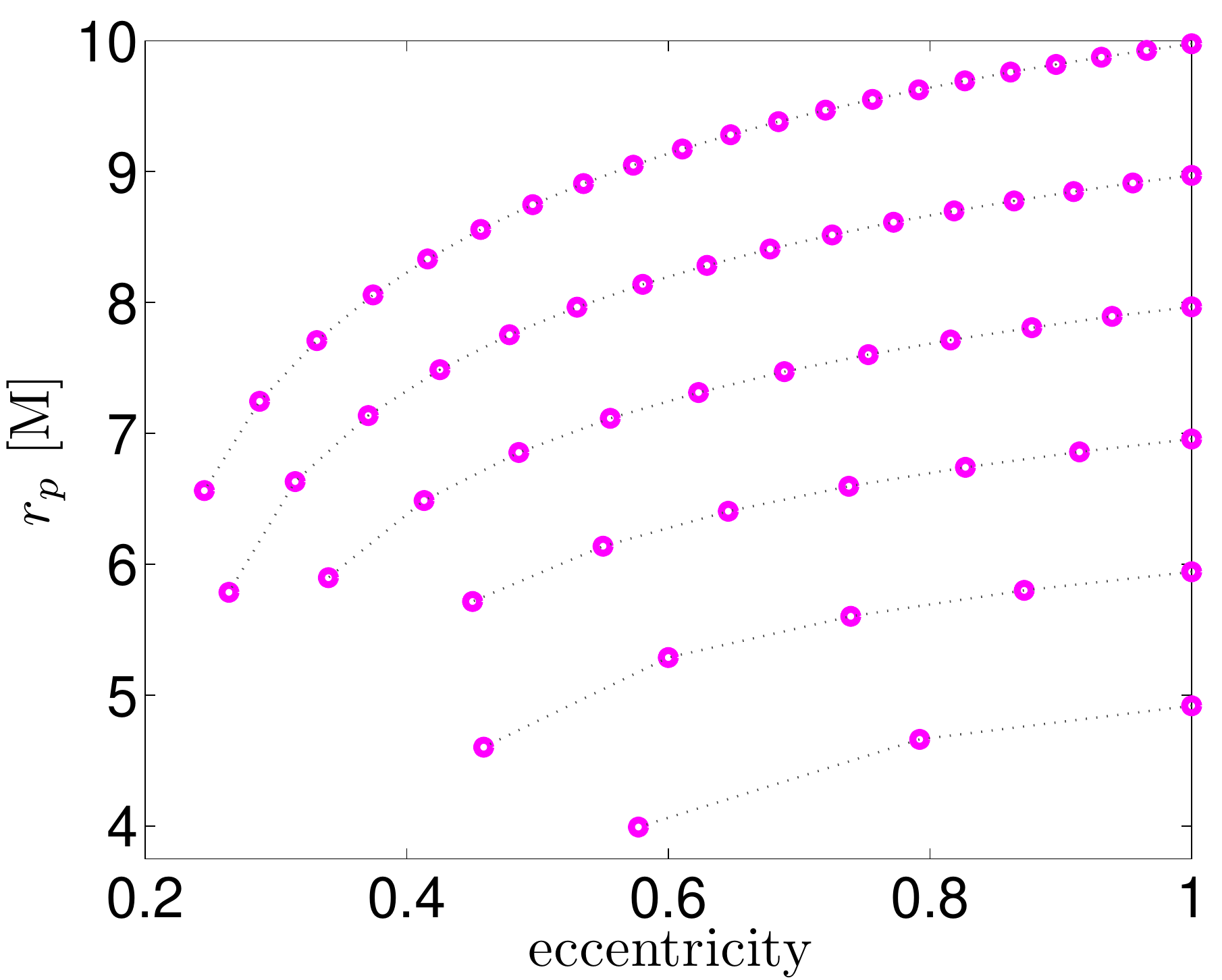}
\caption{
Evolution of orbital parameters for a 4:1 (top panel) and an equal mass ratio (bottom panel) binary.  The effective eccentricity 
is calculated from successive apoapse and periapse distances as $e=(r_a-r_p)/(r_a+r_p)$.
\label{orbit_evo}
}
\end{figure}

Combining the inspiral and merger models allows us to generate complete waveforms
for dynamical capture binaries.  In Fig.~\ref{example_waveform} we show one such example
for a 4:1 mass ratio system with initial orbital parameters corresponding to 
$r_{\rm p}=8\,M$ and $e=1$.  The waveform shows the decreasing time interval between
bursts from close encounters as $r_{\rm p}$ and $e$ decrease
due to gravitational radiation.  The number and timing of the bursts is a sensitive
function of the amount of energy and angular momentum radiated in each close encounter.   
In Fig.~\ref{orbit_evo} we show how $r_{\rm p}$ and $e$
evolve according to this model for some example binaries.  It can be seen that the binaries
considered here, which begin on parabolic orbits with $r_p \leq 10\,M$, still have non-negligible
eccentricity all the way to merger. 

\begin{figure}
\includegraphics[width=.4\textwidth]{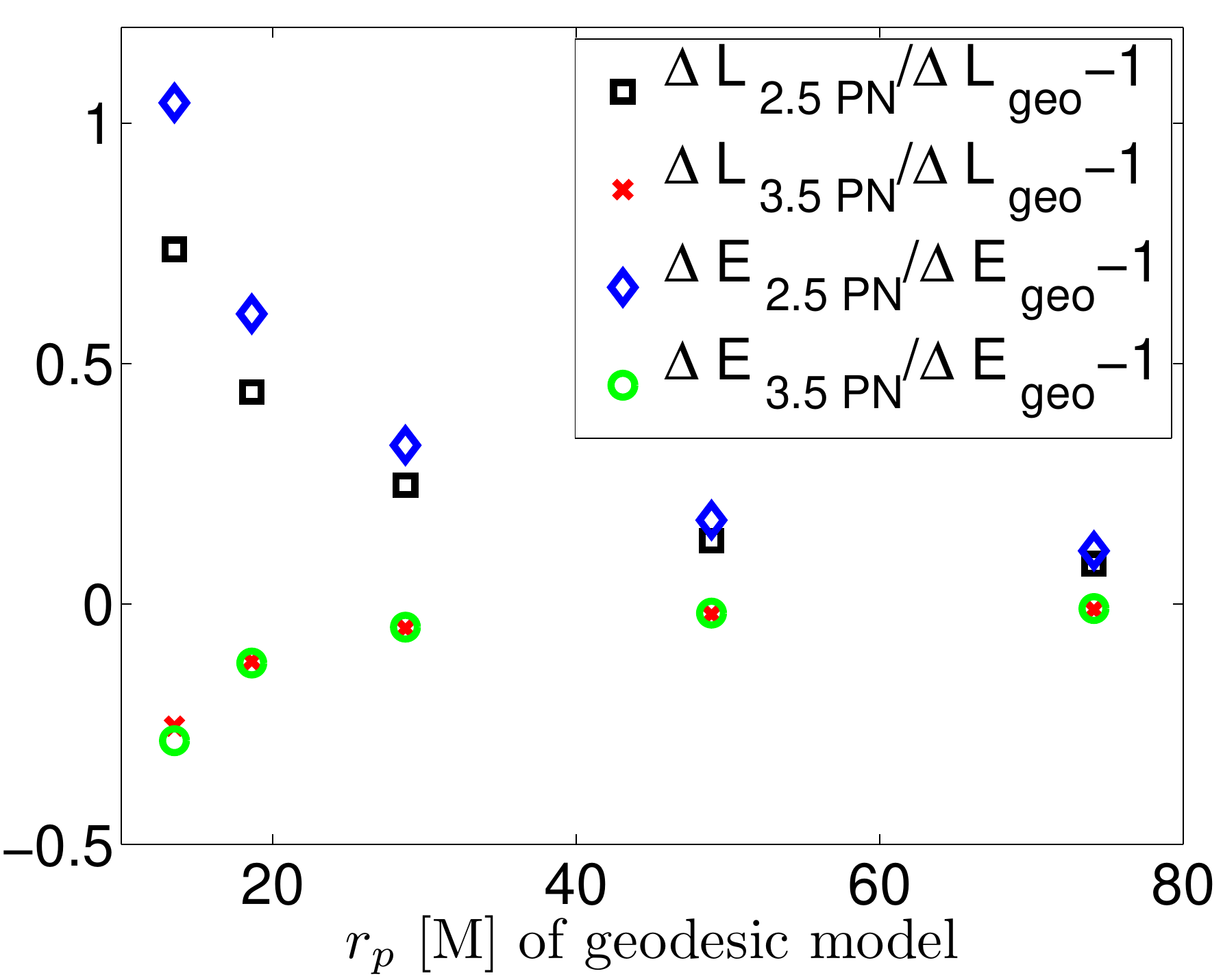}
\caption{
Relative difference in energy and angular momentum lost in a close encounter for the 2.5
and 3.5 PN approximations versus our model.
For this comparison the orbits are chosen to initially have zero energy and the same
value of angular momentum which, for a geodesic with the same initial conditions, 
corresponds to the value of $r_p$ indicated on the $x$ axis.
\label{pn_comp}
}
\end{figure}

We can also compare this model to that given by the 2.5 and 3.5 order PN approximation
as used in~\cite{Kocsis_Levin}.  In Fig.~\ref{pn_comp} we show how the difference in the energy
and angular momentum radiated away in a close encounter for 2.5 or 3.5 PN relative to our
model changes with the initial impact parameter.  The geodesic model predicts less energy and momentum
loss than 2.5 PN but more than the 3.5 PN.  At large impact parameters the three different models converge.
At smaller impact parameters the 2.5 and 3.5 PN approximations begin to diverge. As shown in~\cite{Levin_McWilliams},
the PN approximation fails to converge (or even to provide physically sensible results in the case of 3.5 PN) 
for $r_p\lesssim 10\,M$. 

The gravitational wave model we have outlined in this section is relatively
simple and could be improved upon by, for example, adding more sophisticated conservative dynamics,
including finite-size effects for NSs, as well as going beyond the quadrupole approximation
in determining gravitational radiation.  However, given the decent match between this model
and the full numerical simulations, as well as its consistency with the PN approximation as described above,
it can be used to investigate issues of detectability, as we do in the next section.

\section{Detectability}
\label{sec:det}
\subsection{Detector modeling}
\begin{figure}
\includegraphics[width=.48\textwidth]{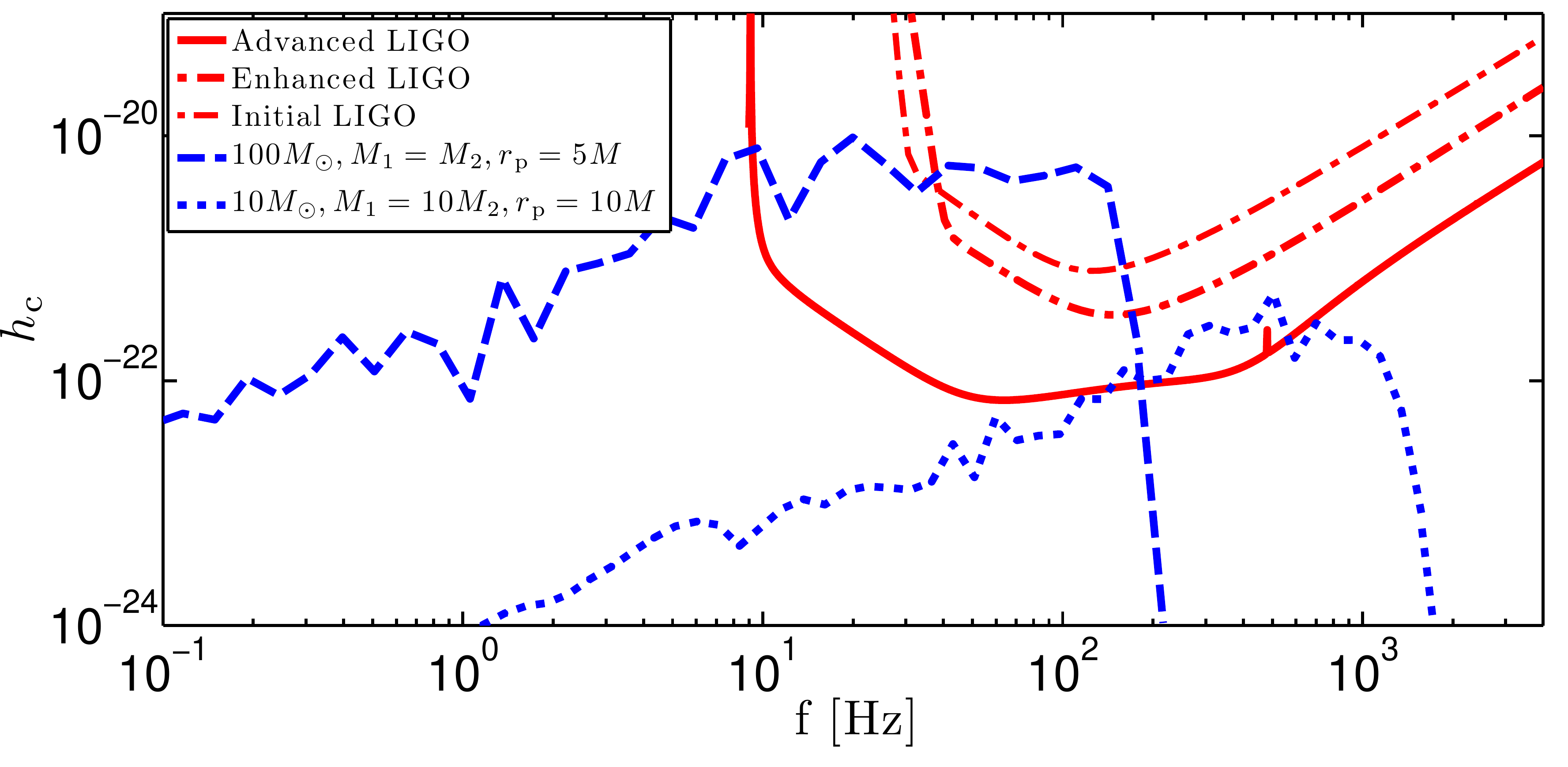}
\caption{
The characteristic strain $h_{\rm c}$ is shown for the initial (thin dash-dotted line), Enhanced (dash-dotted line), and Advanced LIGO (solid line)
detectors, as well as for two example signals at $D_L=1$ Gpc.  The first signal corresponds to an orientation-averaged source with $M=100\,\Msun$,
$q=1$, and $r_{\rm p}=5M$ (dashed line), and the second signal is from a source with $M=10\,\Msun$, $q=0.1$, and $r_{\rm p}=10M$ (dotted line).
Both signal spectra are smoothed to diminish fluctuations and make the trend more clear.  The system with $q=0.1$ has little contribution from  
the merger, so the repeated burst phase dominates the spectra, with $h_{\rm c} \appropto f$, whereas the $q=1$ system signal
comes largely from the merger, where $h_{\rm c} \approx {\rm constant}$ over a small band of frequencies.
\label{hc}
}
\end{figure}

Having developed a model for the gravitational waveforms emitted by high-eccentricity binaries, we 
can now assess the detectability
of these signals for different source parameters and detectors.  The measured strain $h$ is given by
\beq
h = \Re\left[ F \bar{h} \right] = F_{+} h_{+} + F_{\times} h_{\times}\, ,
\eeq
where $F \equiv F_+ - \imagi F_{\times}$ is the sky-dependent detector response.
The signal-to-noise ratio (SNR) $\rho$ using a perfectly matched filter is given by
\beq
\label{eqn:snr_def}
  \rho^2 = \langle h | h \rangle,
\eeq
where $\langle\cdot|\cdot\rangle$ denotes a noise-weighted inner product given by
\beq
\langle h_{1}|h_{2}\rangle \equiv 2\int\limits_{0}^{\infty} df\, \frac{\tilde{h}_{1}^{*}\tilde{h}_{2} + \tilde{h}_{1}\tilde{h}_{2}^{*}}{S_{n}} \, ,
\label{eqn:dotprod}
\eeq
where $S_n(f)$ is the power spectral density of the detector noise, and
$\tilde{h}$ denotes the Fourier transform of the original $h$ time series.
Because we limit our model to the quadrupole component of the signal,
and we focus on detectors (like LIGO) for which the gravitational wavelength is much longer than the detector's armlength, we can trivially relate
the SNR of an optimally oriented and located source to the SNR of an orientation- and sky-location-averaged source.  For such detectors, 
the response function to the two waveform
polarizations, $F_+$ and $F_{\times}$, is simply
the root-mean-squared (rms) average over the sky location and polarization angles, $\sqrt{\langle F_{+,\,\times}^2\rangle }=\sqrt{1/5}$ \cite{Thorne300}.
Likewise, the rms average over source orientations is $\sqrt{ \langle {}_{-2}Y_{2,\,\pm 2}\rangle }=\sqrt{1/5}$,
so that $\sqrt{\langle \rho^2 \rangle } = \rho_{\rm opt}/5$.  We can further define the characteristic strain $h_{\rm c}$ for
both the signal and detector noise.  Given Eq.~\eqref{eqn:dotprod} and the typical practice of plotting sensitivity curves logarithmically,
it is useful to define $h_{\rm c} \equiv \sqrt{\langle {}_{-2}Y_{2,\,\pm 2}\rangle } f \tilde{h}_{\rm opt}$ for signals
and $h_{\rm c} \equiv \sqrt{f S_n/\langle F^2 \rangle}$ for detector noise, so that both signal and noise are characterized
as a dimensionless strain, and the ratio of signal-$h_{\rm c}$ to noise-$h_{\rm c}$ is the square root of the integrand for $\rho^2$
when integrated over logarithmic frequency intervals $df/f$.  We show this characterization of signal and noise in Fig.~\ref{hc}.

\begin{figure}
\includegraphics[trim = 0mm 0mm 0mm 0mm, clip, width=.49\textwidth, angle=0]{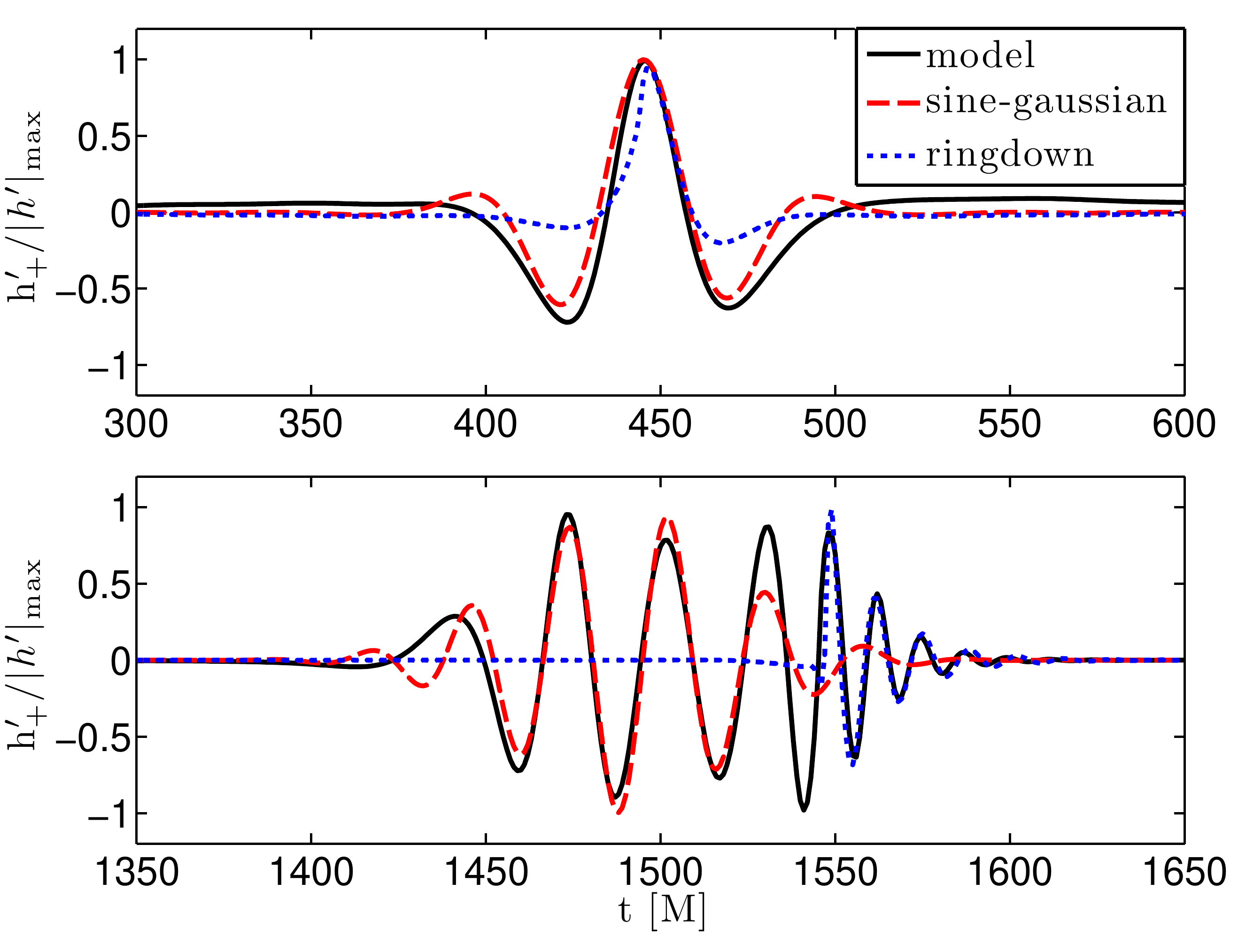}
\caption
{
Whitened waveforms for a 10 $\Msun$ (top panel) and a 100 $\Msun$ (bottom panel) binary with initial $e=1$ and $r_p = 5M$,
along with the (whitened) best--fit template among sine-Gaussian and ringdown templates.
}
\label{fig:hw}
\end{figure}

For assessing the relative contribution of different waveform segments to the SNR,
it is often convenient to work in the time domain by constructing ``whitened'' waveforms \cite{Damour:2000gg},
which weight the amplitude of the waveform as a function of frequency to account for the presence
of noise in the detector,
\beq
h' = \int\limits_{-\infty}^{+\infty} df\, \frac{\tilde{h}}{\sqrt{S_n}}\,e^{-\imagi 2\pi f t}.
\label{eqn:whiten}
\eeq
With these whitened vectors, the noise-weighted inner product \eqref{eqn:dotprod}
can be reexpressed in the time domain:
\beq
\langle h_1 | h_2 \rangle \equiv \int\limits_{-\infty}^{\infty} dt \, h_1'^*(t)\,h_2'(t).
\label{eqn:A_timeprod}
\eeq
Figure \ref{fig:hw} shows portions of the whitened waveform for two example cases with the same mass ratio and initial $r_p$ and $e$,
but different masses.  The upper panel
shows the burst with the largest SNR contribution for a source with total mass $M = 10 \,\Msun$, 
while the lower panel shows the loudest burst for $M = 100 \, \Msun$.  The different masses
change the frequency of the signal, so different bursts are emphasized by the detector 
sensitivity; in particular, for larger masses the final burst and merger 
are emphasized.  We also show best fits for two types of templates that are described below.

\subsection{Templates and detection strategies}
While quasicircular sources are searched for using matched filtering, eccentric
systems are far more susceptible to modeling errors in the relative timing and phase of signal bursts, which is why we focus our attention
on alternative approaches to detection.  For example, a small modeling error in the energy lost during a particular periapse passage $\delta E$
will induce a timing error in the arrival time of the subsequent burst $\delta T$ given by $\delta T \propto \delta E (1-e)^{-5/2}$.  Therefore,
dynamical capture binaries are far more challenging to model with sufficient accuracy to apply matched filtering due to 
their large eccentricities.

We assess detection prospects of GWs from capture binaries for two currently used templates, 
sine-Gaussian (SG) and ringdown (RD) templates, as well as an idealization of
a third strategy based on combining an excess power search with stacking. 
The SG and RD both take the form
\beq
\bar{h} = A \exp \left[ -\left( \frac{t - t_o}{\tau} \right)^{\gamma} + \imagi\omega (t-t_o) - \imagi\phi_o \right],
\eeq
where $\gamma = 1$ and $t \geq t_o$ for the RD templates, and $\gamma = 2$ with $-\infty < t < \infty $ for the SG templates.
Here $A$ is the overall amplitude, $t_o$ and $\phi_o$ are the time and phase of the 
template's amplitude peak, $\tau$ sets the e-folding of the amplitude,
and $\omega$ is the constant frequency.

In addition to assessing the performance of two burst templates, 
we calculate a rough approximation of the potential performance for an excess power search that accumulates power 
from the entire signal~\cite{Kalmus}, which we will call a power-stacking search.
Here, the data would be transformed to a time-frequency (TF) tiling using a basis suitable for capturing
individual bursts within a tile, and
then power from different tiles corresponding to bursts, as informed by our model, would be combined.
Whereas in most existing TF searches an individual element must have enough power to exceed some threshold,
with that threshold being large enough to avoid many false alarms, the approach we describe does not require that the signal be detectable
in any single TF element.  In the case of a monochromatic signal, the SNR
from optimal filtering will accumulate with the number of cycles $N$ as $\sqrt{N}$, while the excess power 
in stacking TF elements (constructed using any basis) overlapping the given frequency 
will accumulate as $N^{1/4}$.  The signals from high-eccentricity binaries are not monochromatic,
but given the typically large number of bursts occurring in band, and the relative flatness of both the source spectrum and the
detector sensitivity across its most sensitive band, we expect the aforementioned scalings to hold approximately for realistic signals.  

This search would be very similar to the stacked search proposed for
combining potential GW counterparts to observed electromagnetic signals from soft gamma repeaters \cite{Kalmus}. There,
TF elements were aligned in time based on the observed bursts, and they demonstrated the $N^{1/4}$ SNR scaling 
when adding power for identical injected signals.
Since we do not have a separate observational trigger, our proposed search would sum power along elements overlapping bursts
as indicated by our waveform model.
We leave it to future work to fully investigate this, though here
we assume we can achieve the $N^{1/4}$ SNR scaling, and thus can estimate
the performance of a power-stacking search
by noting that optimal filtering should outperform power stacking
by $N^{1/4}$. Hence, we can approximate the effective excess power SNR 
as $\rho_{\rm EP} \approx N^{-1/4} \rho$.  This simple estimate will constitute
our third search technique in our subsequent analysis.

We do not employ quasicircular (QC) templates, although they have thus far been the only tool employed
to search for long-lived signals.  QC templates will generically fail to match the performance 
of any of the above methods for the repeated burst phase of
eccentric sources for the following reasons. 
First, during the long intervals between eccentric bursts
a QC template will still be integrating
power from the data, which is predominantly noise. Specifically,
the ratio of the characteristic time scale of an eccentric burst to the period between bursts is roughly
\beq
\frac{\tau_{\rm GW}}{T} \approx \left(1-e\right)^{3/2}.
\label{eq:neonc}
\eeq
In other words, there will be $\sim(1-e)^{-3/2}$ additional cycles between bursts
in a QC signal with the same periapse.
Moreover, even if the QC template is phase aligned to a particular burst,
since the time between bursts is much larger than the GW period,
the rest of the template will effectively have random phase alignment with other bursts
in the sequence and, on average, no additional SNR will be acquired. To summarize, typically 
the best-matched QC template will only integrate signal about the loudest burst, but even
so, the performance will not be as good as a single-burst search due to the larger integrated
noise accumulated over the period of the QC template (expect for the higher mass systems
where only the final merger/ringdown signal is in band).

\subsection{Results}\label{sec_results}
We calculate two useful quantities related to the SNR: the detectability horizon and the detection probability.
Since $h \propto D_L^{-1}$, where $D_L$ is the luminosity distance, we can use (\ref{eqn:snr_def}) to calculate the
distance (which we call the detection horizon) at which a sky- and orientation-averaged source could be 
observed with a SNR of 8 using optimal filtering.
The detection probability for a given strategy is simply the ratio 
of the volume in which the strategy could detect a source with some SNR to
the volume in which the source could be seen with the same SNR using optimal filtering.
In the remainder of this section, we will calculate these quantities for various cases of interest.
We consider the following configurations:
\begin{itemize}
\item three detector sensitivities, corresponding to initial, Enhanced, and Advanced LIGO;
\item three detection strategies, including SGs, RDs, and power stacking, and how they compare to optimal filtering;
\item three intrinsic system parameters:
\begin{itemize}
\item total system mass $M$, ranging from $1\,\Msun$--$2000\,\Msun$;
\item mass ratio $q$ of the binary components, ranging from $0.01$--$1$;
\item initial $r_p$, ranging from $5M$--$10M$ (with initial $e=1$; we exclude $r_p<5M$ simply because in most cases
it is a direct collision qualitatively similar to $r_p=5$, and
see~\cite{Kocsis_Levin} for a study of $r_p>10M$).
\end{itemize}
\end{itemize}

\begin{figure}[t]
\includegraphics[trim = 0mm 0mm 0mm 0mm, clip, width=.45\textwidth, angle=0]{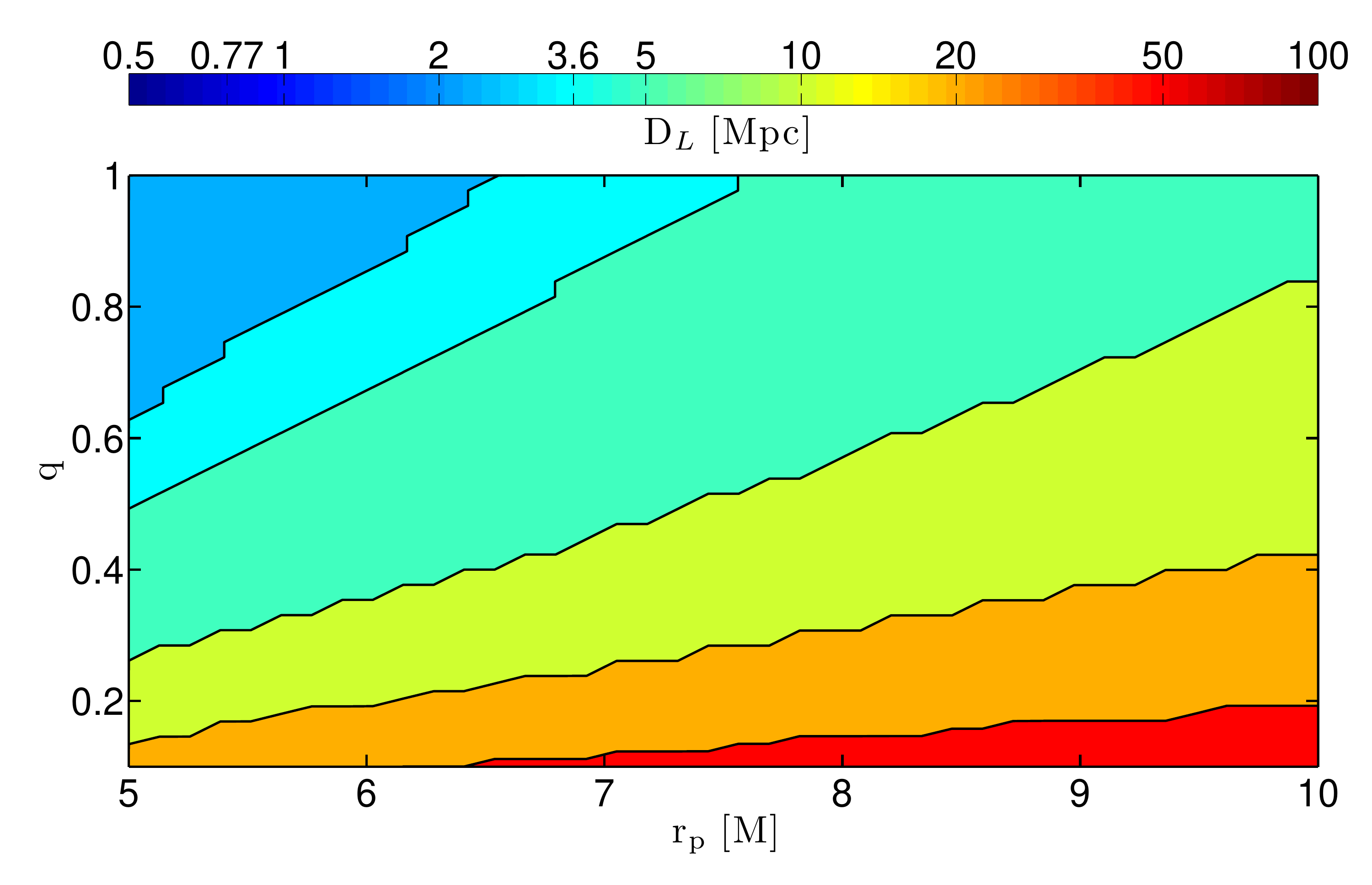}
\includegraphics[trim = 0mm 0mm 0mm 0mm, clip, width=.45\textwidth, angle=0]{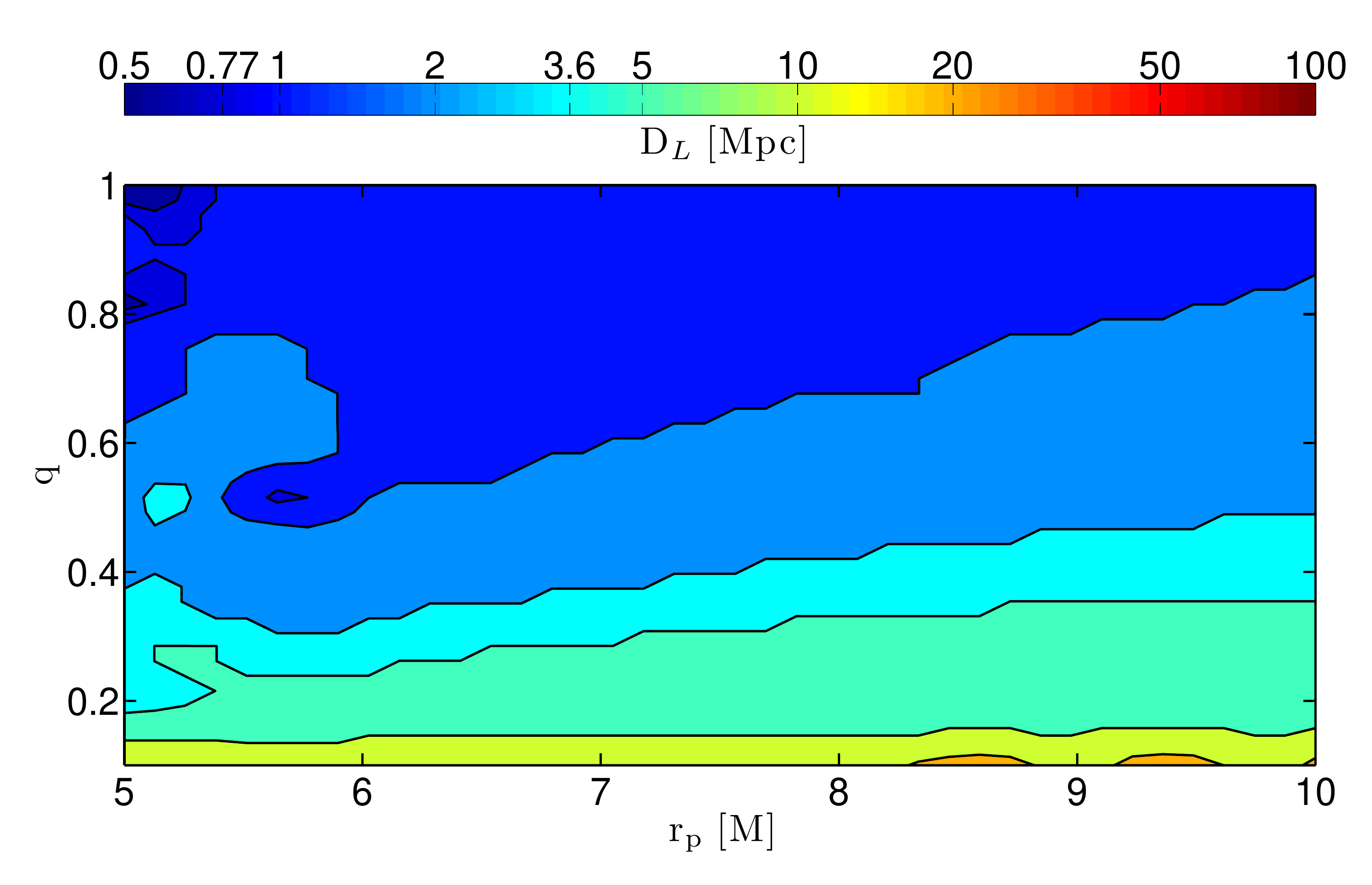}
\includegraphics[trim = 0mm 0mm 0mm 0mm, clip, width=.45\textwidth, angle=0]{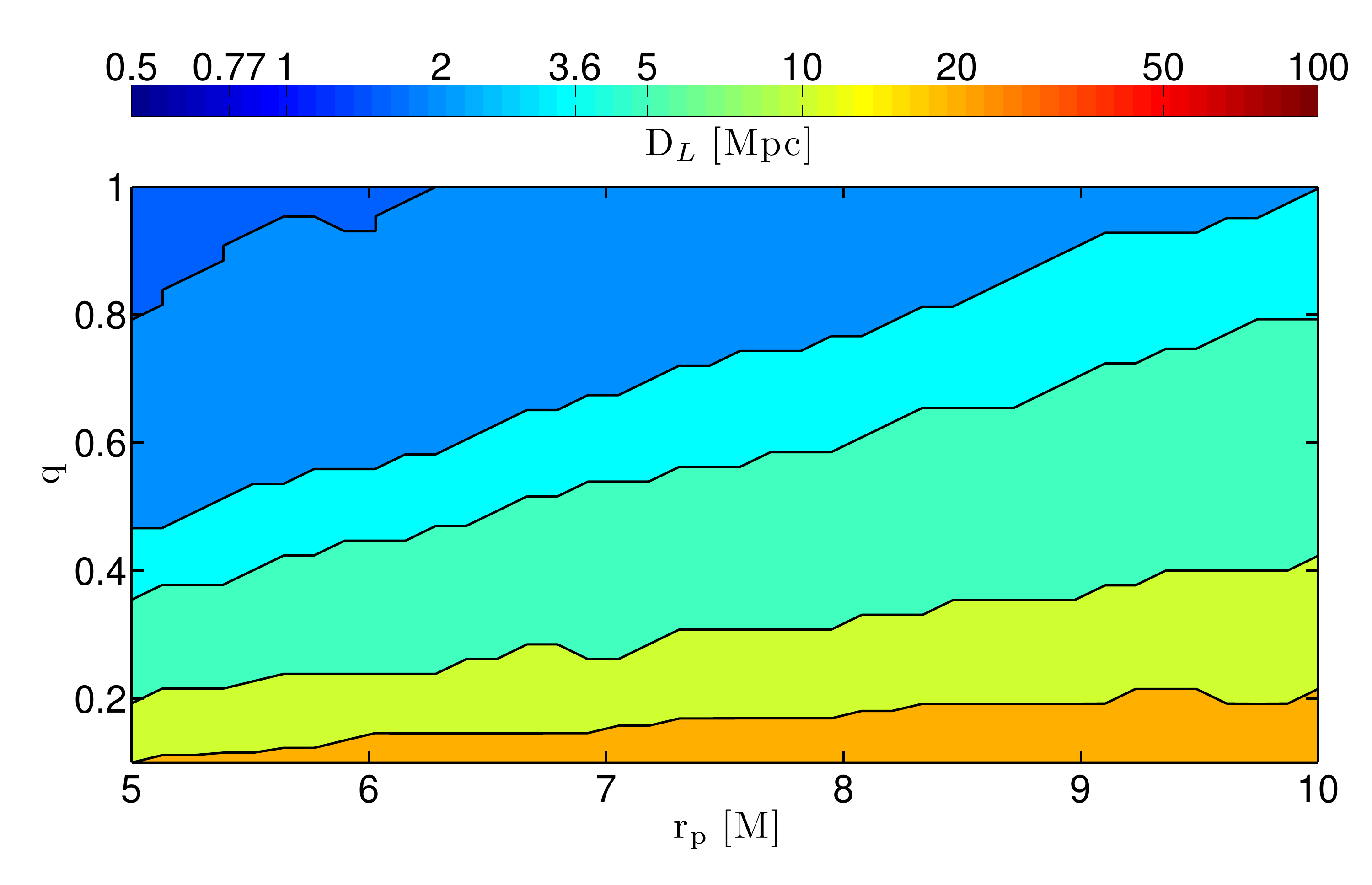}
\caption
{
Contours of horizon distance ($\rho=8$) as a function of mass ratio $q$ and pericenter separation $r_p$ for initial LIGO
using an optimal filter (top panel), 
sine-Gaussian templates (middle panel), and an estimate of a power-stacking search (bottom panel) as described in the text. 
We fix one component to be a 1.35 $\Msun$ neutron star and change the total mass with mass ratio accordingly.
We include a contour at $D_L=0.77$ Mpc and another at 3.6 Mpc, to show the region of parameter space where
existing LIGO searches would not have seen a gravitational wave counterpart to GRB070201 \cite{Mazets} and GRB051103 \cite{Hurley}, respectively.
}
\label{fig:q-r-LIGO-DL}
\end{figure}

In Fig.~\ref{fig:q-r-LIGO-DL}, we show contours of constant horizon distance as a function of $q$ and $r_p$ for initial LIGO,
assuming optimal filtering, SG templates, and power stacking.  Two contours of note, at 0.77 and 3.6 Mpc, correspond to the distances of 
GRB070201 \cite{Mazets} and GRB051103 \cite{Hurley}, respectively.  These were two nearby gamma-ray bursts observed by Swift during the S5 
initial LIGO run, while two interferometers were actively collecting data at or near initial LIGO's design sensitivity.
However, no signal was found in the LIGO data using 
the methods applied (specifically, various burst and quasicircular inspiral templates) for these GRBs, nor for any of the
137 GRBs (35 with measured redshifts) that occurred while initial LIGO was taking science data during its S5 run at or near
design sensitivity \cite{Abbott2010,Abadie2010}.  
Thus in Fig.~\ref{fig:q-r-LIGO-DL} we restrict the mass ratio to the range 0.1--1,
with one of the masses fixed at 1.35 $\Msun$, to focus on systems including a neutron star that are expected to generate GRBs.  In the case
of a dynamical capture binary source at 0.77 Mpc, the signal is sufficiently loud that even suboptimal searches like SG templates would detect them.
However, for a source at 3.6 Mpc, whereas
an optimal filter would have detected a signal from a large region of the parameter space, including all cases with $q<0.5$ or $r_p > 7.5\,M$,
and power stacking would recover signals with $q<0.4$, SG templates are far less effective, and would only recover a small sliver of
parameter space with $q<0.2$.  This suggests the possibility that the searches applied to the LIGO data would not have found the gravitational
wave counterpart to GRB051103 if it was in the form of a dynamical capture binary.  Furthermore, across the full parameter space 
explored, the difference in performance
among these three searches is substantial, with optimal filtering detecting sources as far as $D_L=50$--100 Mpc, while power stacking 
only reaches $D_L\approx 30$ Mpc, and SG templates only reach $D_L\approx 15$ Mpc.

\begin{figure}[t]
\includegraphics[trim = 0mm 0mm 0mm 0mm, clip, width=.45\textwidth, angle=0]{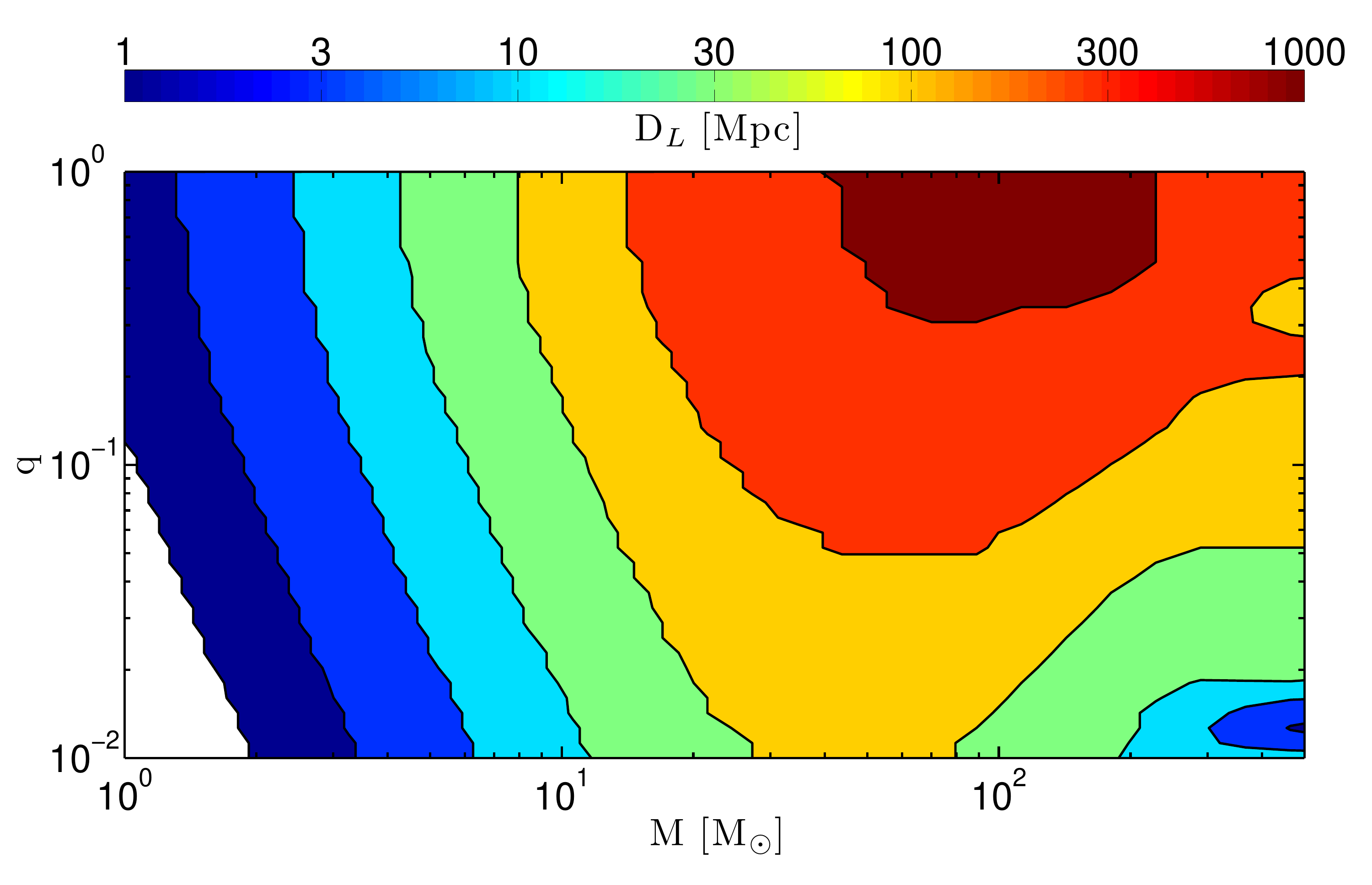}
\includegraphics[trim = 0mm 0mm 0mm 0mm, clip, width=.45\textwidth, angle=0]{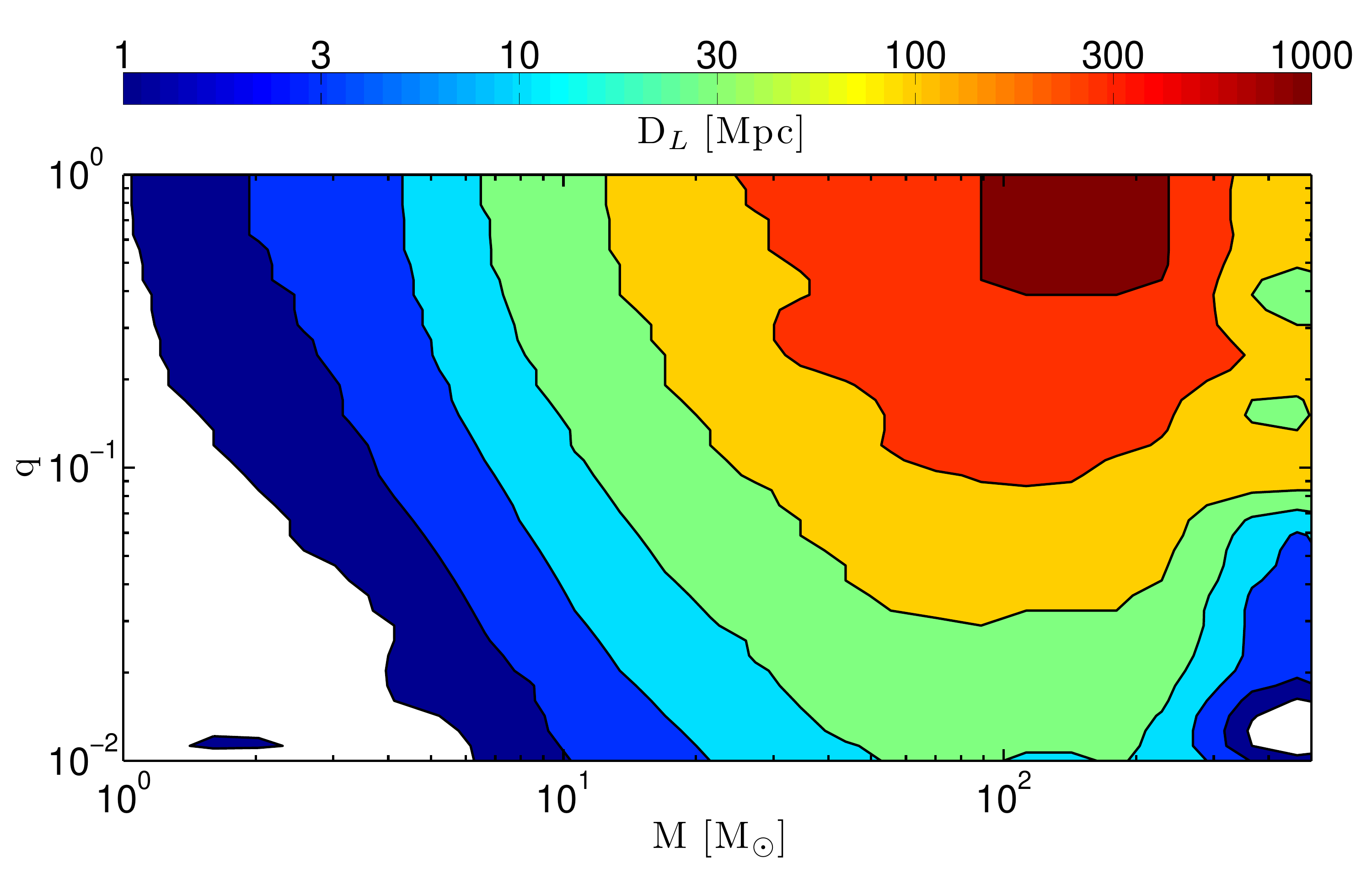}
\includegraphics[trim = 0mm 0mm 0mm 0mm, clip, width=.45\textwidth, angle=0]{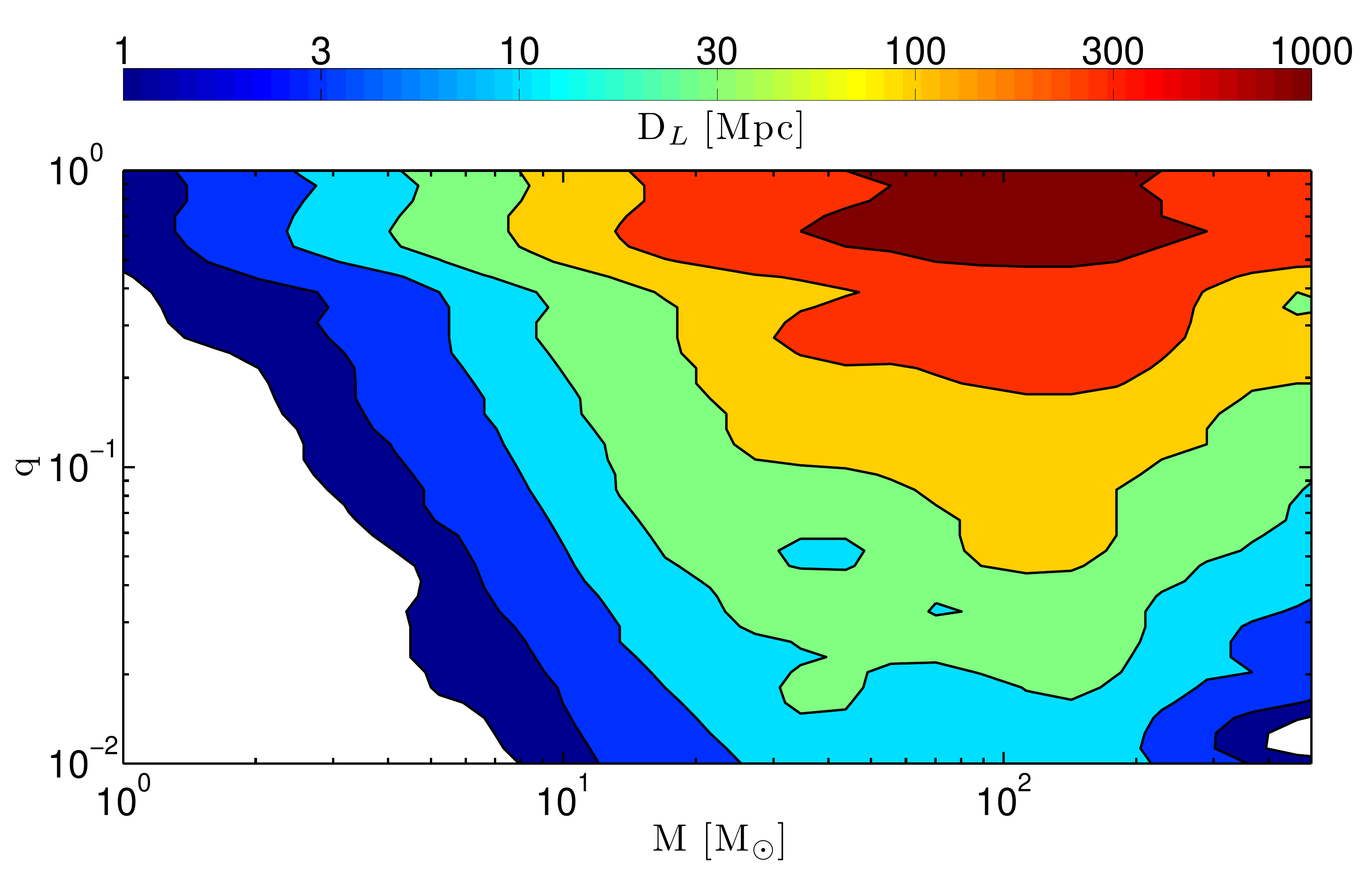}
\caption
{
Contours of horizon distance as a function of rest mass $M$ and mass ratio $q$ for Enhanced LIGO using an optimal filter (top panel), 
sine-Gaussian templates (middle panel), and ringdown templates (bottom panel) for an initial pericenter separation of $r_p=6M$.
}
\label{fig:M-q-ELIGO-DL}
\end{figure}

\begin{figure}[t]
\includegraphics[trim = 0mm 0mm 0mm 0mm, clip, width=.45\textwidth, angle=0]{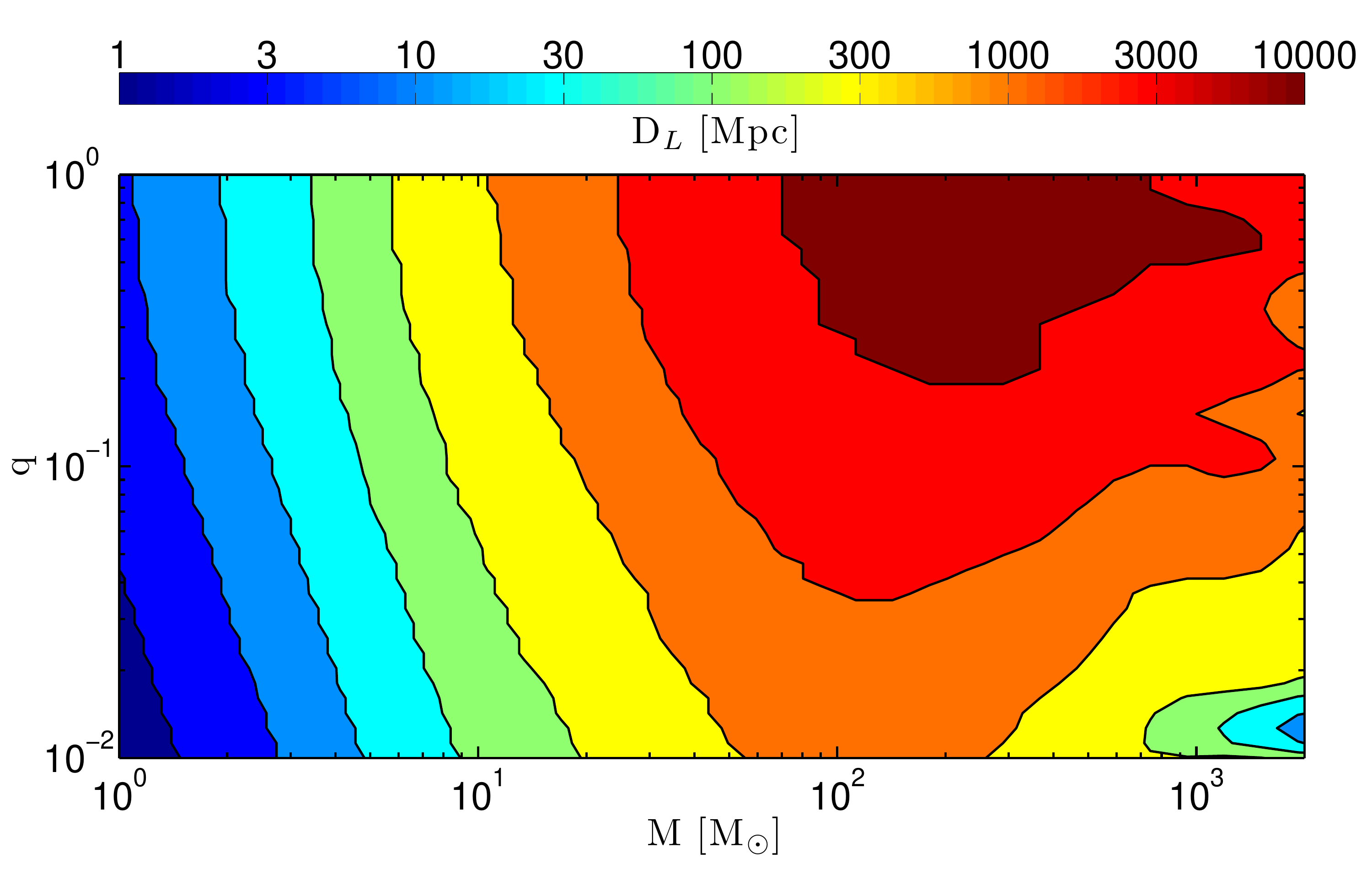}
\includegraphics[trim = 0mm 0mm 0mm 0mm, clip, width=.45\textwidth, angle=0]{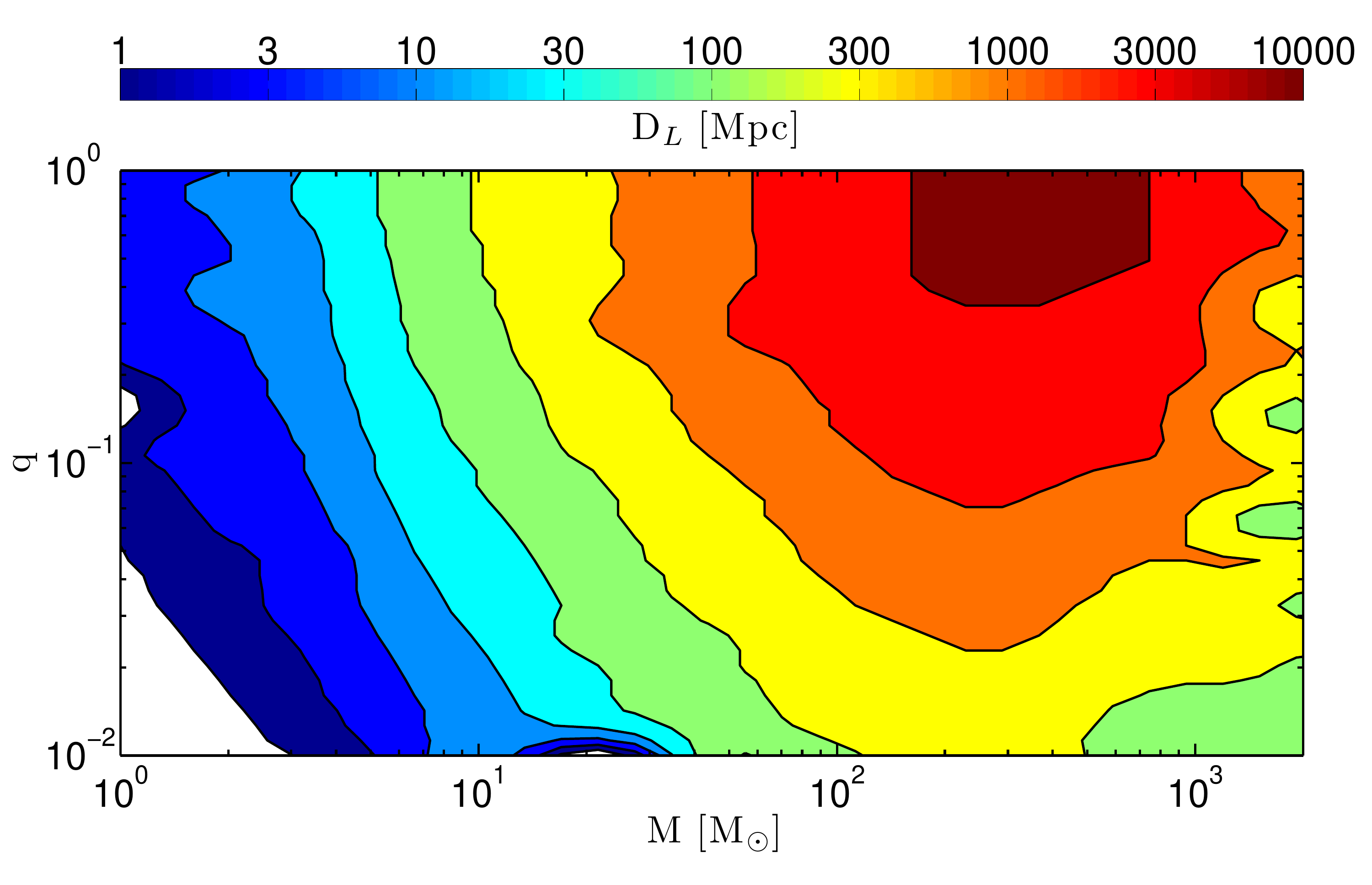}
\includegraphics[trim = 0mm 0mm 0mm 0mm, clip, width=.45\textwidth, angle=0]{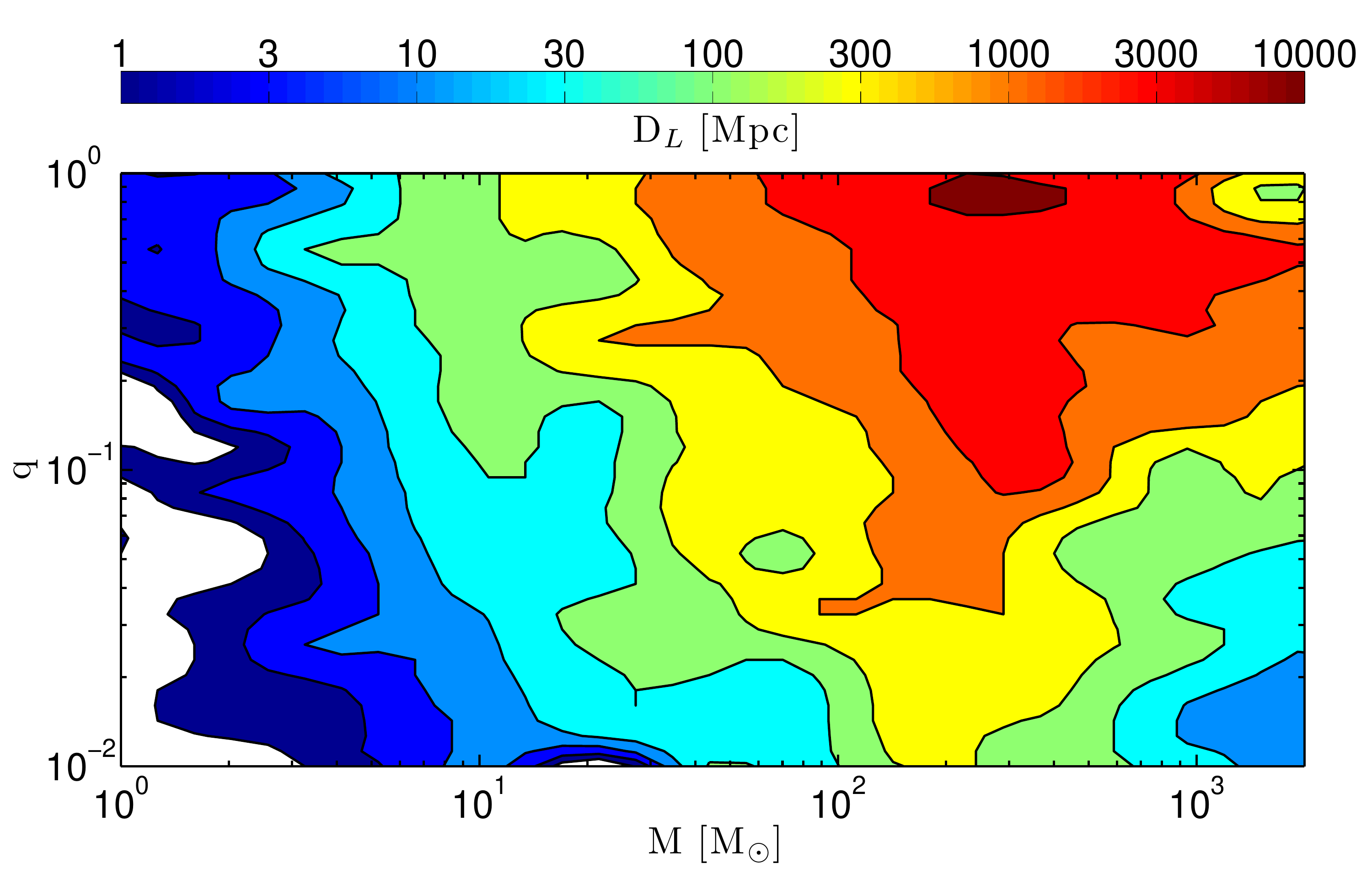}
\caption
{
Contours of horizon distance as a function of rest mass $M$ and mass ratio $q$ for Advanced LIGO using an optimal filter (top panel),
sine-Gaussian templates (middle panel), and ringdown templates (bottom panel) for an initial pericenter separation of $r_p=6M$.
}
\label{fig:M-q-ADVL-DL}
\end{figure}

Figures~\ref{fig:M-q-ELIGO-DL} and \ref{fig:M-q-ADVL-DL} show, for Enhanced and Advanced LIGO respectively, contours
of detection horizon as a function of mass and mass ratio at a fixed $r_p=6\,M$
using an optimal filter and SG and RD templates.  The primary difference in both
cases is the degradation of performance for higher mass ratios (smaller $q$), with the SG performing as well as or better than the RD templates across much of the
parameter space, with the exception of comparable mass ratios, where the ringdown signal is most emphasized. 
For each search,
Enhanced LIGO could detect an equal mass binary with $M=100\,\Msun$ out to $D_L=1$ Gpc, and Advanced LIGO will see the same sources
beyond 10 Gpc.  

\begin{figure}[t]
\includegraphics[trim = 0mm 0mm 0mm 0mm, clip, width=.45\textwidth, angle=0]{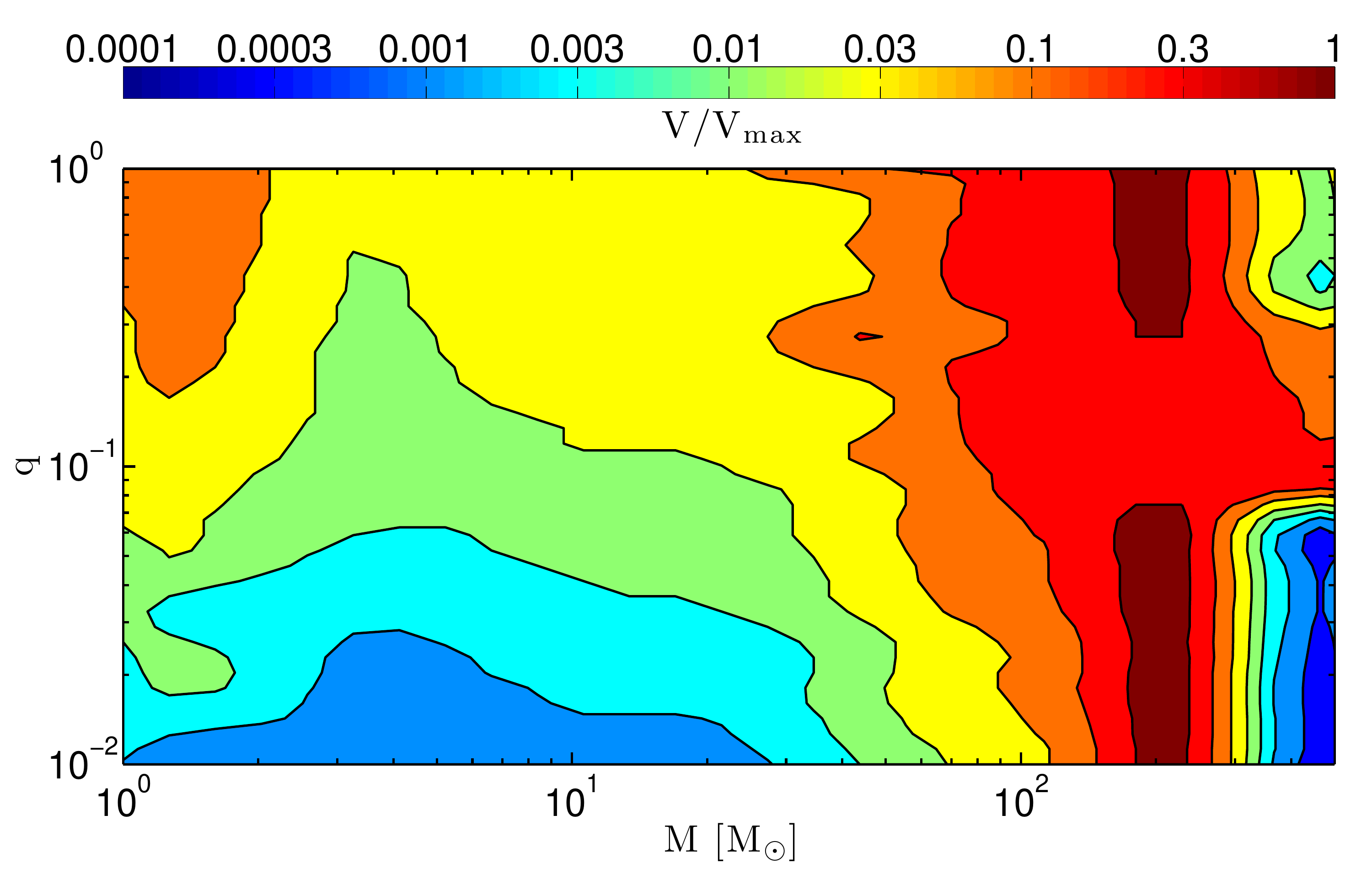}
\includegraphics[trim = 0mm 0mm 0mm 0mm, clip, width=.45\textwidth, angle=0]{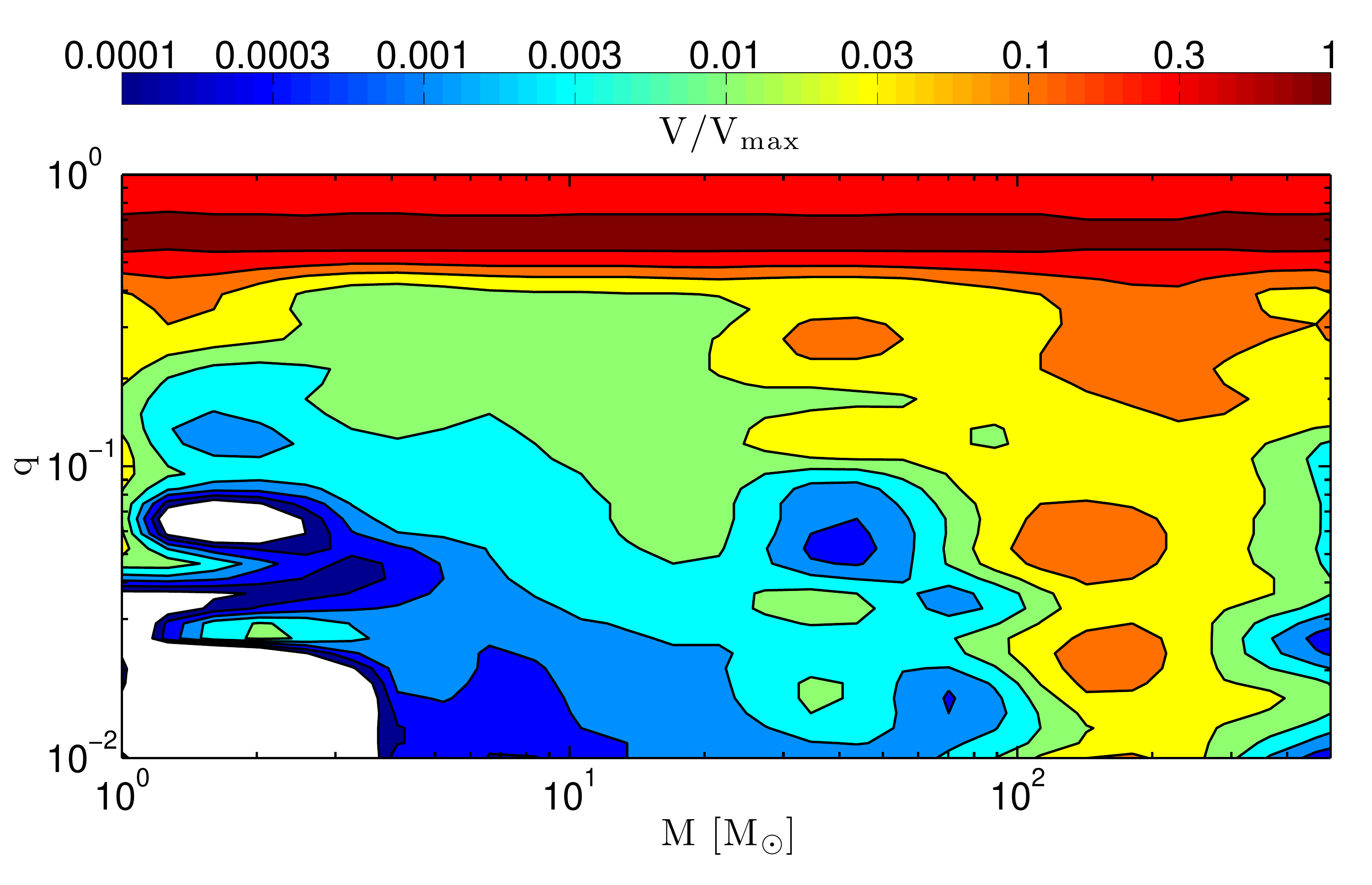}
\caption
{
Contours of detection probability $p\equiv V/V_{\rm max}$ as a function of rest mass $M$ and mass ratio $q$ for Enhanced LIGO
for a source inside the optimal filtering distance horizon, using sine-Gaussian (top panel) and ringdown (bottom panel)
templates for an initial pericenter separation of $r_p=6M$.
}
\label{fig:M-q-ELIGO-V}
\end{figure}

\begin{figure}[t]
\includegraphics[trim = 0mm 0mm 0mm 0mm, clip, width=.45\textwidth, angle=0]{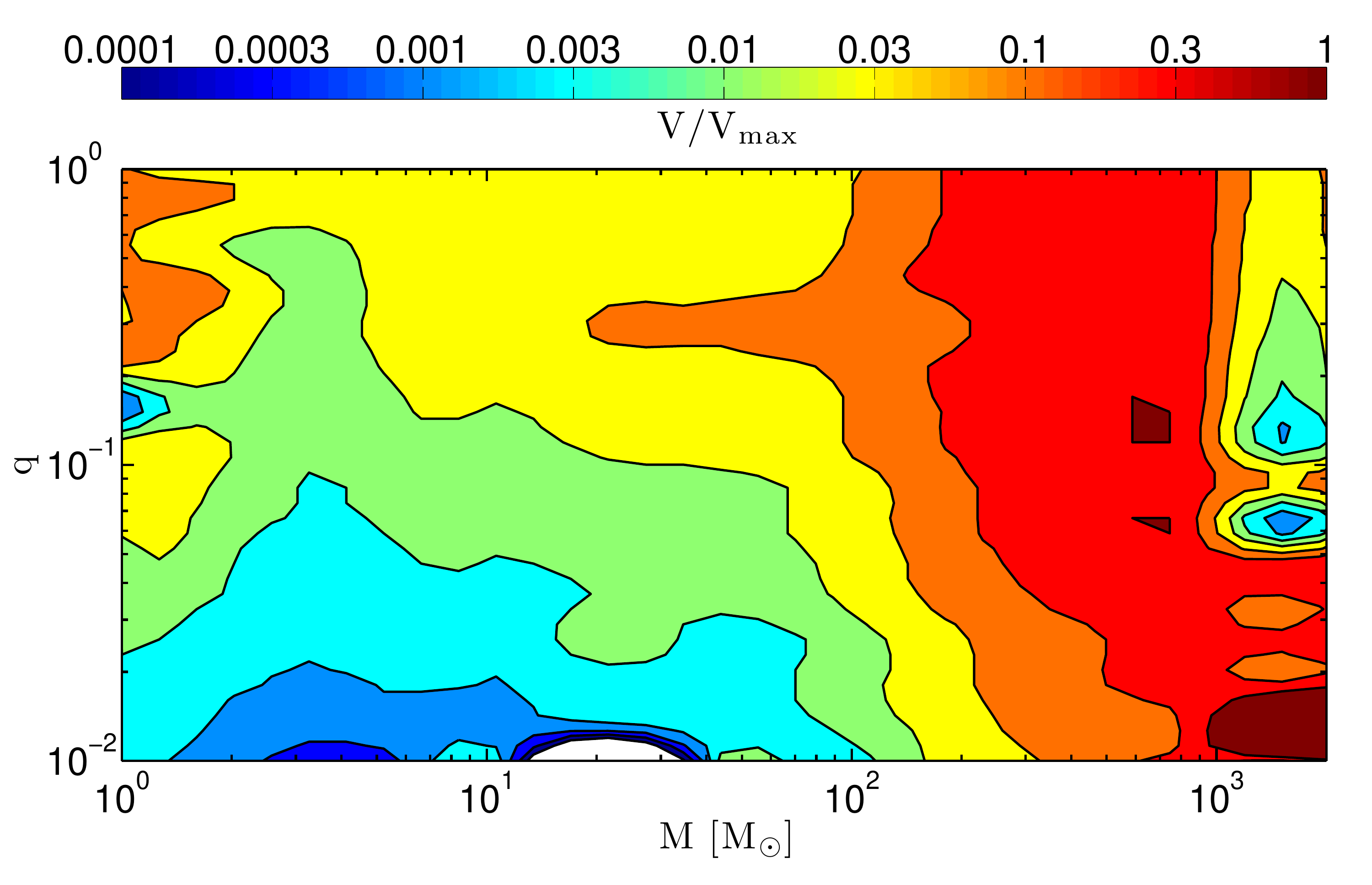}
\includegraphics[trim = 0mm 0mm 0mm 0mm, clip, width=.45\textwidth, angle=0]{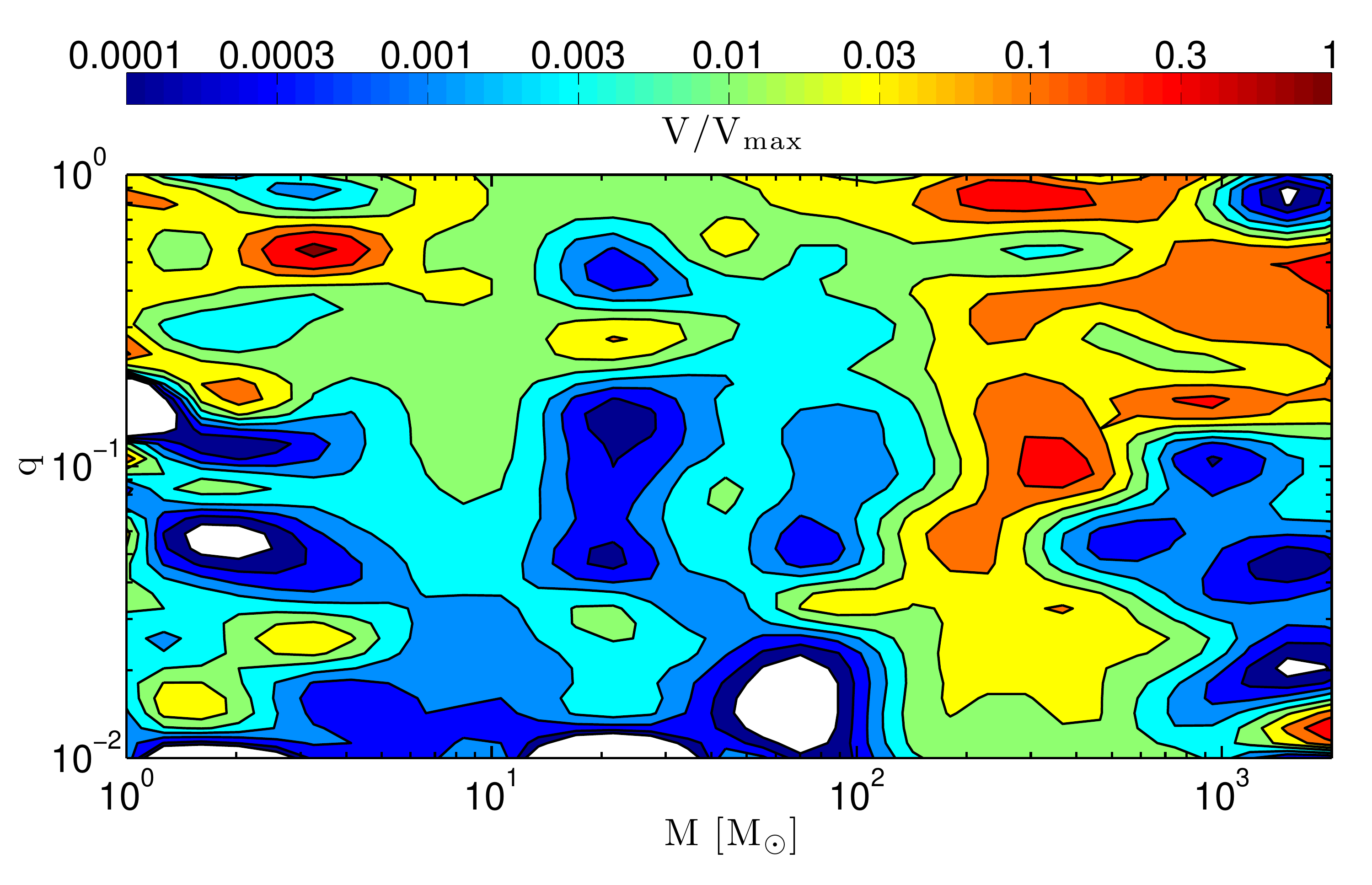}
\caption
{
Contours of detection probability $p\equiv V/V_{\rm max}$ as a function of rest mass $M$ and mass ratio $q$ for Advanced LIGO 
for a source inside the optimal filtering distance horizon, using sine-Gaussian (top panel) and ringdown (bottom panel)
templates for an initial pericenter separation of $r_p=6M$.
}
\label{fig:M-q-ADVL-V}
\end{figure}

The relative performance of SG and RD is further demonstrated
in Figs.~\ref{fig:M-q-ELIGO-V} and \ref{fig:M-q-ADVL-V}, which show the detection probabilities of each template (equivalently,
the ratio of the detectable volume using the templates to the volume using optimal filtering).  SG templates perform best for
$M\approx 200\,\Msun$ systems using Enhanced LIGO and $M\approx 1000\,\Msun$ systems using Advanced LIGO, largely independent
of the mass ratio.  Interestingly, RD templates perform best for comparable mass binaries regardless of total mass for Enhanced LIGO,
whereas no such clear general behavior is observed for Advanced LIGO.  This can be understood because Enhanced LIGO always has fewer
cycles in band than Advanced LIGO, so that the merger-ringdown constitutes a larger fraction of the total SNR, with that fraction
further enhanced for comparable masses (since $\rho \propto \eta = m_1m_2/M^2$ for inspirals, but $\rho \propto \sqrt{\eta}$ for ringdowns \cite{McWilliams:2010eq}).
Advanced LIGO shows no such behavior because the number of inspiral cycles is so large that the merger-ringdown rarely dominates the total SNR.

\begin{figure}[t]
\includegraphics[trim = 0mm 0mm 0mm 0mm, clip, width=.45\textwidth, angle=0]{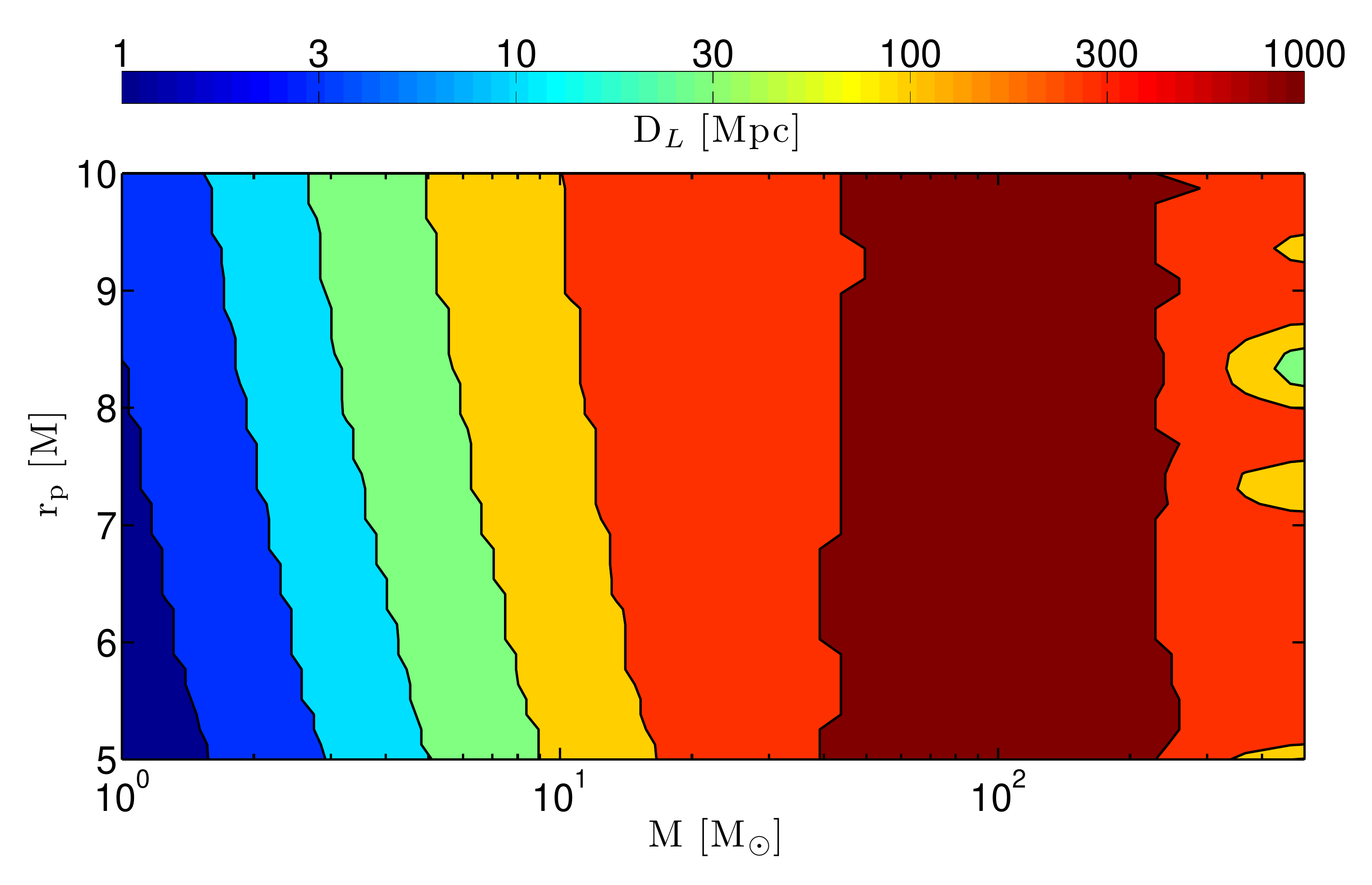}
\includegraphics[trim = 0mm 0mm 0mm 0mm, clip, width=.45\textwidth, angle=0]{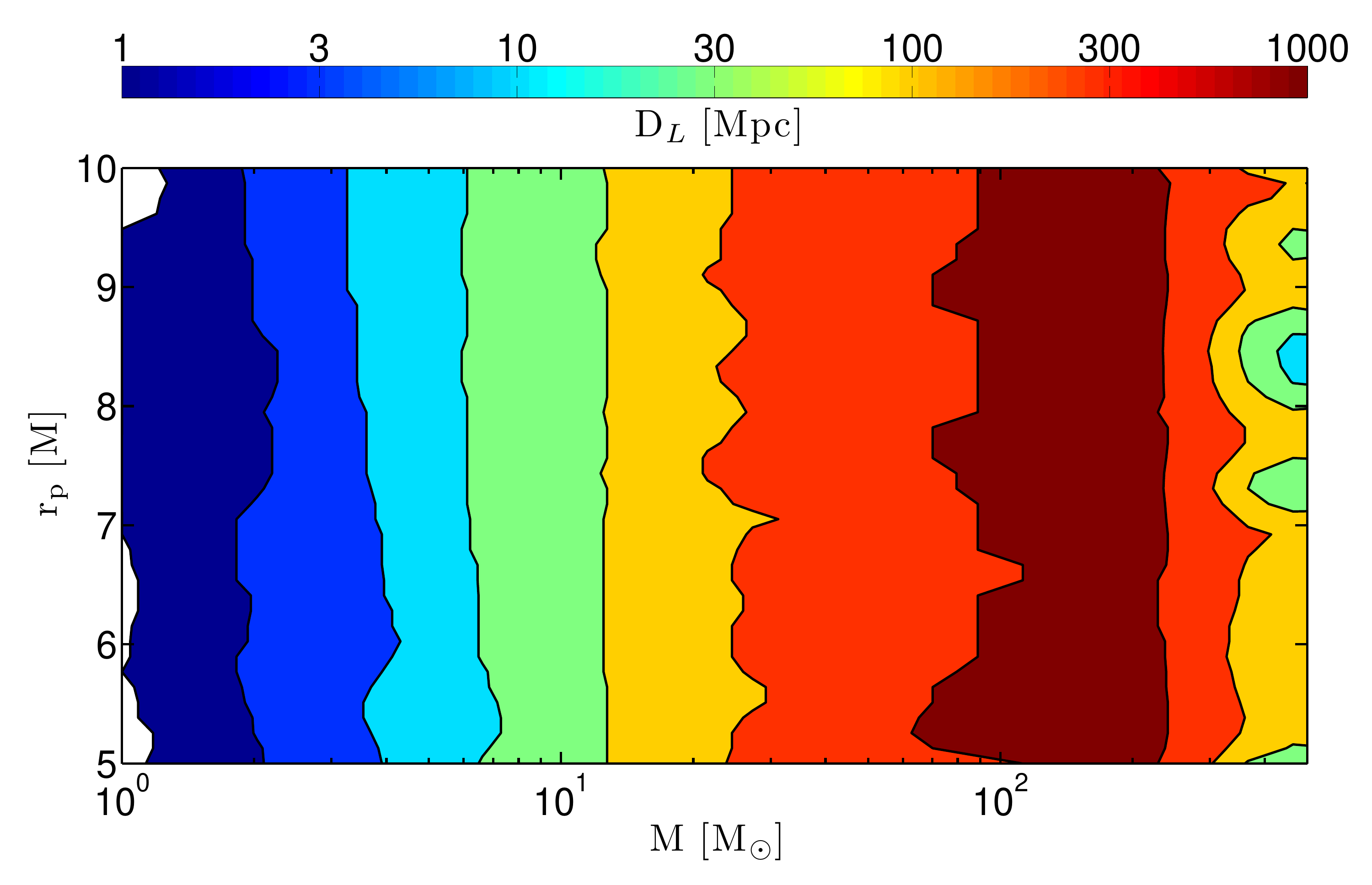}
\includegraphics[trim = 0mm 0mm 0mm 0mm, clip, width=.45\textwidth, angle=0]{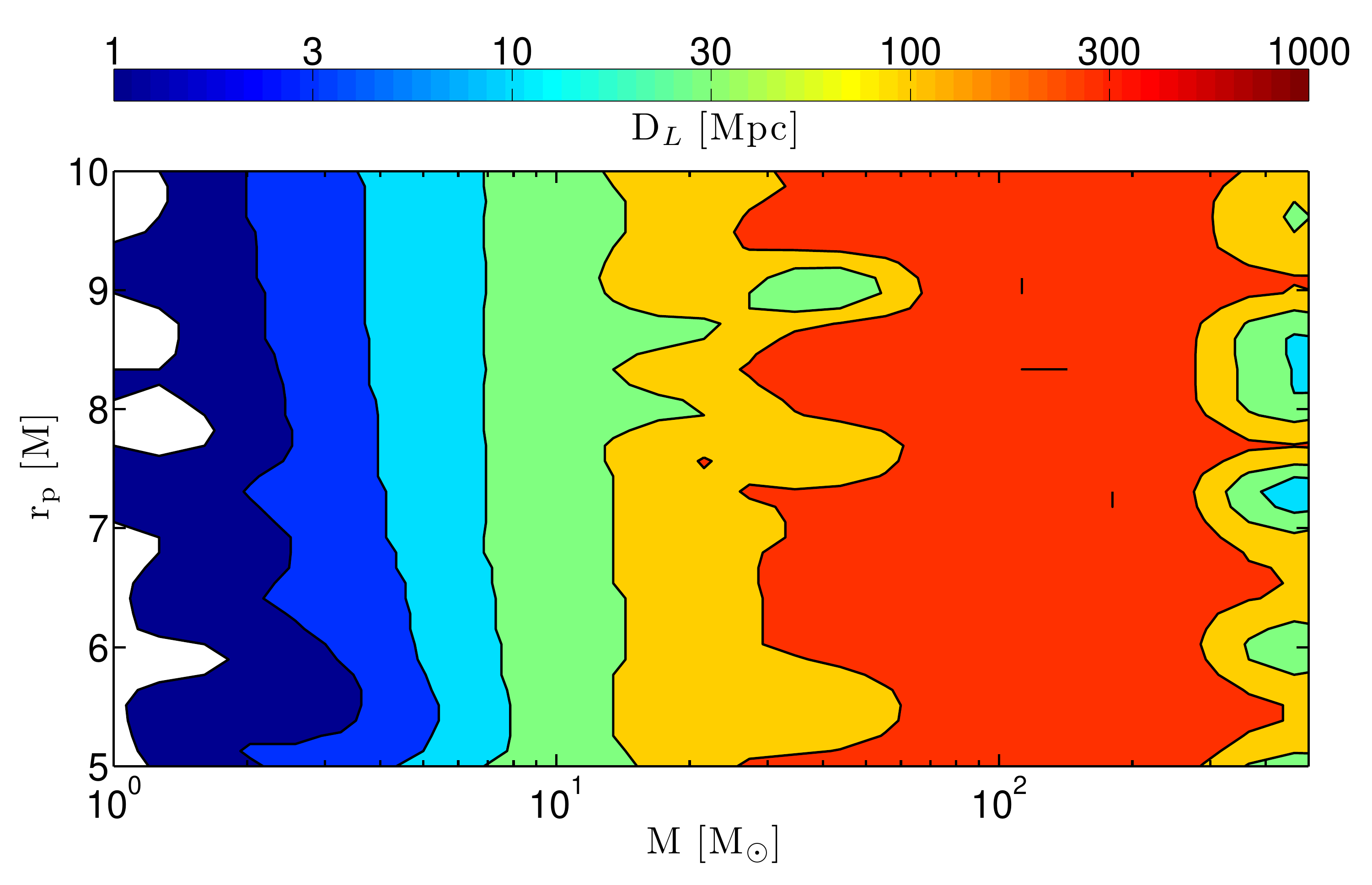}
\caption
{
Contours of horizon distance as a function of rest mass $M$ and pericenter separation $r_p$ for Enhanced LIGO using an optimal filter (top panel), 
sine-Gaussian templates (middle panel), and ringdown templates (bottom panel).  The mass ratio is $q=1$.
}
\label{fig:M-r-ELIGO-DL}
\end{figure}

\begin{figure}[t]
\includegraphics[trim = 0mm 0mm 0mm 0mm, clip, width=.45\textwidth, angle=0]{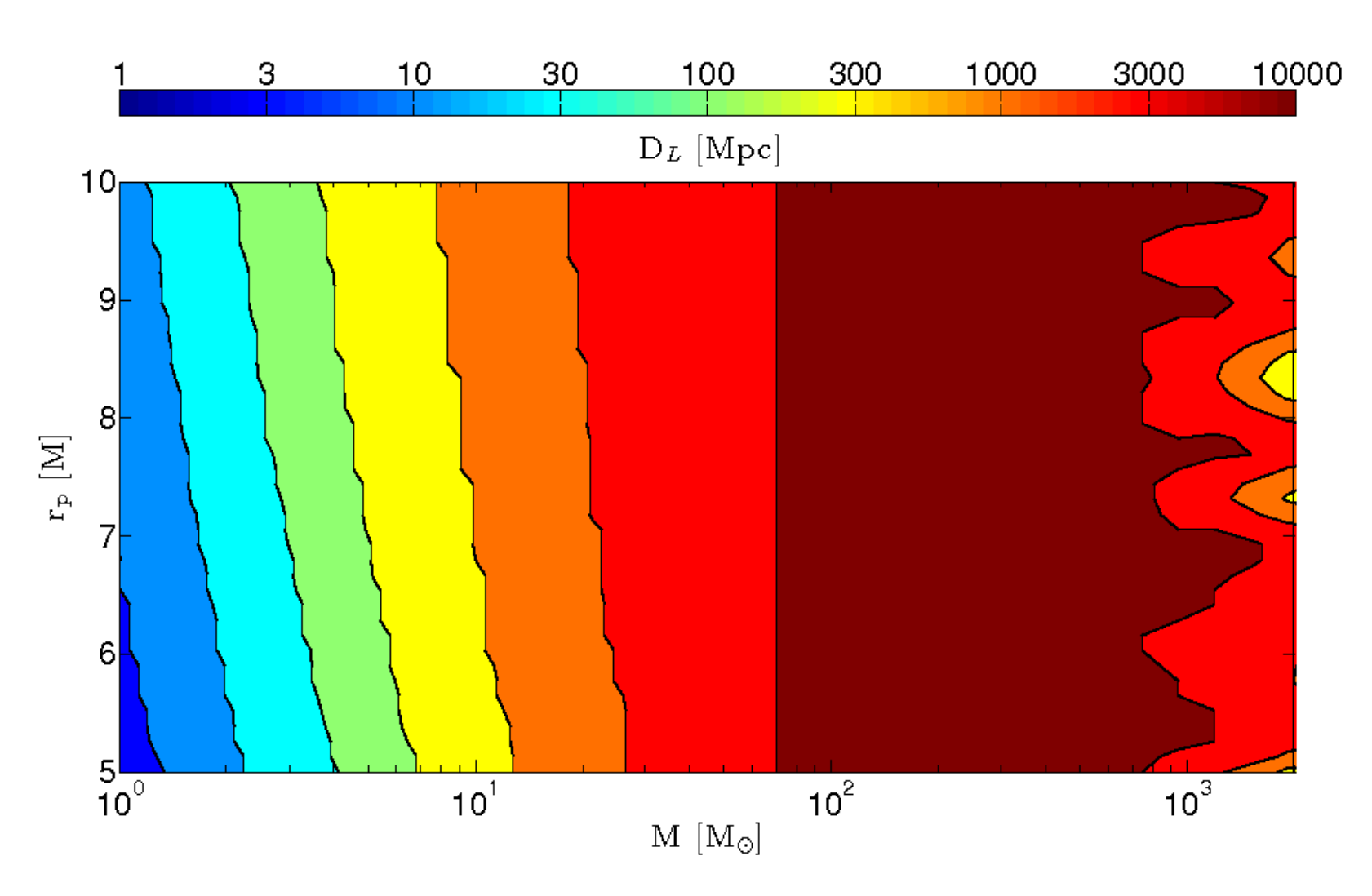}
\includegraphics[trim = 0mm 0mm 0mm 0mm, clip, width=.45\textwidth, angle=0]{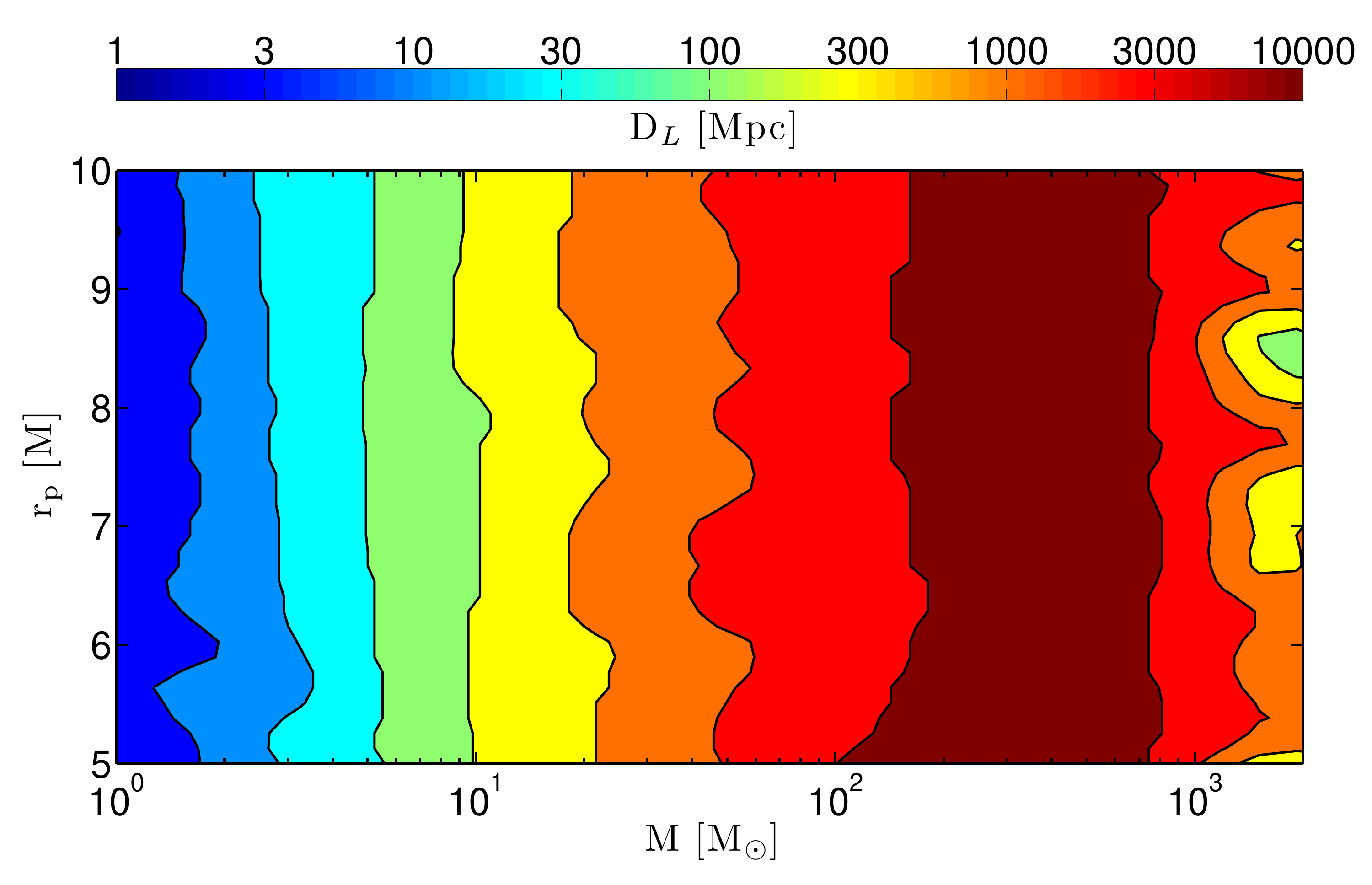}
\includegraphics[trim = 0mm 0mm 0mm 0mm, clip, width=.45\textwidth, angle=0]{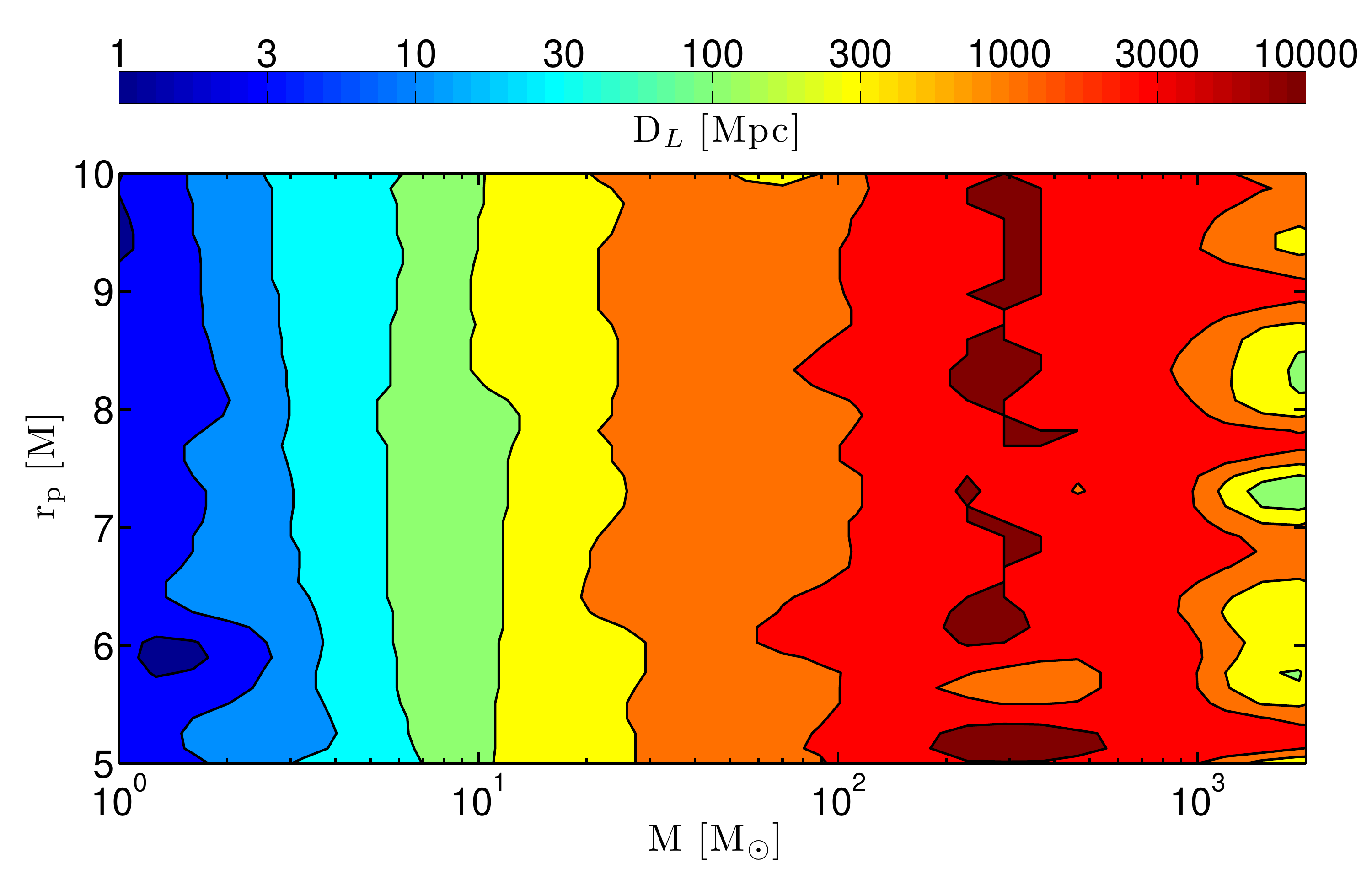}
\includegraphics[trim = 0mm 0mm 0mm 0mm, clip, width=.45\textwidth, angle=0]{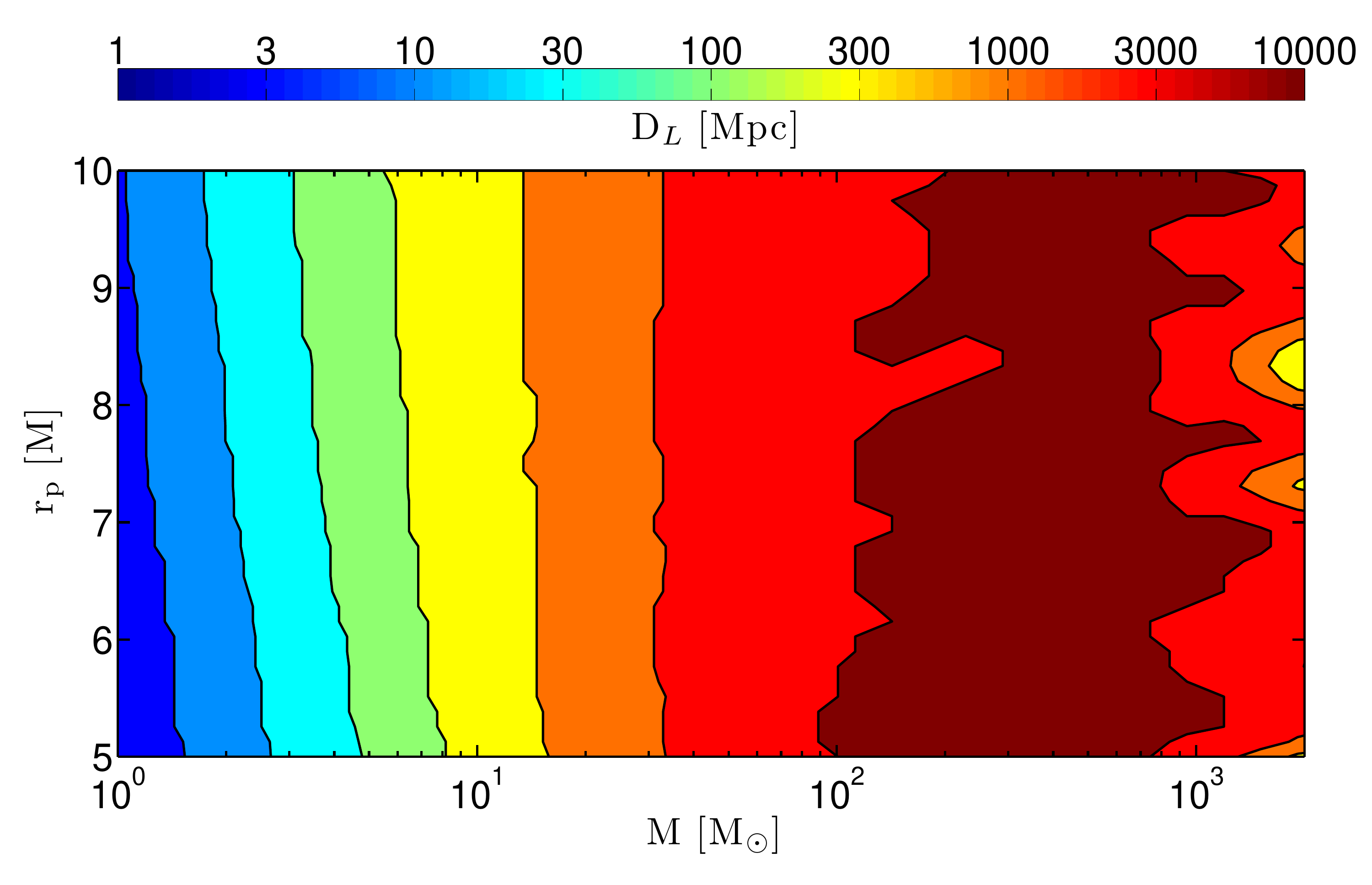}
\caption
{
Contours of horizon distance as a function of rest mass $M$ and pericenter separation $r_p$ for Advanced LIGO using, from top to bottom,
an optimal filter,
sine-Gaussian templates, ringdown templates, and a power-stacking search. The mass ratio is $q=1$.
}
\label{fig:M-r-ADVL-DL}
\end{figure}

\begin{figure}[t]
\includegraphics[trim = 0mm 0mm 0mm 0mm, clip, width=.45\textwidth, angle=0]{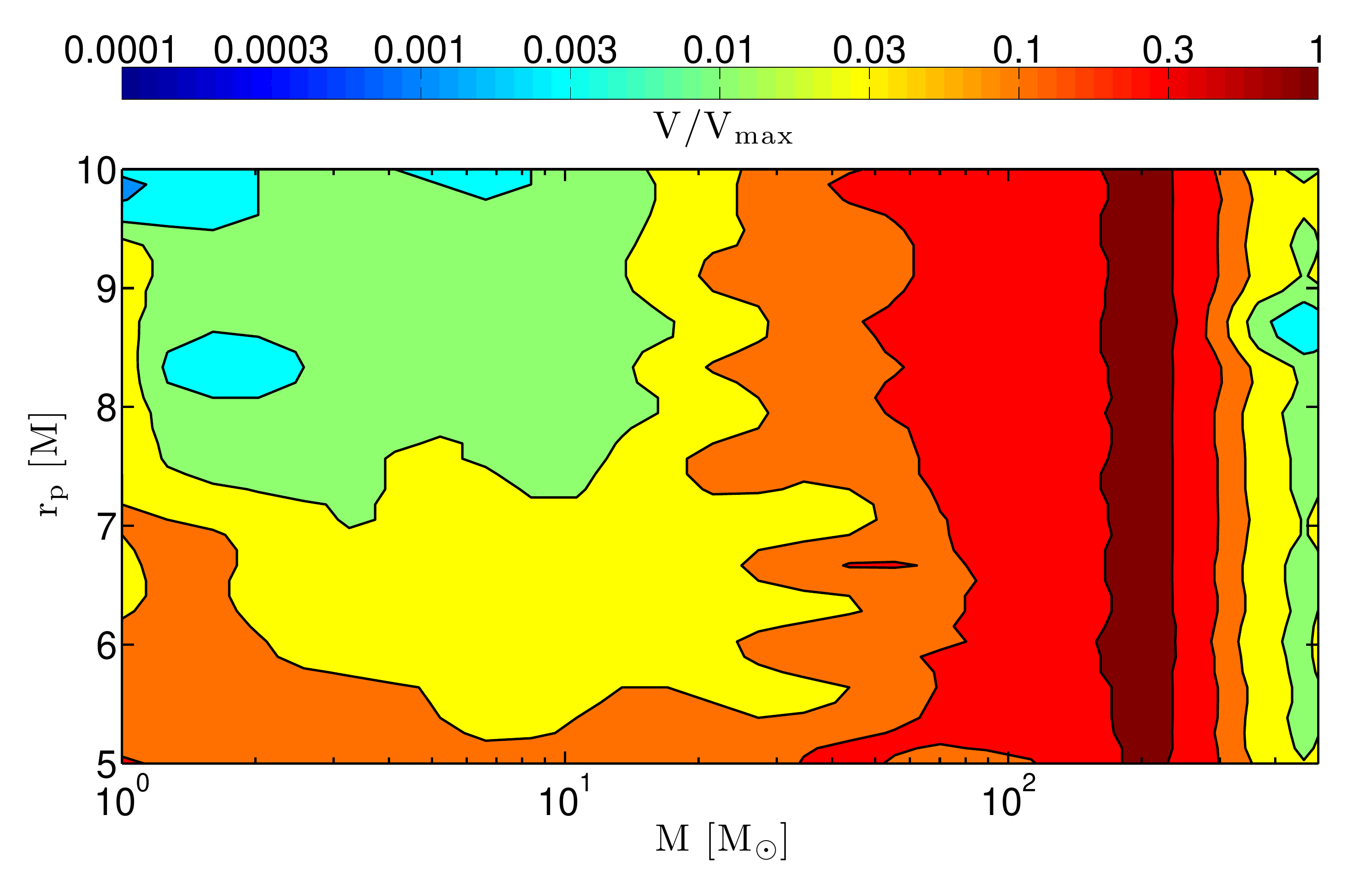}
\includegraphics[trim = 0mm 0mm 0mm 0mm, clip, width=.45\textwidth, angle=0]{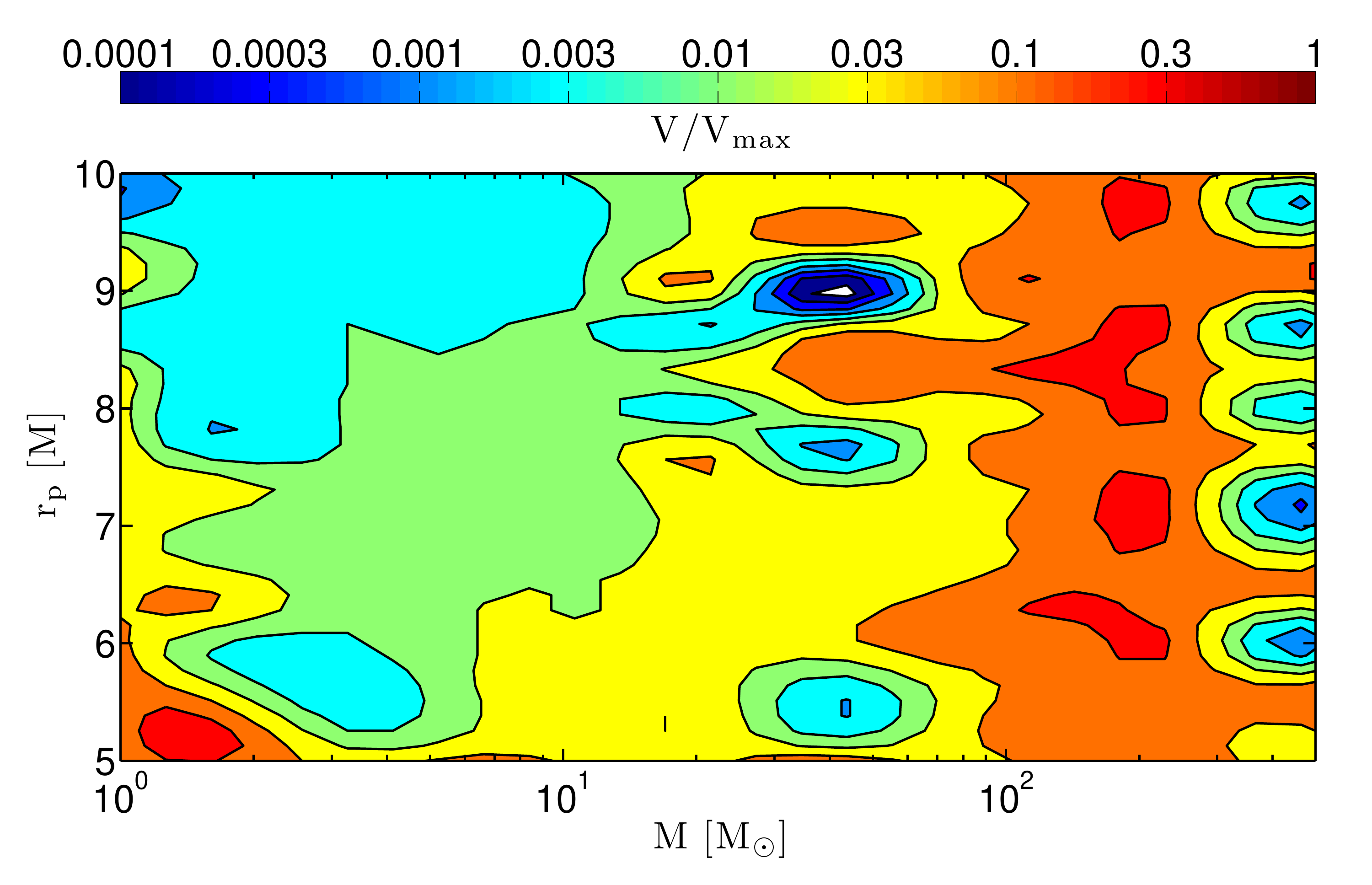}
\caption
{
Contours of detection probability as a function of rest mass $M$ and pericenter separation $r_p$ for Enhanced LIGO
for a source inside the optimal filtering distance horizon, using sine-Gaussian (top panel) and ringdown (bottom panel)
templates. The mass ratio is $q=1$.
}
\label{fig:M-r-ELIGO-V}
\end{figure}

\begin{figure}[t]
\includegraphics[trim = 0mm 0mm 0mm 0mm, clip, width=.45\textwidth, angle=0]{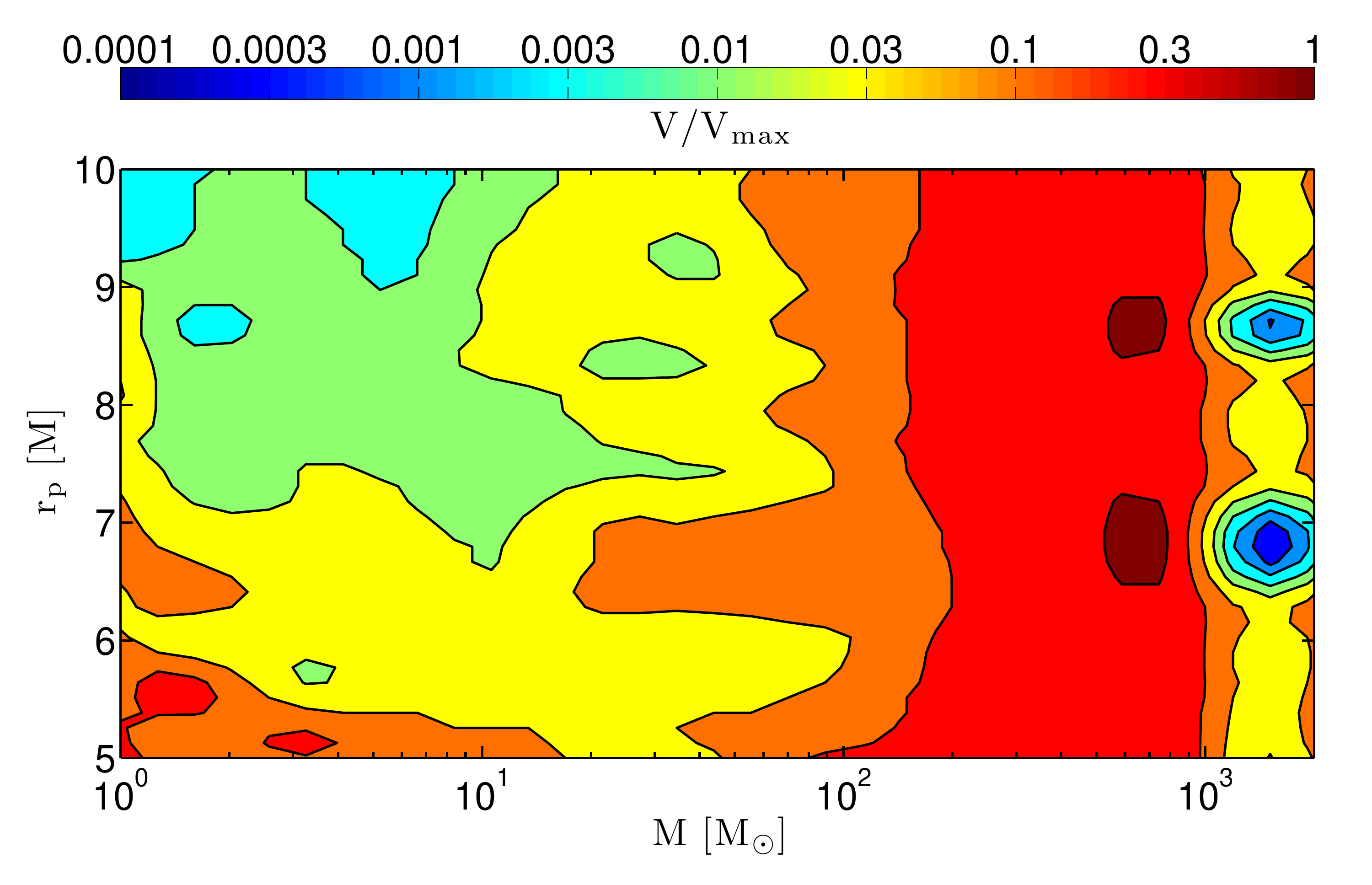}
\includegraphics[trim = 0mm 0mm 0mm 0mm, clip, width=.45\textwidth, angle=0]{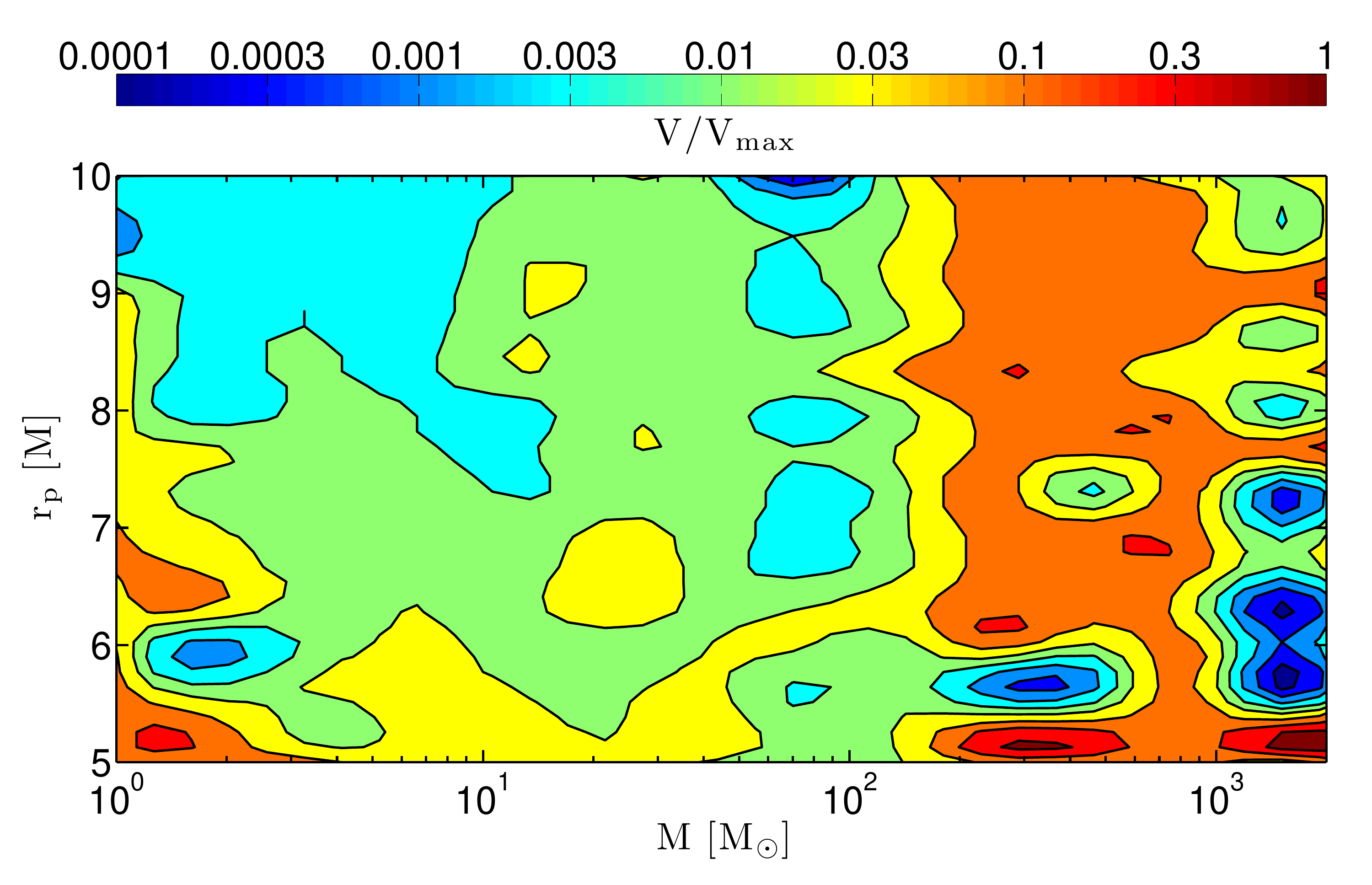}
\includegraphics[trim = 0mm 0mm 0mm 0mm, clip, width=.45\textwidth, angle=0]{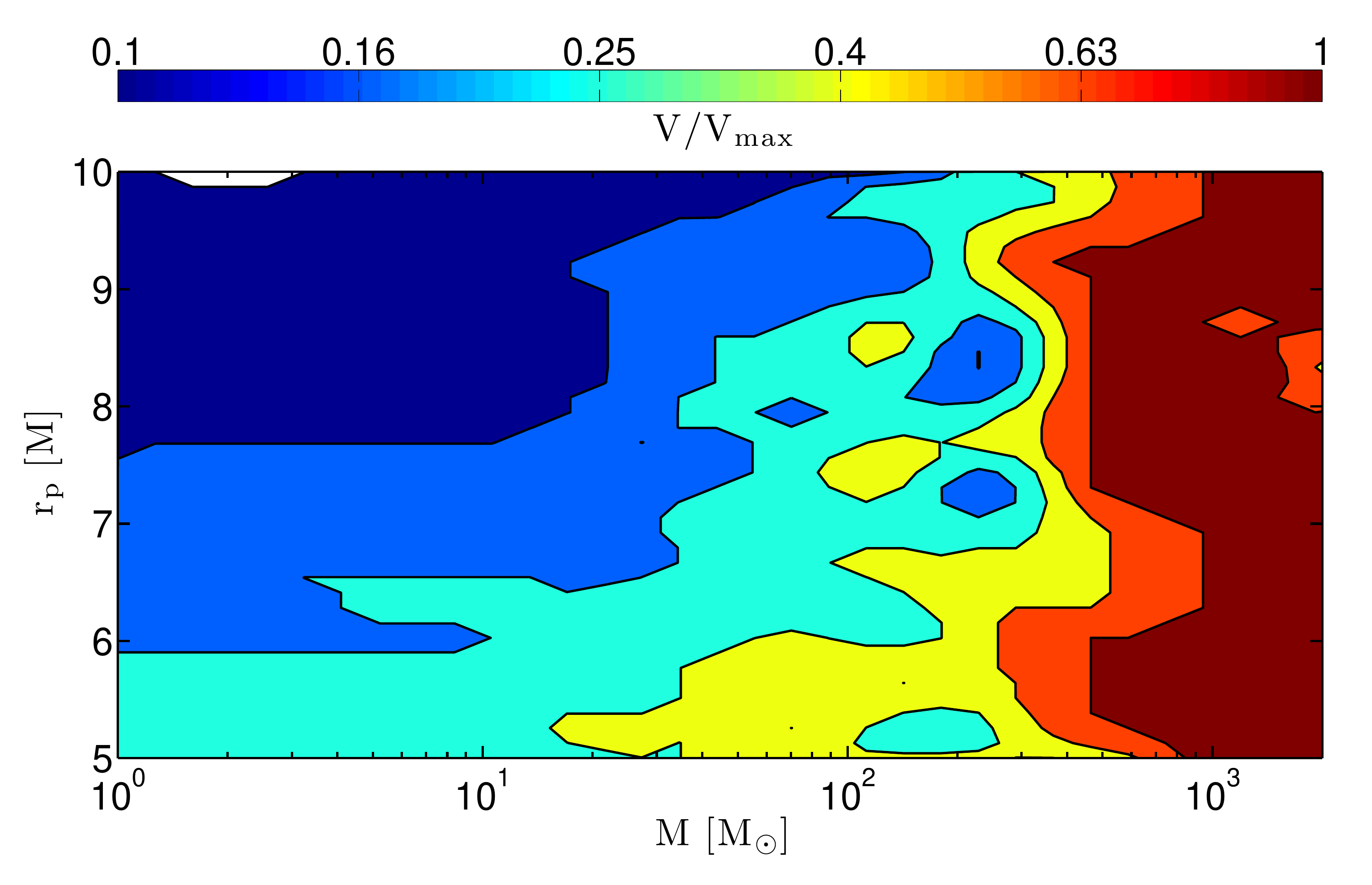}
\caption
{
Contours of detection probability as a function of rest mass $M$ and initial pericenter separation $r_p$ for Advanced LIGO for a source inside 
the optimal-filtering horizon distance, using sine-Gaussian (top panel) and ringdown (middle panel) templates and 
a power-stacking search (bottom panel). The mass ratio is $q=1$. Note the different scale in the bottom figure.
}
\label{fig:M-r-ADVL-V}
\end{figure}

Figures~\ref{fig:M-r-ELIGO-DL} and \ref{fig:M-r-ADVL-DL} again show contours of horizon distance for Enhanced and Advanced LIGO, but as a function
of total mass and initial pericenter distance at fixed $q=1$.  Both $q$ and $M$ rather strongly affect the detectability
of sources over the range of masses considered. $r_p$ moderately affects the detectability for lower
mass systems ($M\lesssim 20\,\Msun$), though very little for higher mass systems (which is expected since
the number of bursts varies significantly with $r_p$ in the range $5M<r_p<10M$, but as the
mass increases fewer of the initial bursts are in band).
SG templates outperform RD templates for all but
the extremely high-mass systems, and a small region of extremely low-mass systems with very small $r_p$, that merge after $\mathcal{O}$(1)
orbit.  This is also clear in Figs.~\ref{fig:M-r-ELIGO-V} and \ref{fig:M-r-ADVL-V}, which show the corresponding
detection probabilities.  In addition to SG and RD templates,
Figs.~\ref{fig:M-r-ADVL-DL} and \ref{fig:M-r-ADVL-V} show the relative performance of a stacked power search, which readily outperforms
burst template searches for the full range of parameters.  Since this is the case for $q=1$, it will apply more so for cases with $q<1$, as they
experience more cycles, so that we can conclude that a power-stacking search will always outperform a burst search and is likely
to be the optimal search approach in the absence of a matched-filter bank. 
SG and RD templates perform best for $M$ in the range $100\,\Msun$--$200\,\Msun$ for both Enhanced and
Advanced LIGO, with the range of horizon distances being the same as in the $M-q$ plots. This is as
expected given that setting $q=1$ maximizes the signal power at fixed
$M$ and $r_p$.  

As shown in Fig.~\ref{fig:M-r-ADVL-V},
all three search methods approach optimal-filter performance for large masses, $M\gsim 500\,\Msun$, since
all three methods benefit from having the SNR concentrated in a small number of cycles.  However,
for lower masses and therefore a larger number of in-band cycles, the SG and RD template performances
degrade much more rapidly than power stacking.  Whereas SG and RD templates reach detection probabilities as low as 0.01\%, power stacking remains above 10\% for the full parameter space considered.
Since our power-stacking estimate is an idealization, uncertainties in
the timing and frequencies of eccentric bursts may degrade the performance
of a true TF power-stacked search. On the other hand, the results of~\cite{Kalmus}
suggest this method is rather robust to timing uncertainties that are smaller
than the characteristic time of each burst .

\section{Conclusions}
\label{sec:conc}

We have developed a novel waveform model for eccentric binary gravitational waveforms which can be 
applied for $r_p \leq 10\, M$, where conventional
post-Newtonian waveforms fail. Such binaries may form through dynamical capture
in dense stellar environments.
Our model is not sufficiently accurate to generate a matched-filter bank, 
and doing so will be very challenging for large eccentricities. 
However, the model is adequate to supply mock signals to explore the performance of 
existing LIGO searches in detecting highly eccentric binary systems. Of existing
search strategies, the ringdown and burst searches are best adapted to these
systems. However, we find that a large fraction of the parameter space,
where we included the impact parameter $5\le r_{p} \le 10 M$ (see
~\cite{Kocsis_Levin} for a complementary study of $r_{p} \ge 10M$), total mass $M\in [1,2000] \Msun$ and 
mass ratio $q\in[0.01,1]$, has a significantly
smaller horizon distance than what is, in principle, achievable with a
matched-filter search. This implies that a corresponding volume of sources could have
been missed in prior searches and may be missed in future searches if better adapted strategies
are not employed. 

Though it may be impractical to construct templates in the near future (via numerical
or analytical methods) for these systems that are accurate enough for optimal
searches, a refinement of the waveform model presented here should be adequate
for informing a power-stacking search. This method has the potential to increase SNR
by $\approx N^{1/4}$ for an $N$-burst event compared to a single-burst search. Though less than
the effective $N^{1/2}$ scaling of a full template search, this would still be a significant improvement.
Note also that even for systems with larger impact parameters that do evolve to an essentially quasicircular
inspiral following the burst phase, for most expected binary parameters the burst phase will
be within the band of ground-based detectors. Thus, the quasicircular inspiral phase will be truncated compared to a primordial 
quasicircular inspiral, and though such a system may still be detectable with a quasicircular template,
it would of course be misidentified, and a bias would be introduced in the estimation of the binary parameters.

For future work, we intend to implement a power-stack search using this waveform model to fully
explore the efficacy of this method and its (in)sensitivity to timing errors, as well as continue
to refine the model to include (for example) spin precession and finite body effects for neutron stars.
We mentioned that the standard PN equations are ill suited to studying the late stages of mergers,
in particular for high-eccentricity binaries, motivating our development of the effective Kerr with radiation-reaction
model described here. However, the EOB approach~\cite{Buonanno:1998gg} is an alternative
to the PN expansion that is well behaved all the way to merger for quasicircular orbits. This approach has 
recently been extended to generic orbits~\cite{Bini:2012ji}, and it will be interesting to explore
EOB as the basis for a repeated burst waveform model.

\acknowledgments
We would like to thank the participants
of the KITP ``Rattle and Shine'' conference (July 2012) for useful discussions.
This research was supported by NSF Grants No. PHY-0745779 (F.P.) and No. AST-0908365 (J.L.), 
the Alfred P. Sloan Foundation (F.P.),
the Simons Foundation (F.P.), and a KITP Scholarship under Grant No. NSF PHY05-51164 (J.L.).
Simulations were run using XSEDE resources provided by NICS under
Grant No. TG-PHY100053.

\clearpage

\bibliographystyle{h-physrev.bst}

\bibliography{references}

\end{document}